\documentclass[11pt,english]{article}

\usepackage[latin9]{inputenc}
\usepackage{amsmath}
\usepackage{amssymb}
\usepackage{graphicx}
\usepackage{geometry}
\usepackage{setspace}
\makeatletter

\providecommand{\tabularnewline}{\\}

\@ifundefined{date}{}{\date{}}
\usepackage{array}
\usepackage{mathrsfs}
\usepackage{color}

\usepackage{rotating}

\usepackage[table]{xcolor}
\usepackage{colortbl}

\makeatother

\usepackage{babel}
\begin{document}
\title{\textbf{Assessing Risk Heterogeneity through}\\
\textbf{Heavy-Tailed Frequency and Severity Mixtures}}
\author{Michael R. Powers\thanks{Corresponding author; Department of Finance, School of Economics and
Management, Tsinghua University, Beijing, China 100084; email: powers@sem.tsinghua.edu.cn.} \ and Jiaxin Xu\thanks{Organization Department, CPC Yichang Municipal Committee, Yichang,
Hubei, China; email: jiaxinxucq@163.com.}}
\date{November 5, 2025}
\maketitle
\begin{abstract}
\begin{singlespace}
\noindent The analysis of risk typically involves dividing a random
damage-generation process into separate frequency (event-count) and
severity (damage-magnitude) components. In the present article, we
construct canonical families of mixture distributions for each of
these components, based on a Negative Binomial kernel for frequencies
and a Gamma kernel for severities. These mixtures are employed to
assess the heterogeneity of risk factors underlying an empirical distribution
through the shape of the implied mixing distribution. From the duality
of the Negative Binomial and Gamma distributions, we first derive
necessary and sufficient conditions for heavy-tailed (i.e., inverse
power-law) canonical mixtures. We then formulate flexible 4-parameter
families of mixing distributions for Geometric and Exponential kernels
to generate heavy-tailed 4-parameter mixture models, and extend these
mixtures to arbitrary Negative Binomial and Gamma kernels, respectively,
yielding 5-parameter mixtures for detecting and measuring risk heterogeneity.
To check the robustness of such heterogeneity inferences, we show
how a fitted 5-parameter model may be re-expressed in terms of alternative
Negative Binomial or Gamma kernels whose associated mixing distributions
form a ``calibrated'' family.\medskip{}

\end{singlespace}
\end{abstract}

\section{Introduction}

\noindent The analysis of risk -- that is, any random process resulting
in potential financial or other damage -- typically involves dividing
the process into separate \emph{frequency} and \emph{severity} components.
Frequencies represent the numbers of damage-causing events to occur
within specified time periods, and generally are modeled by nonnegative
discrete random variables (i.e., $X\in\mathbb{Z}_{\geq0}$). Severities,
on the other hand, measure the individual damage amounts associated
with these events, and typically are modeled by nonnegative continuous
random variables (i.e., $Y\in\mathbb{R}_{\geq0}$) denoting losses
in money, years of human life/productivity, land area, etc. The terms
``frequency'' and ``severity'' are commonly used in operational
risk management and actuarial finance, whereas other disciplines (e.g.,
the medical and geophysical sciences) often use alternatives such
as \emph{rate}, \emph{probability}, or \emph{likelihood} for frequency
(where the last two words essentially imply $X\sim\textrm{Bernoulli}\left(p\right)$
during some limited time period) and \emph{intensity}, \emph{impact},
or \emph{consequence} for severity. Although modeled by distinct random
processes, the frequency and severity components are not necessarily
statistically independent.

For the simplest frequency and severity distributions, the associated
probability mass functions (PMFs) and probability density functions
(PDFs) are strictly decreasing. In such cases, one can imagine modeling
them as continuous mixtures of 1-parameter $\textrm{Geometric}\left(q\right)$
and $\textrm{Exponential}\left(\theta\right)$ kernels, respectively.
In other words, the frequency PMF would be given by
\begin{equation}
f_{X}\left(x\right)={\displaystyle \int_{0}^{1}}f_{X\mid q}^{\left(\textrm{G}\right)}\left(x\right)g_{q}\left(q\right)dq,\:x\in\mathbb{Z}_{\geq0},
\end{equation}
where $f_{X\mid q}^{\left(\textrm{G}\right)}\left(x\right)=\left(1-q\right)q^{x}$
is the $\textrm{Geometric}\left(q\right)$ PMF and $g_{q}\left(q\right)$
denotes the mixing PDF for $q\in\left(0,1\right)$; and the severity
PDF would be given by
\begin{equation}
f_{Y}\left(y\right)={\displaystyle \int_{0}^{\infty}}f_{Y\mid\theta}^{\left(\textrm{E}\right)}\left(y\right)g_{\theta}\left(\theta\right)d\theta,\:y\in\mathbb{R}_{\geq0},
\end{equation}
where $f_{Y\mid\theta}^{\left(\textrm{E}\right)}\left(y\right)=\tfrac{e^{-y/\theta}}{\theta}$
is the $\textrm{Exponential}\left(\theta\right)$ PDF and $g_{\theta}\left(\theta\right)$
denotes the mixing PDF for $\theta\in\mathbb{R}_{>0}$.\footnote{We parameterize the $\textrm{Geometric}\left(q\right)$ distribution
using $q$ to represent the probability of a ``failure'' prior to
the $r^{\textrm{th}}$ ``success'', and the $\textrm{Exponential}\left(\theta\right)$
distribution using $\theta$ to denote the mean. This is so the means
of both distributions are increasing functions of the indicated parameters.}

Apposite illustrations are provided by the world of insurance. First,
suppose a commercial policyholder\textquoteright s medical-expense
loss frequency is given by $X\mid q\sim\textrm{Geometric}\left(q\right)$,
and the mean frequency, $\textrm{E}_{X\mid q}\left[X\right]=\tfrac{q}{1-q}$,
varies with the policyholder's number of employees, such that $q\mid a=1,b=1\sim\textrm{Beta}\left(a=1,b=1\right)\equiv\textrm{Uniform}\left(0,1\right)$.
In this case, the unconditional medical-expense frequency, $X$, is
a $\textrm{Waring}\left(a=1,b=1\right)\equiv\textrm{Yule}\left(b=1\right)$
random variable (with $f_{X\mid a=1,b=1}^{\left(\textrm{W}\right)}\left(x\right)=\tfrac{1}{\left(x+1\right)\left(x+2\right)}$
and\linebreak{}
$\textrm{E}_{X\mid a=1,b=1}^{\left(\textrm{W}\right)}\left[X\right]=\infty$).
Next, suppose a policyholder\textquoteright s liability loss severity
is modeled as $Y\mid\theta\sim\textrm{Exponential}\left(\theta\right)$,
but that, because of underwriting problems, the insurance company
cannot estimate $\theta$ accurately. Suppose further that the company
knows this parameter varies randomly among its policyholders like
an Exponential random variable with mean $\mu$. In that case, the
unconditional distribution of the liability severity, $Y$, is a $\textrm{Pareto 2}\left(\alpha=1,\beta=1\right)$
random variable (with $f_{Y\mid\alpha=1,\beta=1}^{\left(\textrm{P2}\right)}\left(y\right)=\tfrac{1}{\left(y+1\right)^{2}}$
and $\textrm{E}_{Y\mid\alpha=1,\beta=1}^{\left(\textrm{P2}\right)}\left[Y\right]=\infty$).

Naturally, not all frequency PMFs and severity PDFs are strictly decreasing
in the manner of mixtures (1) and (2). In many cases, such as Poisson
frequencies, the relevant distribution possesses a non-zero mode.
Nevertheless, the great majority of frequency and severity distributions
tend to be unimodal, and so it is reasonable to consider the following
generalizations of (1) and (2) in which the 1-parameter $\textrm{Geometric}\left(q\right)$
and $\textrm{Exponential}\left(\theta\right)$ kernels are replaced
by the 2-parameter $\textrm{Negative Binomial}\left(r,q\right)$ and
$\textrm{Gamma}\left(r,\theta\right)$ kernels, with fixed shape parameter
($r$), respectively:
\begin{equation}
f_{X\mid r}\left(x\right)={\displaystyle \int_{0}^{1}}f_{X\mid r,q}^{\left(\textrm{NB}\right)}\left(x\right)g_{q}\left(q\right)dq,
\end{equation}
where $f_{X\mid r,q}^{\left(\textrm{NB}\right)}\left(x\right)=\tfrac{\Gamma\left(r+x\right)}{\Gamma\left(r\right)\Gamma\left(x+1\right)}\left(1-q\right)^{r}q^{x}$;
and
\begin{equation}
f_{Y\mid r}\left(y\right)={\displaystyle \int_{0}^{\infty}}f_{Y\mid r,\theta}^{\left(\Gamma\right)}\left(y\right)g_{\theta}\left(\theta\right)d\theta,
\end{equation}
where $f_{Y\mid r,\theta}^{\left(\Gamma\right)}\left(x\right)=\tfrac{y^{r-1}e^{-y/\theta}}{\Gamma\left(r\right)\theta^{r}}$.
These models, in which the $r$ parameter explicitly allows for an
interior mode, are reasonable for most frequency and severity distributions
encountered in practice, and constitute the principal focus of the
present research. We therefore refer to (3) and (4) as families of
\emph{canonical} frequency and severity mixtures, respectively; and
to the $\textrm{Negative Binomial}\left(r,q\right)$ and $\textrm{Gamma}\left(r,\theta\right)$
distributions as the corresponding canonical kernels.

In the present article, we employ canonical mixture distributions
to assess the heterogeneity of risk factors underlying an empirical
frequency or severity distribution by estimating the shape of the
implied mixing distribution, $g_{q}\left(q\right)$ or $g_{\theta}\left(\theta\right)$.
This involves paying particular attention to any relatively large
collection of weight (i.e., probability mass or density) at the upper
end of a mixing distribution's sample space, which can have a profound
impact on the tail of the resulting mixture. By their nature, mixing
processes almost always magnify the volatility of the relevant frequency
or severity kernel;\footnote{This is reflected in the inequality $\textrm{Var}{}_{W}\left[W\right]=\textrm{E}_{\upsilon}\left[\textrm{Var}{}_{W\mid\upsilon}\left[W\right]\right]+\textrm{Var}{}_{\upsilon}\left[\textrm{E}_{W\mid\upsilon}\left[W\right]\right]\geq\textrm{E}_{\upsilon}\left[\textrm{Var}{}_{W\mid\upsilon}\left[W\right]\right]$
for any mixture $f_{W}\left(w\right)={\textstyle \int_{\Upsilon}f_{W\mid\upsilon}\left(w\right)}d\upsilon$,
as long as all indicated moments are well defined. In other words,
the variance of the mixture distribution tends to be larger than that
of the kernel on average.} and the two insurance examples above reveal how easily the resulting
distribution can be heavy-tailed in the sense of following an inverse
power law (i.e., $\textrm{E}_{X}\left[X^{\kappa}\right]$ or $\textrm{E}_{Y}\left[Y^{\kappa}\right]$
is infinite for some $\kappa\in\mathbb{Z}_{\geq1}$). We therefore
will reserve the term ``heavy-tailed'' for mixture models whose
tails are characterized by an inverse power law, and focus exclusively
on such models.\footnote{In this context, it is important to note that terms such as ``heavier-tailed''
and ``relatively heavy'' may be used to compare the different amounts
of weight in the tails of two or more distributions without suggesting
that any particular distribution follows an inverse power law.} In doing so, we would note that families of heavy-tailed probability
distributions can provide reasonably good fits to lighter-tailed data
simply by selecting the inverse power-law parameter so that $\textrm{E}_{X}\left[X^{\kappa}\right]=\infty$
or $\textrm{E}_{Y}\left[Y^{\kappa}\right]=\infty$ for only very large
values of $\kappa\in\mathbb{Z}_{\geq1}$. However, lighter-tailed
distributions generally cannot model heavy-tailed data effectively.

We begin, in Section 2, by considering the close mathematical connections
between the Negative Binomial and Gamma distributions. This duality,
which is apparent from the similar mathematical forms of their Laplace
transform and other generating functions, facilitates the derivation
of necessary and sufficient conditions for the canonical mixtures
to possess heavy tails. In Section 3, we construct two flexible 4-parameter
families of mixing distributions for producing heavy-tailed frequency
mixtures from a Geometric kernel. Then, in Section 4, we transform
these models to analogous 4-parameter mixing families for generating
heavy-tailed severity mixtures from an Exponential kernel. These mixtures
are extended to arbitrary Negative Binomial and Gamma kernels, respectively,
in Section 5, yielding 5-parameter families for detecting and measuring
risk heterogeneity. To check the robustness of such heterogeneity
inferences, we show how a fitted 5-parameter model may be re-expressed
in terms of alternative Negative Binomial or Gamma kernels whose associated
mixing distributions form a ``calibrated'' family.

\section{Canonical Mixture Distributions}

\subsection{Duality of Negative Binomial and Gamma Distributions}

\noindent It is well known that the $\textrm{Negative Binomial}\left(r,q\right)$
and $\textrm{Gamma}\left(r,\theta\right)$ distributions provide analogous
waiting-time models for the $r^{\textrm{th}}$ event in a (discrete-time)
Bernoulli and (continuous-time) Poisson process, respectively. Therefore,
it is not surprising that their corresponding generating functions
(i.e., moment-generating function, Laplace transform, characteristic
function, etc.) share similar features. For example, if $X\mid r,q\sim\textrm{Negative Binomial}\left(r,q\right)$
and $Y\mid r,\theta\sim\textrm{Gamma}\left(r,\theta\right)$, then
the corresponding Laplace transforms,\footnote{We will work with Laplace transforms, rather than alternative generating
functions, because they are most convenient for characterizing nonnegative
random variables with potentially heavy tails (which can arise in
various mixture distributions).}
\begin{equation}
\mathcal{L}_{X\mid r,q}\left(t\right)=\textrm{E}_{X\mid r,q}\left[e^{-tX}\right]=\left(\dfrac{1-q}{1-qe^{-t}}\right)^{r}
\end{equation}
and
\begin{equation}
\mathcal{L}_{Y\mid r,\theta}\left(u\right)=\textrm{E}_{Y\mid r,\theta}\left[e^{-uY}\right]=\left(\dfrac{1}{1+\theta u}\right)^{r},
\end{equation}
respectively, possess identical functional forms under the transformations
$\theta=\tfrac{q}{1-q}$ and $u=1-e^{-t}$.

The duality implied by (5) and (6) sometimes may be used to transform
results associated with one of the two distributions to comparable,
but more difficult to obtain, results for the other. For example,
Powers and Xu (2024) showed that the two identifiability results stated
in Subsection 2.2 -- for $\textrm{Gamma}\left(r,\theta\right)$ mixtures
with fixed $r$ and $\textrm{Negative Binomial}\left(r,q\right)$
mixtures with fixed, respectively -- immediately imply each other,
although the latter result (by Lüxmann-Ellinghaus, 1987) did not appear
in the literature until more than a quarter century after the former
(by Teicher, 1961). Moreover, Lemma 1 of Subsection 2.3 provides tractable
expressions for the positive-integer moments of $X\mid r,q\sim\textrm{Negative Binomial}\left(r,q\right)$
in terms of corresponding moments of $Y\mid r,\theta\sim\textrm{Gamma}\left(r,\theta\right)$.
This result is used to derive necessary and sufficient conditions
for heavy-tailed Negative Binomial mixtures in Theorem 2.1.

\subsection{Identifiability of Negative Binomial and Gamma Mixtures}

\noindent Let $\mathcal{F}_{\textrm{NB}}$ denote the family of nonnegative
discrete random variables formed as continuous mixtures of a $\textrm{Negative Binomial}\left(r,q\right)$
kernel with fixed $r\in\mathbb{R}_{>0}$ and $q\in\left(0,1\right)$,
and let $\mathcal{F}_{\Gamma}$ denote the family of nonnegative continuous
random variables formed as continuous mixtures of a $\textrm{Gamma}\left(r,\theta\right)$
kernel with fixed $r\in\mathbb{R}_{>0}$ and random $\theta\in\mathbb{R}_{>0}$.
In other words, $\mathcal{F}_{\textrm{NB}}$ contains all $X\mid r\sim f_{X\mid r}\left(x\right)$
satisfying (3) and $\mathcal{F}_{\Gamma}$ contains all $Y\mid r\sim f_{Y\mid r}\left(y\right)$
satisfying (4).

When employing mixture models such as (3) or (4), it often is desirable
to know whether or not the mixed random variable (i.e., $X\mid r\sim f_{X\mid r}\left(x\right)$
or $Y\mid r\sim f_{Y\mid r}\left(y\right)$) can be associated with
a unique mixing distribution ($q\sim g_{q}\left(q\right)$ or $\theta\sim g_{\theta}\left(\theta\right)$,
respectively). This property, known as identifiability, is necessary
if one wishes to estimate the parameters of the mixing distribution
from observations of the mixed random variable (see, e.g., Xekalaki
and Panaretos, 1983). In the present research, identifiability is
crucial because the principal aim is to assess the risk heterogeneity
underlying an empirical frequency or severity distribution through
characteristics of the implied mixing distribution.

The following two results are well known in the research literature.\medskip{}

\noindent\textbf{Theorem 1.1:} Any random variable $X\mid r\in\mathcal{F}_{\textrm{NB}}$
is identifiable.\medskip{}

\noindent\textbf{Proof:} See Lüxmann-Ellinghaus (1987).\medskip{}

\noindent\textbf{Theorem 1.2:} Any random variable $Y\mid r\in\mathcal{F}_{\Gamma}$
is identifiable.\medskip{}

\noindent\textbf{Proof:} See Teicher (1961).\medskip{}

For fixed values of their scale parameters, both the $\textrm{Negative Binomial}\left(r,q\right)$
and $\textrm{Gamma}\left(r,\theta\right)$ distributions are additively
closed with respect to the shape parameter, $r$. Consequently, it
follows from Teicher (1961) that mixtures formed from these kernels
also are identifiable. However, such mixtures are not immediately
useful for the problem at hand because frequency and severity mixture
models generally treat variation in the scale parameter as the principal
source of risk heterogeneity.

Naturally, this does not mean that variation in the shape parameter
is precluded from contributing to heterogeneity; and indeed, bivariate
mixtures such as
\begin{equation}
X\sim f_{X}\left(x\right)=\int_{0}^{\infty}\int_{0}^{1}f_{X\mid r,q}^{\left(\textrm{NB}\right)}\left(x\right)g_{r,q}\left(r,q\right)dqdr
\end{equation}
and
\begin{equation}
Y\sim f_{Y}\left(y\right)=\int_{0}^{\infty}\int_{0}^{\infty}f_{Y\mid r,\theta}^{\left(\Gamma\right)}\left(y\right)g_{r,\theta}\left(r,\theta\right)d\theta dr,
\end{equation}
for joint mixing PDFs $g_{r,q}\left(r,q\right)$ and $g_{r,\theta}\left(r,\theta\right)$,
are reasonable and attractive models. Unfortunately, however, such
models often are not identifiable. For example, if there exists a
set of unique mixing PDFs, $g_{q\mid r}\left(q\right)$, such that
$f_{X}\left(x\right)=f_{X\mid r}\left(x\right)=\int_{0}^{1}f_{X\mid r,q}^{\left(\textrm{NB}\right)}\left(x\right)g_{q\mid r}\left(q\right)dq$
is invariant for all $r$ in some interval $\left(A,B\right)\subset\mathbb{R}^{+}$,
then $g_{r,q}\left(r,q\right)$ cannot be unique because it may by
expressed as $g_{r,q}\left(r,q\right)=g_{q\mid r}\left(q\right)g_{r}\left(r\right)$
for \emph{any} PDF $g_{r}\left(r\right),\:r\in\left(A,B\right)$;
and the same problem arises in the case of $g_{\theta\mid r}\left(\theta\right)$.

In fact, for any specified value of the shape parameter, $r$, and
mixing distribution, $g_{q\mid r}\left(q\right)$ or $g_{\theta\mid r}\left(\theta\right)$,
it is possible to derive sets of unique mixing PDFs, $g_{q\mid s}\left(q\right)$
and $g_{\theta\mid s}\left(\theta\right)$, such that
\[
f_{X}\left(x\right)=\int_{0}^{1}f_{X\mid r,q}^{\left(\textrm{NB}\right)}\left(x\right)g_{q\mid r}\left(q\right)dq
\]
\begin{equation}
=\int_{0}^{1}f_{X\mid s,q}^{\left(\textrm{NB}\right)}\left(x\right)g_{q\mid s}\left(q\right)dq
\end{equation}
and
\[
f_{Y}\left(y\right)=\int_{0}^{\infty}f_{Y\mid r,\theta}^{\left(\Gamma\right)}\left(y\right)g_{\theta\mid r}\left(\theta\right)d\theta
\]
\begin{equation}
=\int_{0}^{\infty}f_{Y\mid s,\theta}^{\left(\Gamma\right)}\left(y\right)g_{\theta\mid s}\left(\theta\right)d\theta
\end{equation}
are invariant over $s\in\left[r,\infty\right)$. As will be shown
in Section 4, the pairs $\left[f_{X\mid s,q}^{\left(\textrm{NB}\right)}\left(x\right),g_{q\mid s}\left(q\right)\right]$
and\linebreak{}
$\left[f_{Y\mid s,\theta}^{\left(\Gamma\right)}\left(y\right),g_{\theta\mid s}\left(\theta\right)\right]$
(to be called ``calibrated'' families for $f_{X}\left(x\right)$
and $f_{Y}\left(y\right)$, respectively) may be used to explore the
dependency of risk heterogeneity on the choice of $r$.

The more general problem of determining whether the bivariate mixtures
of (7) and (8) are identifiable for certain restricted classes of
the joint mixing PDFs ($g_{r,q}\left(r,q\right)$ and $g_{r,\theta}\left(r,\theta\right)$,
respectively) is left for future research.

\subsection{Heavy-Tailed Negative Binomial and Gamma Mixtures}

\noindent Consider the well-known expression for the raw moments of
$Y\mid r,\theta\sim\textrm{Gamma}\left(r,q\right)$: 
\begin{equation}
\textrm{E}_{Y\mid r,\theta}^{\left(\Gamma\right)}\left[Y^{\kappa}\right]=\dfrac{\Gamma\left(r+\kappa\right)\theta^{\kappa}}{\Gamma\left(r\right)},
\end{equation}
for $\kappa\in\mathbb{Z}_{\geq0}$. Although corresponding moments
of $X\mid r,q\sim\textrm{Negative Binomial}\left(r,q\right)$ can
be found in published sources (see, e.g., Johnson, Kemp, and Kotz,
2005 and Weisstein, 2023), their clear connection to (11) is not widely
disseminated. In the following lemma, we employ the duality described
by (5) and (6) to derive expressions for the Negative Binomial raw
moments directly from those in (11). The moments of both distributions
are used to provide necessary and sufficient conditions for the canonical
frequency and severity mixtures to be heavy-tailed (in Theorems 2.1
and 2.2 below).\medskip{}

\noindent\textbf{Lemma 1:} For all $\kappa\in\mathbb{Z}_{\geq0}$,
\[
\textrm{E}_{X\mid r,q}\left[X^{\kappa}\right]={\displaystyle \sum_{i=1}^{\kappa}S\left(\kappa,i\right)\textrm{E}_{Y\mid r,\tfrac{q}{1-q}}\left[Y^{i}\right]}
\]
\[
={\displaystyle \sum_{i=1}^{\kappa}S\left(\kappa,i\right)\dfrac{\Gamma\left(r+i\right)}{\Gamma\left(r\right)}\left(\dfrac{q}{1-q}\right)^{i}},
\]
where the $S\left(\kappa,i\right)$ are Stirling numbers of the second
kind.\medskip{}

\noindent\textbf{Proof:} See Subsection A.1 of Appendix A.\medskip{}

Now let $\mathcal{F}_{\textrm{NB}}^{\textrm{H}}\subset\mathcal{F}_{\textrm{NB}}$
and $\mathcal{F}_{\Gamma}^{\textrm{H}}\subset\mathcal{F}_{\Gamma}$
denote, respectively, the families of canonical mixture random variables
characterized by heavy tails (i.e., an inverse power law). That is,
$X\in\mathcal{F}_{\textrm{NB}}^{\textrm{H}}\Longrightarrow\textrm{E}_{X}\left[X^{\kappa}\right]=\infty$
and $Y\in\mathcal{F}_{\Gamma}^{\textrm{H}}\Longrightarrow\textrm{E}_{Y}\left[Y^{\kappa}\right]=\infty$
for some for some $\kappa\in\mathbb{Z}_{\geq1}$. Furthermore, let
$\mathcal{G}_{\textrm{NB}}^{\textrm{H}}$ and $\mathcal{G}_{\Gamma}^{\textrm{H}}$
denote, respectively, the families of continuous random variables,
$q\sim g_{q}\left(q\right)$ and $\theta\sim g_{\theta}\left(\theta\right)$,
that generate the mixtures in $\mathcal{F}_{\textrm{NB}}^{\textrm{H}}$
and $\mathcal{F}_{\Gamma}^{\textrm{H}}$. For simplicity, we will
restrict attention to differentiable PDFs $g_{q}\left(q\right)$ and
$g_{\theta}\left(\theta\right)$ that do not oscillate at the bounds
of their sample spaces (i.e., as $q$ approaches $0$ or $1$ and
$\theta$ approaches $0$ or $\infty$, respectively). The family
of such PDFs, for which $g_{q}^{\prime}\left(q\right)$ and $g_{\theta}^{\prime}\left(\theta\right)$
are well defined and change sign only a finite number of times, will
be denoted by $\mathcal{D}_{\textrm{FIN}}$.

The following results provide necessary and sufficient conditions
for $q\in\mathcal{G}_{\textrm{NB}}^{\textrm{H}}$ and $\theta\in\mathcal{G}_{\Gamma}^{\textrm{H}}$.\medskip{}

\noindent\textbf{Theorem 2.1:} For $q\in\mathcal{D}_{\textrm{FIN}}$,
the following three statements are equivalent:

\noindent (a) $q\in\mathcal{G}_{\textrm{NB}}^{\textrm{H}}$;

\noindent (b) $\underset{q\uparrow1}{\lim}\:g_{q}\left(q\right)\left(1-q\right)^{1-\rho}=\infty$
for some $\rho\in\mathbb{R}_{>0}$; and

\noindent (c) $\underset{q\uparrow1}{\lim}\:\dfrac{\ln\left(g_{q}\left(q\right)\right)}{\ln\left(1-q\right)}<\infty$.\medskip{}

\noindent\textbf{Proof:} See Subsection A.2 of Appendix A.\medskip{}

\noindent\textbf{Theorem 2.2:} For $\theta\in\mathcal{D}_{\textrm{FIN}}$,
the following three statements are equivalent:

\noindent (a) $\theta\in\mathcal{G}_{\Gamma}^{\textrm{H}}$;

\noindent (b) $\underset{\theta\rightarrow\infty}{\lim}g_{\theta}\left(\theta\right)\theta^{1+\rho}=\infty$
for some $\rho\in\mathbb{R}_{>0}$; and

\noindent (c) $\underset{\theta\rightarrow\infty}{\lim}\dfrac{\ln\left(g_{\theta}\left(\theta\right)\right)}{\ln\left(\theta\right)}>-\infty$.\medskip{}

\noindent\textbf{Proof:} See Subsection A.3 of Appendix A.\medskip{}

In each of the above theorems, statement (b) means that the mixing
PDF follows an inverse power law in the limit as the argument approaches
the upper bound of its sample space. The value of statement (c) derives
from its parsimony; that is, the condition can be expressed without
reference to any specific values, $\rho\in\mathbb{R}_{>0}$.

\section{Heavy-Tailed Frequency Mixtures with Geometric Kernel}

\subsection{The Generalized Waring 2$\boldsymbol{\left(a,b,c\right)}$ Mixture}

\noindent One well-studied frequency mixing model for which $q\in\mathcal{G}_{\textrm{NB}}^{\textrm{H}}$
is the 3-parameter $\textrm{Generalized Beta 1}$\linebreak{}
$\left(a,b,c\right)$ distribution, with PDF
\begin{equation}
g_{q\mid a,b,c}^{\left(\textrm{G}\mathcal{B}1\right)}\left(q\right)=\dfrac{c}{\mathcal{B}\left(a,b\right)}q^{ca-1}\left(1-q^{c}\right)^{b-1}
\end{equation}
for $a,b,c\in\mathbb{R}_{>0}$, where $\mathcal{B}\left(v,w\right)=\tfrac{\Gamma\left(v\right)\Gamma\left(w\right)}{\Gamma\left(v+w\right)}$
denotes the beta function. This distribution, defined by McDonald
(1984) with an arbitrary positive scale factor, contains $\textrm{Beta}\left(a,b\right)\equiv\textrm{Generalized Beta 1}\left(a,b,c=1\right)$
and $\textrm{Kumaraswamy}\left(b,c\right)\equiv\textrm{Generalized Beta 1}\left(a=1,b,c\right)$
as special cases. When applied to the simplest canonical frequency
kernel, $\textrm{Geometric}\left(q\right)\equiv\textrm{Negative Binomial}\left(r=1,q\right)$,
it generates the mixture
\[
X\mid a,b,c\sim f_{X\mid a,b,c}\left(x\right)={\displaystyle \int_{0}^{1}}f_{X\mid q}^{\left(\textrm{G}\right)}\left(x\right)g_{q\mid a,b,c}^{\left(\textrm{G}\mathcal{B}1\right)}\left(q\right)dq
\]
\begin{equation}
=\dfrac{\mathcal{B}\left(a+\dfrac{x}{c},b\right)-\mathcal{B}\left(a+\dfrac{\left(x+1\right)}{c},b\right)}{\mathcal{B}\left(a,b\right)},
\end{equation}
which we will call the $\textrm{Generalized Waring 2}\left(a,b,c\right)$
model.\footnote{This name is chosen to distinguish the indicated distribution from
the $\textrm{Generalized Waring}\left(r,a,b\right)$ model, introduced
by Irwin (1968), with $f_{X\mid r,a,b}^{\left(\textrm{GW}\right)}\left(x\right)={\textstyle \int_{0}^{1}}f_{X\mid r,q}^{\left(\textrm{NB}\right)}\left(x\right)g_{q\mid a,b,c=1}^{\left(\textrm{G}\mathcal{B}1\right)}\left(q\right)dq=\tfrac{\mathcal{B}\left(x+a,b+r\right)}{\mathcal{B}\left(a,b\right)x\mathcal{B}\left(x,r\right)}$.} Applying the $\textrm{Beta}\left(a,b\right)$ and $\textrm{Kumaraswamy}$\linebreak{}
$\left(b,c\right)$ mixing distributions to the $\textrm{Geometric}\left(q\right)$
kernel yields the special cases:

\noindent$X\mid a,b\sim\textrm{Waring}\left(a,b\right)\equiv\textrm{Generalized Waring 2}\left(a,b,c=1\right)$,
with
\[
f_{X\mid a,b}^{\left(\textrm{W}\right)}\left(x\right)={\displaystyle \int_{0}^{1}}f_{X\mid q}^{\left(\textrm{G}\right)}\left(x\right)g_{q\mid a,b,c=1}^{\left(\textrm{G}\mathcal{B}1\right)}\left(q\right)dq
\]
\[
=\dfrac{\mathcal{B}\left(a+x,b+1\right)}{\mathcal{B}\left(a,b\right)};
\]
and

\noindent$X\mid b,c\sim\textrm{K-Mix}\left(b,c\right)\equiv\textrm{Generalized Waring 2}\left(a=1,b,c\right)$,\footnote{The distribution is named ``K-Mix'' to indicate its provenance as
a mixture formed by the Kumaraswamy mixing distribution. } with
\[
f_{X\mid b,c}^{\left(\textrm{KM}\right)}\left(x\right)={\displaystyle \int_{0}^{1}}f_{X\mid q}^{\left(\textrm{G}\right)}\left(x\right)g_{q\mid a=1,b,c}^{\left(\textrm{G}\mathcal{B}1\right)}\left(q\right)dq
\]
\[
=b\left[\mathcal{B}\left(\dfrac{x}{c}+1,b\right)-\mathcal{B}\left(\dfrac{\left(x+1\right)}{c}+1,b\right)\right].
\]

To show that $q\mid a,b,c\sim\textrm{Generalized Beta 1}\left(a,b,c\right)$
belongs to $\mathcal{G}_{\textrm{NB}}^{\textrm{H}}$, and therefore
that $X\mid a,b,c\sim\textrm{Generalized Waring 2}\left(a,b,c\right)$
belongs to $\mathcal{F}_{\textrm{NB}}^{\textrm{H}}$, one can employ
condition (b) of Theorem 2.1. Since\linebreak{}
$\underset{q\uparrow1}{\lim}\:\tfrac{1-q^{c}}{1-q}=\underset{q\uparrow1}{\lim}\:cq^{c-1}=c>0$
by a straightforward application of L'Hôpital's rule, it follows that
\[
\underset{q\uparrow1}{\lim}\:g_{q\mid a,b,c}^{\left(\textrm{G}\mathcal{B}1\right)}\left(q\right)\left(1-q\right)^{-\rho+1}=\underset{q\uparrow1}{\lim}\:\dfrac{c}{\mathcal{B}\left(a,b\right)}q^{ca-1}\left(1-q^{c}\right)^{b-1}\left(1-q\right)^{-\rho+1}
\]
\[
=\dfrac{c}{\mathcal{B}\left(a,b\right)}\underset{q\uparrow1}{\lim}\:\left(\dfrac{1-q^{c}}{1-q}\right)^{b-1}\left(1-q\right)^{b-1}\left(1-q\right)^{-\rho+1}
\]
\[
=\dfrac{c^{b}}{\mathcal{B}\left(a,b\right)}\underset{q\uparrow1}{\lim}\:\left(1-q\right)^{b-\rho},
\]
which is greater than 0 for all $\rho\geq b$.

\subsection{Incorporating the Zeta$\boldsymbol{\left(s\right)}$ Mixture Distribution}

\noindent Two of the simplest and best-known random variables belonging
to $\mathcal{F}_{\textrm{NB}}^{\textrm{H}}$ are:

\noindent$X\mid b\sim\textrm{Zeta}\left(b\right)$, with
\begin{equation}
f_{X\mid b}^{\left(\textrm{Z}\right)}\left(x\right)=\dfrac{\left(x+1\right)^{-\left(b+1\right)}}{\zeta\left(b+1\right)},
\end{equation}
for $b\in\left(0,\infty\right)$, where $\zeta\left(\sigma\right)={\textstyle \sum_{k=0}^{\infty}}\left(k+1\right)^{-\sigma}$
denotes the Riemann zeta function; and

\noindent$X\mid b\sim\textrm{Yule}\left(b\right)$, with
\begin{equation}
f_{X\mid b}^{\left(\textrm{Y}\right)}\left(x\right)=b\mathcal{B}\left(x+1,b+1\right),
\end{equation}
for $b\in\mathbb{R}_{>0}$.\footnote{The $\textrm{Zeta}\left(s\right)$ and $\textrm{Yule}\left(b\right)$
distributions often are defined on the sample space $x\in\mathbb{Z}_{\geq1}$
rather than $x\in\mathbb{Z}_{\geq0}$. However, we work with the latter
characterization because it is more commonly used in risk-analytic
applications.}

Both (14) and (15) have been proposed to model frequency data,\footnote{See, for example, Doray and Luong (1995) for Zeta models and Irwin
(1968) for Yule models.} and they possess comparable properties, including asymptotically
equivalent tails. Nevertheless, the similarities and differences between
the two models have not been analyzed closely in the risk and actuarial
literatures. In particular, although it is well known that the Yule
distribution (as a special case of the 3-parameter Generalized Waring
distribution) can be expressed as a continuous mixture of Geometric
random variables, no comparable result existed for the Zeta distribution
until recently (see Dai, Huang, Powers, and Xu, 2021). Therefore,
it is interesting and instructive to consider how these two distributions
can be incorporated into a more general mixture framework.

As shown in Dai, Huang, Powers, and Xu (2021), the $\textrm{Zeta}\left(b\right)$
PDF can be expressed as the following mixture of Geometric PDFs:
\[
f_{X\mid s}^{\left(\textrm{Z}\right)}\left(x\right)={\displaystyle \int_{0}^{1}}f_{X\mid q}^{\left(\textrm{G}\right)}\left(x\right)g_{q\mid s}^{\left(1\right)}\left(q\right)dq,
\]
where $g_{q\mid b}^{\left(1\right)}\left(q\right)=\tfrac{\left(-\ln\left(q\right)\right)^{b}}{\zeta\left(b+1\right)\Gamma\left(b+1\right)\left(1-q\right)}$.
Similarly, as a special case of the Generalized Waring distribution,
the $\textrm{Yule}\left(b\right)$ PDF can be expressed as:
\[
f_{X\mid b}^{\left(\textrm{Y}\right)}\left(x\right)={\displaystyle \int_{0}^{1}}f_{X\mid q}^{\left(\textrm{G}\right)}\left(x\right)g_{q\mid b}^{\left(2\right)}\left(q\right)dq,
\]
where $g_{q\mid b}^{\left(2\right)}\left(q\right)=b\left(1-q\right)^{b-1}$.
Noting that $\underset{c\downarrow0}{\lim}\:\tfrac{1-q^{c}}{c}=-\ln\left(q\right)$
for all $q\in\left(0,1\right)$, one can see that the two relevant
mixing PDFs may be written as special cases of the 2-parameter PDF,
\[
g_{q\mid b,c}^{\left(\Sigma\mathcal{B}\right)}\left(q\right)=\dfrac{c}{\Sigma_{\mathcal{B}}\left(\dfrac{1}{c},\dfrac{1}{c},b\right)}\dfrac{\left(1-q^{c}\right)^{b}}{\left(1-q\right)},
\]
where $\Sigma_{\mathcal{B}}\left(\xi,v,w\right)\equiv{\textstyle \sum_{k=0}^{\infty}}\mathcal{B}\left(\xi k+v,w+1\right)$.

We will call the above mixing model the ``$\Sigma\mathcal{B}\left(b,c\right)$''
(read ``sigma-beta'') distribution because of the expression in
the denominator of its normalizing constant. The associated mixture
model, with PMF
\[
f_{X\mid b,c}^{\left(\textrm{ZY}\right)}\left(x\right)=\dfrac{\mathcal{B}\left(\dfrac{\left(x+1\right)}{c},b+1\right)}{\Sigma_{\mathcal{B}}\left(\dfrac{1}{c},\dfrac{1}{c},b\right)},
\]
will be called the ``$\textrm{ZY}\left(b,c\right)$'' distribution
because it generalizes both the Zeta and Yule models. Using now-familiar
arguments based on condition (b) of Theorem 2.1, it is straightforward
to show that $q\mid b,c\sim\Sigma\mathcal{B}\left(b,c\right)$ belongs
to $\mathcal{G}_{\textrm{NB}}^{\textrm{H}}$, implying $X\mid b,c\sim\textrm{ZY}\left(b,c\right)$
belongs to $\mathcal{F}_{\textrm{NB}}^{\textrm{H}}$.

\subsection{The $\mathbf{HGZY}\boldsymbol{\left(a,b,c,d\right)}$ and $\mathbf{HGZY}^{\prime}\boldsymbol{\left(a,b,c,d\right)}$
Mixture Families}

\noindent It is difficult to compare the analytical forms of the heavy-tailed
$\textrm{Waring}\left(a,b\right)$, $\textrm{K-Mix}\left(b,c\right)$,
and $\textrm{ZY}\left(b,c\right)$ PMFs directly because the index
($x$) appears as an argument of beta functions in all three cases.
Alternatively, however, it is quite easy to compare the associated
mixing distributions that give rise to these families as mixtures
of a $\textrm{Geometric}\left(q\right)$ random variable; that is,
the $\textrm{Beta}\left(a,b\right)$, $\textrm{Kumaraswamy}\left(b,c\right)$,
and $\textrm{SigmaBeta}\left(b,c\right)$ PDFs:
\begin{equation}
g_{q\mid a,b}^{\left(\mathcal{B}\right)}\left(q\right)=\dfrac{1}{\mathcal{B}\left(a,b\right)}q^{a-1}\left(1-q\right)^{b-1}\propto q^{a-1}\left(1-q\right)^{b-1},
\end{equation}
\begin{equation}
g_{q\mid b,c}^{\left(\textrm{K}\right)}\left(q\right)=bcq^{c-1}\left(1-q^{c}\right)^{b-1}\propto q^{c-1}\left(1-q^{c}\right)^{b-1},
\end{equation}
and
\begin{equation}
g_{q\mid b,c}^{\left(\Sigma\mathcal{B}\right)}\left(q\right)=\dfrac{c}{\Sigma_{\mathcal{B}}\left(\dfrac{1}{c},\dfrac{1}{c},b\right)}\dfrac{\left(1-q^{c}\right)^{b}}{\left(1-q\right)}\propto\dfrac{\left(1-q^{c}\right)^{b}}{\left(1-q\right)},
\end{equation}
respectively.

From (12), we know that (16) and (17) can be merged into the $\textrm{Generalized Beta 1}\left(a,b,c\right)$
PDF,
\begin{equation}
g_{q\mid a,b,c}^{\left(\textrm{G}\mathcal{B}1\right)}\left(q\right)=\dfrac{c}{\mathcal{B}\left(a,b\right)}q^{ca-1}\left(1-q^{c}\right)^{b-1}\propto q^{ca-1}\left(1-q^{c}\right)^{b-1}.
\end{equation}
Looking closely at (16) and (18), one can see that introducing an
additional parameter similarly brings their functional forms closer
together in a natural way. Specifically, inserting a factor of $q^{ca-1}$
in (18) and replacing $q$ by $q^{c}$ and $a$ by $ac$ in (16) yields
what we will call the ``$\textrm{Generalized }\Sigma\mathcal{B}\left(a,b,c\right)$''
PDF,
\begin{equation}
g_{q\mid a,b,c}^{\left(\textrm{G}\Sigma\mathcal{B}\right)}\left(q\right)=\dfrac{c}{\Sigma_{\mathcal{B}}\left(\dfrac{1}{c},a,b\right)}\dfrac{q^{ca-1}\left(1-q^{c}\right)^{b}}{\left(1-q\right)}\propto\dfrac{q^{ca-1}\left(1-q^{c}\right)^{b}}{\left(1-q\right)}.
\end{equation}
Condition (b) of Theorem 2.1 then can be used to show that $q\mid a,b,c\sim\textrm{Generalized }\Sigma\mathcal{B}\left(a,b,c\right)$,
like $q\mid a,b,c\sim\textrm{Generalized Beta 1}\left(a,b,c\right)$,
belongs to $\mathcal{G}_{\textrm{NB}}^{\textrm{H}}$.

Applying (19) and (20) to construct mixtures of $\textrm{Geometric}\left(q\right)$
random variables yields the following generalizations of the Waring
and ZY PMFs, respectively:
\begin{equation}
f_{X\mid a,b,c}^{\left(\textrm{GW2}\right)}\left(x\right)=\dfrac{\mathcal{B}\left(a+\dfrac{x}{c},b\right)-\mathcal{B}\left(a+\dfrac{\left(x+1\right)}{c},b\right)}{\mathcal{B}\left(a,b\right)},
\end{equation}
of the ``$\textrm{Generalized Waring 2}\left(a,b,c\right)$'' distribution
(previously given by (13)); and
\begin{equation}
f_{X\mid a,b,c}^{\left(\textrm{GZY}\right)}\left(x\right)=\dfrac{\mathcal{B}\left(a+\dfrac{x}{c},b+1\right)}{\Sigma_{\mathcal{B}}\left(\dfrac{1}{c},a,b\right)},
\end{equation}
of what will be called the ``$\textrm{Generalized ZY}\left(a,b,c\right)$''
distribution. Obviously, these two heavy-tailed PMFs are quite similar,
with each of the three infinite series of (22) truncated to its first
term in (21).

To unify (21) and (22) into a single, 4-parameter PMF, we again turn
to the associated mixing PDFs (in (19) and (20), respectively). By
introducing the parameter $d\in\mathbb{R}_{>0}$ as an exponent of
$q$ in the denominator, the two mixing models are subsumed into the
PDF of what we will call the ``$\textrm{Hyper-Generalized }\Sigma\mathcal{B}$\linebreak{}
$\left(a,b,c,d\right)$'' distribution,
\begin{equation}
g_{q\mid a,b,c,d}^{\left(\textrm{HG}\Sigma\mathcal{B}\right)}\left(q\right)=\dfrac{c}{\Sigma_{\mathcal{B}}\left(\dfrac{d}{c},a,b\right)}\dfrac{q^{ca-1}\left(1-q^{c}\right)^{b}}{\left(1-q^{d}\right)}\propto\dfrac{q^{ca-1}\left(1-q^{c}\right)^{b}}{\left(1-q^{d}\right)},
\end{equation}
where (23) and all PDFs comprising special cases of (23) are said
to constitute the Hyper-Generalized $\Sigma\mathcal{B}$ family of
distributions, $\mathcal{G}^{\textrm{HG}\Sigma\mathcal{B}}\subset\mathcal{G}_{\textrm{NB}}^{\textrm{H}}$.
Using (23) to form a $\textrm{Geometric}\left(q\right)$ mixture then
extends the Generalized ZY and Generalized Waring 2 PMFs to that of
the heavy-tailed 4-parameter ``$\textrm{Hyper-Generalized}$\linebreak{}
$\textrm{ZY}\left(a,b,c,d\right)$'' distribution,
\begin{equation}
f_{X\mid a,b,c,d}^{\left(\textrm{HGZY}\right)}\left(x\right)=\dfrac{\Sigma_{\mathcal{B}}\left(\dfrac{d}{c},a+\dfrac{x}{c},b\right)-\Sigma_{\mathcal{B}}\left(\dfrac{d}{c},a+\dfrac{\left(x+1\right)}{c},b\right)}{\Sigma_{\mathcal{B}}\left(\dfrac{d}{c},a,b\right)},
\end{equation}
where (24) and all PMFs comprising special cases of (24) are said
to form the Hyper-Generalized ZY family of distributions, $\mathcal{F}^{\textrm{HGZY}}$.
The constant of integration in (23) is derived in Subsections A.4
of Appendix A, and the functional form of (24) is provided by Subsection
A.5 (setting $s=1$).

Figure 1 clarifies the parametric hierarchy among the nine members
of the $\mathcal{F}^{\textrm{HGZY}}$ family and their $\mathcal{G}^{\textrm{HG}\Sigma\mathcal{B}}$
counterparts. For completeness, one further 1-parameter model is included:
the ``$\textrm{Quadratic}\left(c\right)$'' distribution, whose
PMF is given by
\[
f_{X\mid c}^{\left(\textrm{Q}\right)}\left(x\right)=\dfrac{c}{\left(x+c\right)\left(x+c+1\right)},
\]
with an associated $\textrm{Kumaraswamy}\left(b=1,c\right)$ mixing
PDF. Detailed summaries of the $\mathcal{G}^{\textrm{HG}\Sigma\mathcal{B}}$
PDFs and $\mathcal{F}^{\textrm{HGZY}}$ PMFs are provided in Table
B1 of Appendix B.

Although $\mathcal{G}^{\textrm{HG}\Sigma\mathcal{B}}$, in conjunction
with the Geometric kernel, offers a flexible framework for modeling
frequency data, it suffers from two obvious shortcomings. First, any
frequency mixture based on a Geometric kernel must have a strictly
decreasing PMF, a clearly unrealistic limitation. Second, most mixing
distributions within $\mathcal{G}^{\textrm{HG}\Sigma\mathcal{B}}$
are asymmetric on the unit interval; that is, for a given parameter
vector $\left[a,b,c,d\right]$, there does not exist a corresponding
vector $\left[a^{\prime},b^{\prime},c^{\prime},d^{\prime}\right]$
such that $g_{q\mid a,b,c,d}^{\left(\textrm{HG}\Sigma\mathcal{B}\right)}\left(q\right)=g_{q\mid a^{\prime},b^{\prime},c^{\prime},d^{\prime}}^{\left(\textrm{HG}\Sigma\mathcal{B}\right)}\left(1-q\right)$
for all $q\in\left(0,1\right)$.\footnote{A notable exception is the $\textrm{Beta}\left(a,b\right)$ distribution,
for which $g_{q\mid a,b}^{\left(\mathcal{B}\right)}\left(q\right)=g_{q\mid b,a}^{\left(\mathcal{B}\right)}\left(1-q\right)$.} This imposes a further undesirable restriction.

\begin{singlespace}
\begin{center}
\includegraphics[scale=0.22]{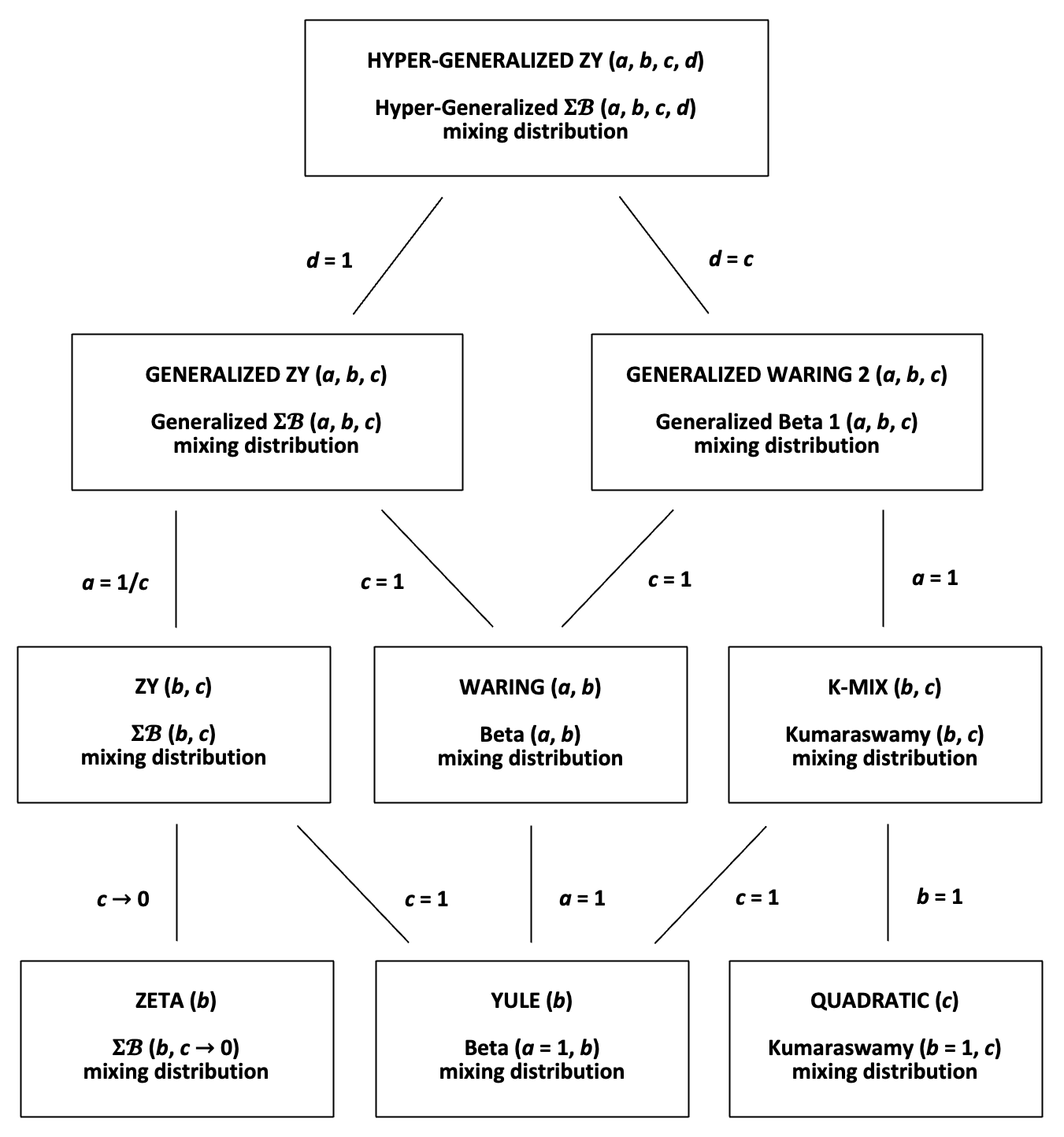}
\par\end{center}

\begin{center}
Figure 1. Hierarchy of Distributions within the Hyper-Generalized
ZY Family\pagebreak{}
\par\end{center}
\end{singlespace}

As noted in the Introduction, the first issue may be addressed by
extending our approach to the (canonical) Negative Binomial kernel,
thereby permitting unimodal mixtures with arbitrarily large interior
modes. This is carried out in Section 5. To address the second issue,
we simply construct a complementary family of mixing distributions
to ``mirror'' $\mathcal{G}^{\textrm{HG}\Sigma\mathcal{B}}$ by substituting
$1-q$ for $q$ in the PDF of each of its elements. The ``$\textrm{Complementary Hyper-Generalized }\Sigma\mathcal{B}\left(a,b,c,d\right)$''
distribution is then characterized by the PDF
\[
g_{q\mid a,b,c,d}^{\left(\textrm{CHG}\Sigma\mathcal{B}\right)}\left(q\right)=\dfrac{c}{\Sigma_{\mathcal{B}}\left(\dfrac{d}{c},a,b\right)}\dfrac{\left(1-q\right)^{ca-1}\left[1-\left(1-q\right)^{c}\right]^{b}}{\left[1-\left(1-q\right)^{d}\right]}\left|\dfrac{d\left(1-q\right)}{dq}\right|
\]
\begin{equation}
=\dfrac{c}{\Sigma_{\mathcal{B}}\left(\dfrac{d}{c},a,b\right)}\dfrac{\left(1-q\right)^{ca-1}\left[1-\left(1-q\right)^{c}\right]^{b}}{\left[1-\left(1-q\right)^{d}\right]},
\end{equation}
with the corresponding Complementary Hyper-Generalized $\Sigma\mathcal{B}$
family denoted by $\mathcal{G}^{\textrm{CHG}\Sigma\mathcal{B}}\subset\mathcal{G}_{\textrm{NB}}^{\textrm{H}}$.
In conjunction with the Geometric kernel, this yields the heavy-tailed
``$\textrm{Hyper-Generalized ZY Prime}\left(a,b,c,d\right)$'' distribution,
with PMF
\begin{equation}
f_{X\mid a,b,c,d}^{\left(\textrm{HGZY}^{\prime}\right)}\left(x\right)=\sum_{j=0}^{x}\binom{x}{j}\left(-1\right)^{j}\dfrac{\Sigma_{\mathcal{B}}\left(\dfrac{d}{c},a+\dfrac{\left(j+1\right)}{c},b\right)}{\Sigma_{\mathcal{B}}\left(\dfrac{d}{c},a,b\right)}
\end{equation}
and corresponding Hyper-Generalized ZY Prime family, $\mathcal{F}^{\textrm{HGZY}^{\prime}}$.
The functional form of (26) is provided by Subsection A.6 of Appendix
A (setting $s=1$).

Figure 2 presents the parametric hierarchy among the nine members
of the $\mathcal{F}^{\textrm{HGZY}^{\prime}}$ family and their $\mathcal{G}^{\textrm{CHG}\Sigma\mathcal{B}}$
counterparts. Summaries of the $\mathcal{G}^{\textrm{CHG}\Sigma\mathcal{B}}$
PDFs and $\mathcal{F}^{\textrm{HGZY}^{\prime}}$ PMFs are given in
Table B2 of Appendix B.

\section{Heavy-Tailed Severity Mixtures with Exponential Kernel}

\noindent In this section, we construct families of severity mixture
distributions analogous to $\mathcal{F}^{\textrm{HGZY}}$ and $\mathcal{F}^{\textrm{HGZY}^{\prime}}$
of the frequency case, with the notable exception that the analogue
of the latter family is \emph{not} heavy-tailed. As in Section 3,
the mixture families will be formed by applying corresponding sets
of mixing distributions (analogous to $\mathcal{G}^{\textrm{HG}\Sigma\mathcal{B}}$
and $\mathcal{G}^{\textrm{CHG}\Sigma\mathcal{B}}$, respectively)
to the simplest canonical kernel, $\textrm{Exponential}\left(\theta\right)\equiv\textrm{Gamma}\left(r=1,\theta\right)$.
However, unlike the previous section, we do not assemble the mixing
families piecewise by joining various 2- and 3-parameter families
together. Rather, we take advantage of the mathematical duality between
the Negative Binomial and Gamma distributions to derive explicit analogues
of $g_{q\mid a,b,c,d}^{\left(\textrm{HG}\Sigma\mathcal{B}\right)}\left(q\right)$
and $g_{q\mid a,b,c,d}^{\left(\textrm{CHG}\Sigma\mathcal{B}\right)}\left(q\right)$
by substituting $\theta$ for $\tfrac{q}{1-q}$ in the respective
4-parameter PDFs. For reasons that will become clear later, the corresponding
severity mixing PDFs will be denoted by $g_{\theta\mid\alpha,\beta,\gamma,\delta}^{\left(\textrm{IHG}\Sigma\Gamma\right)}\left(\theta\right)$
and $g_{\theta\mid\alpha,\beta,\gamma,\delta}^{\left(\textrm{HG}\Sigma\Gamma\right)}\left(\theta\right)$,
respectively.

\begin{singlespace}
\begin{center}
\includegraphics[scale=0.185]{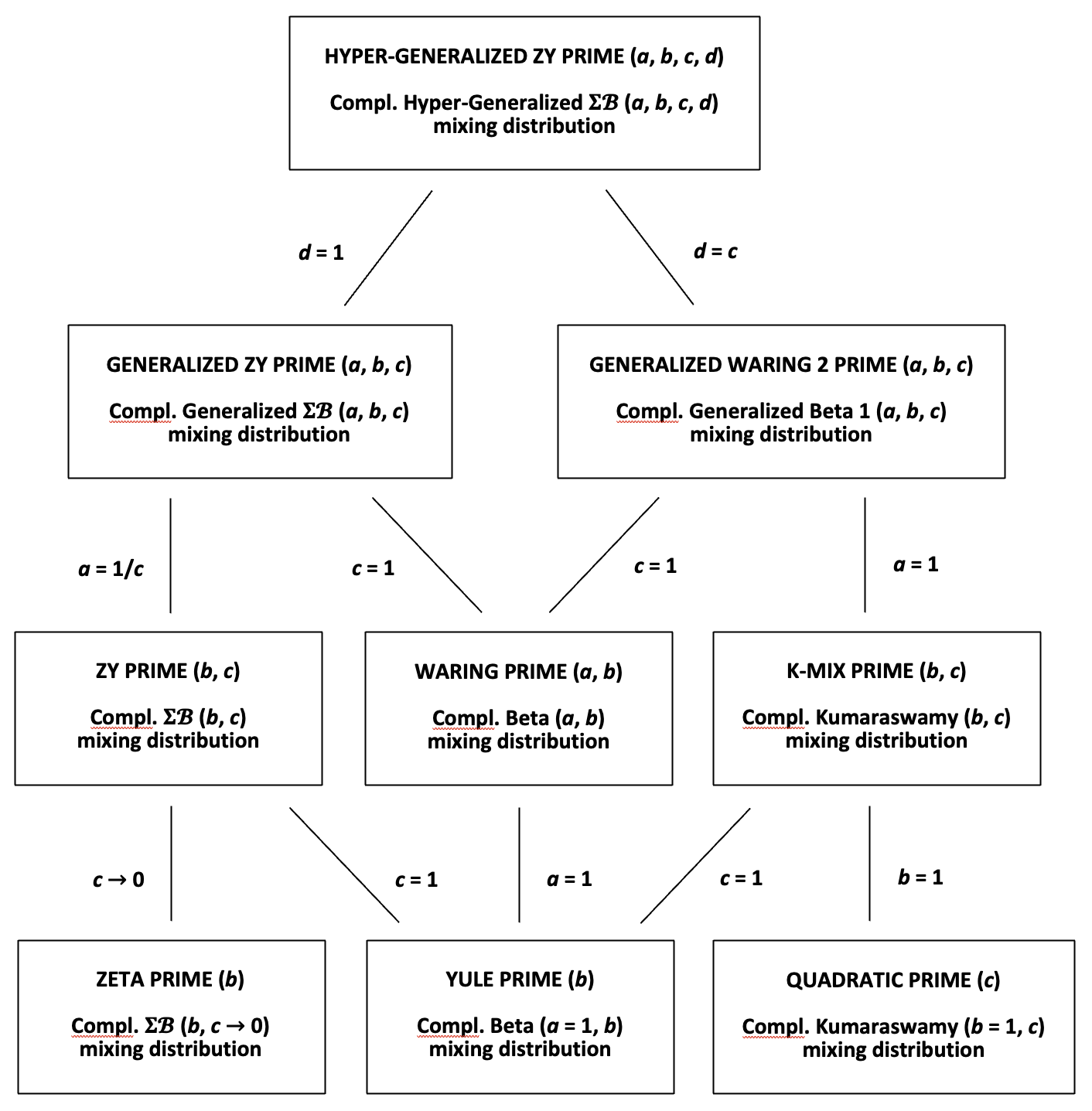}
\par\end{center}

\begin{center}
Figure 2. Hierarchy of Distributions within the Hyper-Generalized
ZY Prime Family\medskip{}
\par\end{center}
\end{singlespace}

Beginning with
\[
g_{q\mid a,b,c,d}^{\left(\textrm{HG}\Sigma\mathcal{B}\right)}\left(q\right)\propto\dfrac{q^{ca-1}\left(1-q^{c}\right)^{b}}{\left(1-q^{d}\right)}
\]
from (23), we write
\[
g_{\theta\mid\alpha,\beta,\gamma,\delta}^{\left(\textrm{HG}\Sigma\Gamma\right)}\left(\theta\right)\propto\dfrac{\left(\dfrac{\theta}{\theta+1}\right)^{\gamma\alpha-1}\left[1-\left(\dfrac{\theta}{\theta+1}\right)^{\gamma}\right]^{\beta}}{\left[1-\left(\dfrac{\theta}{\theta+1}\right)^{\delta}\right]}\left|\dfrac{d\left(\dfrac{\theta}{\theta+1}\right)}{d\theta}\right|
\]
\begin{equation}
=\dfrac{\left(\dfrac{\theta}{\theta+1}\right)^{\gamma\alpha-1}\left[1-\left(\dfrac{\theta}{\theta+1}\right)^{\gamma}\right]^{\beta}\left(\dfrac{1}{\theta+1}\right)^{2}}{\left[1-\left(\dfrac{\theta}{\theta+1}\right)^{\delta}\right]},
\end{equation}
where the Greek-lettered parameters play roles roughly comparable
to those of their Roman-lettered counterparts. For ease of exposition,
we first derive the PDF $g_{\theta\mid a,b,c,d}^{\left(\textrm{HG}\Sigma\Gamma\right)}\left(\theta\right)$
based on the limiting behavior of (27) as $\theta\rightarrow0^{+}$,
and then address the right tail of the sample space (i.e., $\theta\rightarrow\infty$)
by taking $g_{\theta\mid a,b,c,d}^{\left(\textrm{IHG}\Sigma\Gamma\right)}\left(\theta\right)$
to be the ``inverse'' PDF.

In a small positive neighborhood of 0, (27) may be approximated as
\[
g_{\theta\mid\alpha,\beta,\gamma,\delta}^{\left(\textrm{HG}\Sigma\Gamma\right)}\left(\theta\right)\propto\dfrac{\theta^{\gamma\alpha-1}\left(1-\theta^{\gamma}\right)^{\beta}}{\left(1-\theta^{\delta}\right)}
\]
\[
\approx\theta^{\gamma\alpha-1}\exp\left(-\beta\theta^{\gamma}\right)\exp\left(\theta^{\delta}\right)
\]
\begin{equation}
\approx\theta^{\gamma\alpha-1}\exp\left(-\beta\theta^{\gamma}\right)e^{\theta/\delta},
\end{equation}
where the substitution in the argument of the second exponential function
is based on the assumption
\begin{equation}
\theta^{\delta}\approx\dfrac{\theta}{\delta},
\end{equation}
a somewhat ad hoc approximation chosen primarily to facilitate the
tractability of the PDF. Naturally, (29) is valid for fixed values
of $\delta$ as $\theta\rightarrow0^{+}$. In addition,
\[
\textrm{sgn}\left(\dfrac{\partial}{\partial\delta}\left(\theta^{\delta}\right)\right)=\textrm{sgn}\left(\dfrac{\partial}{\partial\delta}\left(\dfrac{\theta}{\delta}\right)\right)
\]
for fixed $\theta$ sufficiently close to 0. As shown in Subsection
A.7 of Appendix A, one may integrate the right-hand side of (28) over
$\theta\in\mathbb{R}_{>0}$ to solve for the constant of integration,
yielding
\begin{equation}
g_{\theta\mid\alpha,\beta,\gamma,\delta}^{\left(\textrm{HG}\Sigma\Gamma\right)}\left(\theta\right)=\dfrac{\gamma\beta^{\alpha}}{\Sigma_{\Gamma}\left(\dfrac{1}{\gamma},\alpha,\dfrac{\beta^{-1/\gamma}}{\delta}\right)}\theta^{\gamma\alpha-1}\exp\left(-\beta\theta^{\gamma}\right)e^{\theta/\delta},
\end{equation}
where $\Sigma_{\Gamma}\left(\xi,v,w\right)\equiv{\textstyle \sum_{k=0}^{\infty}}\dfrac{w^{k}}{k!}\Gamma\left(\xi k+v\right)$,
and one of the following conditions must hold for the PDF to integrate
to 1: $\gamma>1\:\vee\:\left(\gamma=1\:\wedge\:\beta>\tfrac{1}{\delta}\right)$.

Interestingly, the PDF in (30) never satisfies condition (b) of Theorem
2.2, which means $\theta\mid\alpha,\beta,\gamma,\delta\notin\mathcal{G}_{\Gamma}^{\textrm{H}}$
for this distribution. Therefore, we must look to the complementary
PDF formed by substituting $\tfrac{1}{\theta}$ for $\theta$ in $g_{\theta\mid\alpha,\beta,\gamma,\delta}^{\left(\textrm{HG}\Sigma\Gamma\right)}\left(\theta\right)$
to model heavy-tailed mixtures, $Y\mid\alpha,\beta,\gamma,\delta\in\mathcal{F}_{\Gamma}^{\textrm{H}}$.
This is given by
\[
g_{\theta\mid\alpha,\beta,\gamma,\delta}^{\left(\textrm{IHG}\Sigma\Gamma\right)}\left(\theta\right)=\dfrac{\gamma\beta^{\alpha}}{\Sigma_{\Gamma}\left(\dfrac{1}{\gamma},\alpha,\dfrac{\beta^{-1/\gamma}}{\delta}\right)}\dfrac{\exp\left(-\beta\theta^{-\gamma}\right)e^{1/\left(\delta\theta\right)}}{\theta^{\gamma\alpha-1}}\left|\dfrac{d\left(1/\theta\right)}{d\theta}\right|
\]
\begin{equation}
=\dfrac{\gamma\beta^{\alpha}}{\Sigma_{\Gamma}\left(\dfrac{1}{\gamma},\alpha,\dfrac{\beta^{-1/\gamma}}{\delta}\right)}\dfrac{\exp\left(-\beta\theta^{-\gamma}\right)e^{1/\left(\delta\theta\right)}}{\theta^{\gamma\alpha+1}},
\end{equation}
where the integrability conditions $\gamma>1\:\vee\:\left(\gamma=1\:\wedge\:\beta>\tfrac{1}{\delta}\right)$
again must hold. This PDF always satisfies condition (b) of Theorem
2.2, implying that $\theta\mid\alpha,\beta,\gamma,\delta\in\mathcal{G}_{\Gamma}^{\textrm{H}}$.
The qualitatively distinct behavior of $g_{\theta\mid\alpha,\beta,\gamma,\delta}^{\left(\textrm{HG}\Sigma\Gamma\right)}\left(\theta\right)$
and $g_{\theta\mid\alpha,\beta,\gamma,\delta}^{\left(\textrm{IHG}\Sigma\Gamma\right)}\left(\theta\right)$
in terms of generating heavy tails is thus substantially different
from the relationship between $g_{q\mid a,b,c,d}^{\left(\textrm{HG}\Sigma\mathcal{B}\right)}\left(q\right)$
and $g_{q\mid a,b,c,d}^{\left(\textrm{CHG}\Sigma\mathcal{B}\right)}\left(q\right)$
in the frequency case (where both mixing distributions are elements
of $\mathcal{G}_{\textrm{NB}}^{\textrm{H}}$). The denominator of
the expression in (30) provides the rationale for naming it the ``$\textrm{Hyper-Generalized }\Sigma\Gamma\left(\alpha,\beta,\gamma,\delta\right)$''
PDF (with corresponding Hyper-Generalized $\Sigma\Gamma$ family,
$\mathcal{G}^{\textrm{HG}\Sigma\Gamma}\nsubseteq\mathcal{G}_{\Gamma}^{\textrm{H}}$).
In naming (31), we add the term ``Inverse'' (rather than ``Complementary'')
-- yielding the ``$\textrm{Inverse Hyper-Generalized }\Sigma\Gamma\left(\alpha,\beta,\gamma,\delta\right)$''
PDF (and corresponding Inverse Hyper-Generalized $\Sigma\Gamma$ family,
$\mathcal{G}^{\textrm{IHG}\Sigma\Gamma}\subset\mathcal{G}_{\Gamma}^{\textrm{H}}$)
-- because the indicated expression is formed by substituting $\tfrac{1}{\theta}$
for $\theta$ (rather than $1-q$ for $q$, as in the frequency case).

As previously noted, we will view $\mathcal{G}^{\textrm{IHG}\Sigma\Gamma}$
as the counterpart of $\mathcal{G}^{\textrm{HG}\Sigma\mathcal{B}}$,
and $\mathcal{G}^{\textrm{HG}\Sigma\Gamma}$ as the counterpart of
$\mathcal{G}^{\textrm{CHG}\Sigma\mathcal{B}}$. This is primarily
because of the functional characteristics of the corresponding mixture
distributions. In conjunction with the Exponential kernel, $\mathcal{G}^{\textrm{IHG}\Sigma\Gamma}$
yields the ``$\textrm{Hyper-Generalized }\Sigma\Sigma\left(\alpha,\beta,\gamma,\delta\right)$''
PDF
\begin{equation}
f_{Y\mid\alpha,\beta,\gamma,\delta}^{\left(\textrm{HG}\Sigma\Sigma\right)}\left(y\right)=\beta^{-1/\gamma}\dfrac{\Sigma_{\Gamma}\left(\dfrac{1}{\gamma},\alpha+\dfrac{1}{\gamma},\left(\dfrac{1}{\delta}-y\right)\beta^{-1/\gamma}\right)}{\Sigma_{\Gamma}\left(\dfrac{1}{\gamma},\alpha,\dfrac{\beta^{-1/\gamma}}{\delta}\right)}
\end{equation}
(and corresponding Hyper-Generalized $\Sigma\Sigma$ family, $\mathcal{F}^{\textrm{HG}\Sigma\Sigma}$),
whereas $\mathcal{G}^{\textrm{HG}\Sigma\Gamma}$ yields the ``$\textrm{Hyper-Generalized }\Sigma\Sigma\textrm{ Prime}\left(\alpha,\beta,\gamma,\delta\right)$''
PDF
\begin{equation}
f_{Y\mid\alpha,\beta,\gamma,\delta}^{\left(\textrm{HG}\Sigma\Sigma^{\prime}\right)}\left(y\right)=\beta^{1/\gamma}{\displaystyle \sum_{j=0}^{\infty}}\dfrac{\left(-y\beta^{1/\gamma}\right)^{j}}{j!}\dfrac{\Sigma_{\Gamma}\left(\dfrac{1}{\gamma},\alpha-\dfrac{\left(j+1\right)}{\gamma},\dfrac{\beta^{-1/\gamma}}{\delta}\right)}{\Sigma_{\Gamma}\left(\dfrac{1}{\gamma},\alpha,\dfrac{\beta^{-1/\gamma}}{\delta}\right)}
\end{equation}
(and corresponding Hyper-Generalized $\Sigma\Sigma$ Prime family,
$\mathcal{F}^{\textrm{HG}\Sigma\Sigma^{\prime}}$). Consistent with
the frequency case, the functional forms characterizing the ``Prime''
mixture family are more complicated than those of the ``non-Prime''
family. Moreover, the PDF in (33) is not well-defined if $\alpha$
is an integer and $\gamma=1$ because these parameters entail the
evaluation of the gamma function at one or more nonpositive integers.
Fortunately, when $\gamma=1$, one can use the alternative derivation,
\[
f_{Y\mid\alpha,\beta,\gamma=1,\delta}^{\left(\textrm{HG}\Sigma\Sigma^{\prime}\right)}\left(y\right)={\displaystyle \int_{0}^{\infty}}\dfrac{e^{-y/\theta}}{\theta}\dfrac{\beta^{\alpha}}{{\displaystyle \sum_{i=0}^{\infty}}\dfrac{\left[1/\left(\beta\delta\right)\right]^{i}}{i!}\Gamma\left(\alpha+i\right)}\theta^{\alpha-1}e^{-\left(\beta-1/\delta\right)\theta}d\theta
\]
\[
={\displaystyle \int_{0}^{\infty}}\dfrac{\beta^{\alpha}\left[1-1/\left(\beta\delta\right)\right]^{\alpha}}{\Gamma\left(\alpha\right)}\theta^{\alpha-2}e^{-y/\theta}e^{-\left(\beta-1/\delta\right)\theta}d\theta
\]
\[
={\displaystyle \int_{0}^{\infty}}\dfrac{\left(\beta-1/\delta\right)^{\alpha}}{\Gamma\left(\alpha\right)}\theta^{\alpha-2}e^{-y/\theta}e^{-\left(\beta-1/\delta\right)\theta}d\theta
\]
\[
=\dfrac{2\left(\beta-1/\delta\right)^{\left(\alpha+1\right)/2}}{\Gamma\left(\alpha\right)}y^{\left(\alpha-1\right)/2}K_{\alpha-1}\left(2\sqrt{\left(\beta-1/\delta\right)y}\right),
\]
where $K_{\upsilon}\left(z\right)$ denotes the modified Bessel function
of the second kind.

\begin{singlespace}
\begin{center}
\includegraphics[scale=0.22]{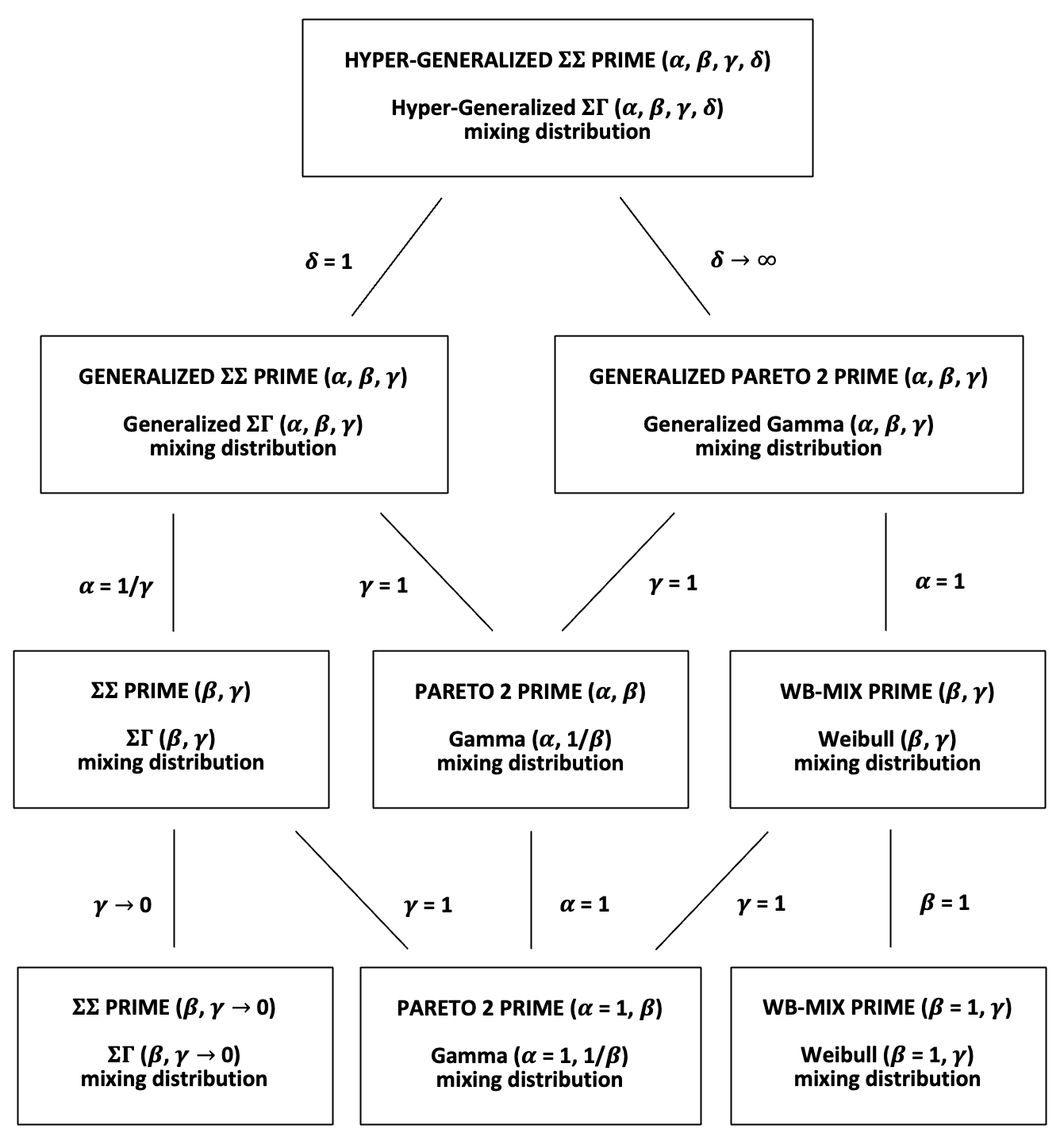}
\par\end{center}

\begin{center}
Figure 3. Hierarchy of Distributions within the Hyper-Generalized
$\Sigma\Sigma$ Prime Family\pagebreak{}
\par\end{center}

\begin{center}
\includegraphics[scale=0.22]{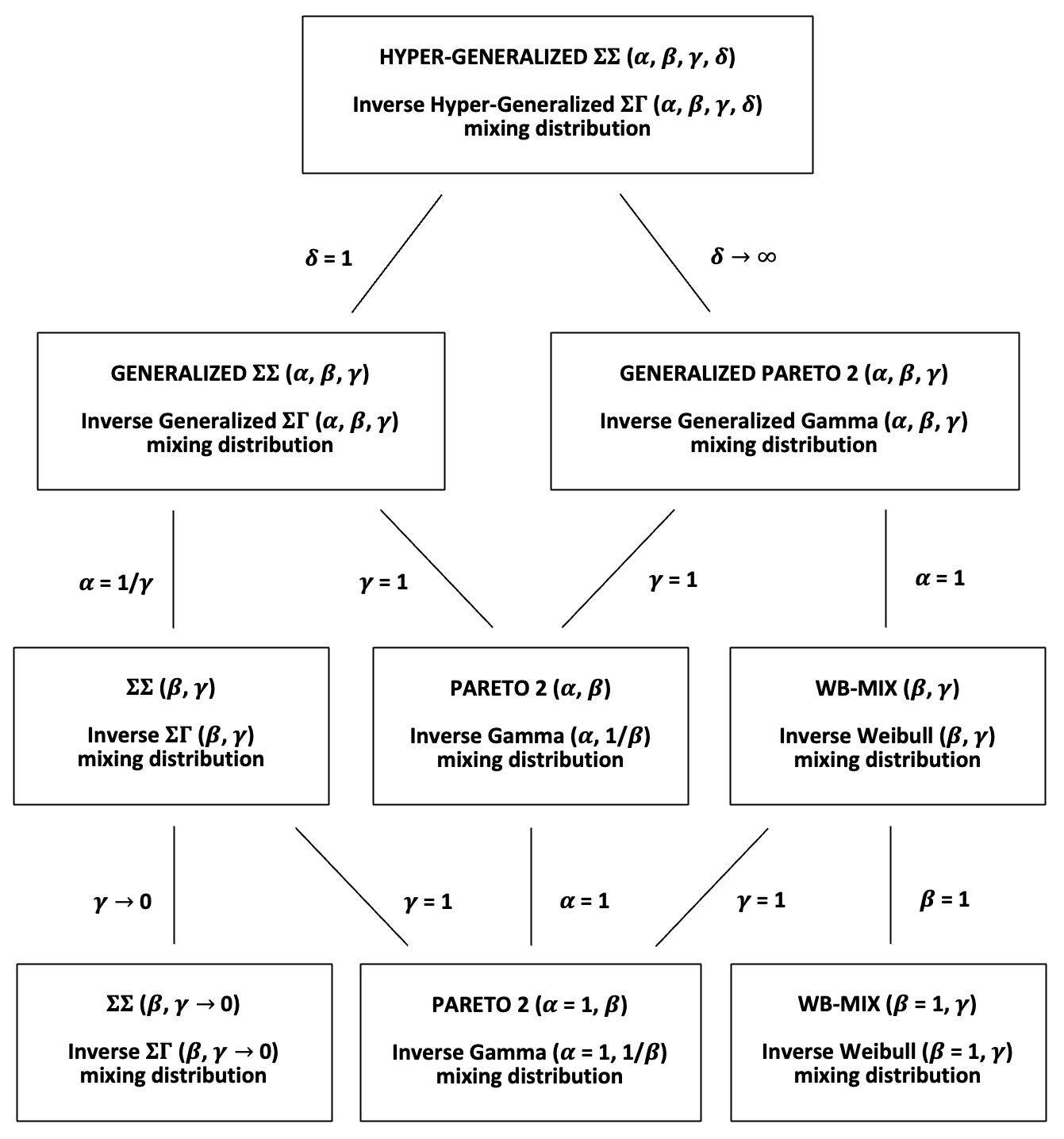}
\par\end{center}

\begin{center}
Figure 4. Hierarchy of Distributions within the Hyper-Generalized
$\Sigma\Sigma$ Family\medskip{}
\par\end{center}
\end{singlespace}

The expressions in (32) and (33) are given by Subsections A.8 and
A.9, respectively, of Appendix A (setting $s=1$), and all PDFs associated
with $\mathcal{G}^{\textrm{IHG}\Sigma\Gamma}$, $\mathcal{G}^{\textrm{HG}\Sigma\Gamma}$,
$\mathcal{F}^{\textrm{HG}\Sigma\Sigma}$, and $\mathcal{F}^{\textrm{HG}\Sigma\Sigma^{\prime}}$
can be derived as special cases of (31), (30), (32), and (33), respectively.
Figure 3 presents the parametric hierarchy of the $\mathcal{F}^{\textrm{HG}\Sigma\Sigma}$
family and its $\mathcal{G}^{\textrm{IHG}\Sigma\Gamma}$ counterparts,
and Table B3 of Appendix B summarizes the $\mathcal{G}^{\textrm{IHG}\Sigma\Gamma}$
and $\mathcal{F}^{\textrm{HG}\Sigma\Sigma}$ PDFs. Although we will
not work directly with the (non-heavy-tailed) $\mathcal{F}^{\textrm{HG}\Sigma\Sigma^{\prime}}$
family and its $\mathcal{G}^{\textrm{HG}\Sigma\Gamma}$ counterpart,
Figure 4 provides its parametric hierarchy and Table B4 of Appendix
B the corresponding $\mathcal{G}^{\textrm{HG}\Sigma\Gamma}$ and $\mathcal{F}^{\textrm{HG}\Sigma\Sigma^{\prime}}$
PDFs. 

\section{Arbitrary Canonical Kernels}

\subsection{5-Parameter Mixture Distributions}

\noindent Clearly, each member of the mixture families derived in
Sections 3 and 4 -- $\mathcal{F}^{\textrm{HGZY}}$, $\mathcal{F}^{\textrm{HGZY}^{\prime}}$,
$\mathcal{F}^{\textrm{HG}\Sigma\Sigma}$, and the (non-heavy-tailed)
$\mathcal{F}^{\textrm{HG}\Sigma\Sigma^{\prime}}$ -- can be generalized
along an additional dimension by applying its underlying mixing PDF
to the relevant canonical kernel (Negative Binomial or Gamma) with
arbitrary $r\in\mathbb{R}_{>0}$. This yields four new families in
which each member is characterized by five parameters ($\left[r,a,b,c,d\right]$
or $\left[r,\alpha,\beta,\gamma,\delta\right]$) and denoted by appending
``$r$'' to the superscript of the relevant symbol (e.g., $\mathcal{F}^{\textrm{HGZY}\left(r\right)}$).
The 5-parameter PMFs/PDFs associated with these four families are
given by:
\[
f_{X\mid r,a,b,c,d}^{\left(\textrm{HGZY}\left(r\right)\right)}\left(x\right)=\dfrac{\Gamma\left(r+x\right)}{\Gamma\left(r\right)\Gamma\left(x+1\right)}{\displaystyle \sum_{\ell=0}^{\infty}\binom{r}{\ell}\left(-1\right)^{\ell}}\dfrac{\Sigma_{\mathcal{B}}\left(\dfrac{d}{c},a+\dfrac{\left(x+\ell\right)}{c},b\right)}{\Sigma_{\mathcal{B}}\left(\dfrac{d}{c},a,b\right)},
\]
\[
f_{X\mid r,a,b,c,d}^{\left(\textrm{HGZY}^{\prime}\left(r\right)\right)}\left(x\right)=\dfrac{\Gamma\left(r+x\right)}{\Gamma\left(r\right)\Gamma\left(x+1\right)}\sum_{j=0}^{x}\binom{x}{j}\left(-1\right)^{j}\dfrac{\Sigma_{\mathcal{B}}\left(\dfrac{d}{c},a+\dfrac{\left(r+j\right)}{c},b\right)}{\Sigma_{\mathcal{B}}\left(\dfrac{d}{c},a,b\right)},
\]
\[
f_{Y\mid r,\alpha,\beta,\gamma,\delta}^{\left(\textrm{HG}\Sigma\Sigma\left(r\right)\right)}\left(y\right)=\dfrac{\beta^{-r/\gamma}y^{r-1}}{\Gamma\left(r\right)}\dfrac{\Sigma_{\Gamma}\left(\dfrac{1}{\gamma},\alpha+\dfrac{r}{\gamma},\left(\dfrac{1}{\delta}-y\right)\beta^{-1/\gamma}\right)}{\Sigma_{\Gamma}\left(\dfrac{1}{\gamma},\alpha,\dfrac{\beta^{-1/\gamma}}{\delta}\right)},
\]
and
\[
f_{Y\mid r,\alpha,\beta,\gamma,\delta}^{\left(\textrm{HG}\Sigma\Sigma^{\prime}\left(r\right)\right)}\left(y\right)=\dfrac{\beta^{r/\gamma}y^{r-1}}{\Gamma\left(r\right)}{\displaystyle \sum_{j=0}^{\infty}}\dfrac{\left(-y\beta^{1/\gamma}\right)^{j}}{j!}\dfrac{\Sigma_{\Gamma}\left(\dfrac{1}{\gamma},\alpha-\dfrac{\left(r+j\right)}{\gamma},\dfrac{\beta^{-1/\gamma}}{\delta}\right)}{\Sigma_{\Gamma}\left(\dfrac{1}{\gamma},\alpha,\dfrac{\beta^{-1/\gamma}}{\delta}\right)},
\]
as shown in Subsections A.5, A.6, A.8, and A.9, respectively, of Appendix
A.

Although these families may be of theoretical interest, our purpose
in formulating them is entirely pragmatic. Basically, we wish to provide
a new and useful approach to analyzing risk heterogeneity in an empirical
frequency or severity distribution by: (i) fitting the empirical distribution
with members of the above mixture families and estimating the relevant
parameter vector ($\left[\hat{r},\hat{a},\hat{b},\hat{c},\hat{d}\right]$
or $\left[\hat{r},\hat{\alpha},\hat{\beta},\hat{\gamma},\hat{\delta}\right]$);
and (ii) detecting and measuring significant risk-heterogeneity patterns
within the associated (fitted) mixing distribution, with particular
attention to dual modes, heavy tails, and other indicators of potentially
large financial losses. For this reason, we will work with both $\mathcal{F}^{\textrm{HGZY}\left(r\right)}$
and $\mathcal{F}^{\textrm{HGZY}^{\prime}\left(r\right)}$ (and the
corresponding $\mathcal{G}^{\textrm{HG}\Sigma\mathcal{B}\left(r\right)}$
and $\mathcal{G}^{\textrm{CHG}\Sigma\mathcal{B}\left(r\right)}$)
for frequencies, but with only $\mathcal{F}^{\textrm{HG}\Sigma\Sigma\left(r\right)}$
(and the corresponding $\mathcal{G}^{\textrm{IHG}\Sigma\Gamma\left(r\right)}$)
for severities, since $\mathcal{F}^{\textrm{HG}\Sigma\Sigma^{\prime}\left(r\right)}$
is not heavy-tailed.

\subsection{Calibrated Families -- A Robustness Check}

\noindent In Subsection 2.2, we noted that for any canonical kernel
with shape parameter $r$ and corresponding mixing distribution, $g_{q\mid r}\left(q\right)$
or $g_{\theta\mid r}\left(\theta\right)$, it is possible to construct
a set of unique mixing PDFs, $g_{q\mid s}\left(q\right)$ or $g_{\theta\mid s}\left(\theta\right)$,
respectively, such that condition (9) or (10) holds for all $s\in\left[r,\infty\right)$.
We will call the resulting pairs, $\left[f_{X\mid s,q}^{\left(\textrm{NB}\right)}\left(x\right),g_{q\mid s}\left(q\right)\right]$
or $\left[f_{Y\mid s,\theta}^{\left(\Gamma\right)}\left(y\right),g_{\theta\mid s}\left(\theta\right)\right]$,
``calibrated'' families for $f_{X}\left(x\right)$ and $f_{Y}\left(y\right)$,
respectively, because they allow one to adjust the value of the parameter
$r$ while maintaining exactly the same quality of fit to a given
empirical distribution. This is particularly useful for checking the
robustness of inferences regarding risk heterogeneity based on the
fitted mixing distribution, because it enables one to investigate
the sensitivity of such inferences to the choice of $r$.

For example, upon fitting an empirical distribution by a particular
member of $\mathcal{F}^{\textrm{HGZY}\left(r\right)}$, $\mathcal{F}^{\textrm{HGZY}^{\prime}\left(r\right)}$,
or $\mathcal{F}^{\textrm{HG}\Sigma\Sigma\left(r\right)}$, one may
find that the associated member of $\mathcal{G}^{\textrm{HG}\Sigma\mathcal{B}}$,
$\mathcal{G}^{\textrm{CHG}\Sigma\mathcal{B}}$, or $\mathcal{G}^{\textrm{IHG}\Sigma\Gamma}$
appears to be characterized by a relatively high-risk sub-population.
In the insurance sector, such sub-populations (of policyholders) often
arise from underwriting and/or risk-classification errors associated
with any of a number of potential causes -- adverse selection, aggressive
portfolio expansion, ineffective modeling, etc. -- that could impart
a rise (and local mode) in the mixing PDF ($g_{q}\left(q\right)$
or $g_{\theta}\left(\theta\right)$). However, evidence of the supposed
high-risk sub-group may vanish for larger values of $r$ within the
relevant calibrated family because the underlying behavior can be
explained equally well by a greater mode in the kernel in conjunction
with a mixing distribution from outside $\mathcal{G}^{\textrm{HG}\Sigma\mathcal{B}}$,
$\mathcal{G}^{\textrm{CHG}\Sigma\mathcal{B}}$, or $\mathcal{G}^{\textrm{IHG}\Sigma\Gamma}$.
In such cases, the original impression of distinct risk heterogeneity
would have to be reconsidered.

\subsection{Calibrated Frequency Mixtures}

\noindent The mathematical formulation of the calibrated frequency
family, $\left[f_{X\mid s,q}^{\left(\textrm{NB}\right)}\left(x\right),g_{q\mid s}\left(q\right)\right]$,
for $s\in\left[r,\infty\right)$, is given by the following theorem.\footnote{Earlier versions of Theorem 3.1 and its two corollaries appeared in
Dai, Huang, Powers, and Xu (2022).}\medskip{}

\noindent\textbf{Theorem 3.1:} For a given frequency, $X\sim f_{X}\left(x\right)$,
if there exists a mixing PDF, $g_{q\mid r}\left(q\right)$, satisfying
$f_{X}\left(x\right)=\int_{0}^{1}f_{X\mid r,q}^{\left(\textrm{NB}\right)}\left(x\right)g_{q\mid r}\left(q\right)dq$
for some $r\in\mathbb{R}_{>0}$, then:

\noindent (a) $g_{q\mid r}\left(q\right)$ is unique;

\noindent (b) for all $s\in\left(r,\infty\right)$, the function
\begin{equation}
g_{q\mid s}\left(q\right)=\dfrac{1}{\mathcal{B}\left(r,s-r\right)\left(1-q\right)^{s}}{\displaystyle \int_{q}^{1}}\left(\dfrac{q}{\omega-q}\right)^{r-1}\left(\dfrac{\omega-q}{\omega}\right)^{s-2}\dfrac{\left(1-\omega\right)^{r}}{\omega}g_{q\mid r}\left(\omega\right)d\omega
\end{equation}
is the unique PDF satisfying $f_{X}\left(x\right)=\int_{0}^{1}f_{X\mid s,q}^{\left(\textrm{NB}\right)}\left(x\right)g_{q\mid s}\left(q\right)dq$;
and

\noindent (c) for all $s\in\left(0,r\right)$, the function
\[
g_{q\mid s}\left(q\right)=\dfrac{1}{\left(s-r\right)\mathcal{B}\left(r,s-r\right)\left(1-q\right)^{s}}
\]
\[
\times\left\{ {\displaystyle \int_{q}^{1}}\left(\dfrac{q}{\omega-q}\right)^{r-1}\left(\dfrac{\omega-q}{\omega}\right)^{s-1}\left(1-\omega\right)^{r-1}\left[\left(s-1\right)\dfrac{\left(1-\omega\right)}{\omega}+r\right]g_{q\mid r}\left(\omega\right)d\omega\right.
\]
\begin{equation}
\left.-{\displaystyle \int_{q}^{1}}\left(\dfrac{q}{\omega-q}\right)^{r-1}\left(\dfrac{\omega-q}{\omega}\right)^{s-1}\left(1-\omega\right)^{r}g_{q\mid r}^{\prime}\left(\omega\right)d\omega\right\} 
\end{equation}
is either the unique PDF or a quasi-PDF (such that $g_{q\mid s}\left(q\right)<0$
for some $q\in\left(0,1\right)$) satisfying $f_{X}\left(x\right)=\int_{0}^{1}f_{X\mid s,q}^{\left(\textrm{NB}\right)}\left(x\right)g_{q\mid s}\left(q\right)dq$.\medskip{}

\noindent\textbf{Proof:} See Subsection A.10 of Appendix A.\medskip{}

Whether or not $g_{q\mid s}\left(q\right)$ is a quasi-PDF when $s<r$
depends on the behavior of the original mixing PDF, $g_{q\mid r}\left(q\right)$.
In particular, a close inspection of the two integrands in (35) reveals
the right-hand side is negative if and only if
\[
\textrm{E}_{q\mid r}\left[\left(s-1\right)\dfrac{\left(1-q\right)}{q}+r-\left(1-q\right)\dfrac{g_{q\mid r}^{\prime}\left(q\right)}{g_{q\mid r}\left(q\right)}\right]<0,
\]
a condition more likely to be true if the elasticity function, $e_{q\mid r}\left(q\right)=\tfrac{qg_{q\mid r}^{\prime}\left(q\right)}{g_{q\mid r}\left(q\right)}$,
tends to be greater than $s-1+\dfrac{rq}{\left(1-q\right)}$. Consequently,
quasi-PDFs occur more often when $r$ and $s$ are small and $g_{q\mid r}\left(q\right)$
tends to be positive sloping. For $r=1$, the next result provides
a simple sufficient condition.

\medskip{}

\noindent\textbf{Corollary 3.1.1: }For a given frequency, $X\sim f_{X}\left(x\right)$,
and mixing PDF, $g_{q\mid r}\left(q\right)$, satisfying $f_{X}\left(x\right)=$\linebreak{}
$\int_{0}^{1}f_{X\mid r=1,q}^{\left(\textrm{NB}\right)}\left(x\right)g_{q\mid r=1}\left(q\right)dq$,
let $\underset{q\downarrow0}{\lim}\:e_{q\mid r=1}\left(q\right)>s-1$
for $s\in\left(0,1\right)$. Then the function
\[
g_{q\mid s}\left(q\right)=\dfrac{1}{\left(1-q\right)^{s}}\left\{ {\displaystyle \int_{q}^{1}}\left(\dfrac{\omega-q}{\omega}\right)^{s-1}\left[\left(s-1\right)\dfrac{\left(1-\omega\right)}{\omega}+1\right]g_{q\mid r=1}\left(\omega\right)d\omega\right.
\]
\begin{equation}
\left.-{\displaystyle \int_{q}^{1}}\left(\dfrac{\omega-q}{\omega}\right)^{s-1}\left(1-\omega\right)g_{q\mid r=1}^{\prime}\left(\omega\right)d\omega\right\} 
\end{equation}
is a quasi-PDF, with $g_{q\mid s}\left(q\right)<0$ for all $q$ in
some neighborhood of 0, satisfying\linebreak{}
$f_{X}\left(x\right)=\int_{0}^{1}f_{X\mid s,q}^{\left(\textrm{NB}\right)}\left(x\right)g_{q\mid s}\left(q\right)dq$.\medskip{}

\noindent\textbf{Proof:} See Subsection A.11 of Appendix A.\medskip{}

It is well known that any Negative Binomial random variable can be
expressed as a unique continuous mixture of Poisson random variables.
Specifically,
\begin{equation}
f_{X\mid r,q}^{\left(\textrm{NB}\right)}\left(x\right)=\int_{0}^{\infty}f_{X\mid\lambda}^{\left(\textrm{P}\right)}\left(x\right)g_{\lambda\mid r,\tfrac{q}{1-q}}^{\left(\Gamma\right)}\left(\lambda\right)d\lambda,
\end{equation}
where $f_{X\mid\lambda}^{\left(\textrm{P}\right)}\left(x\right)=\tfrac{e^{-\lambda}\lambda^{x}}{x!},\:x\in\mathbb{Z}_{\geq0}$
and $g_{\lambda\mid r,\tfrac{q}{1-q}}^{\left(\Gamma\right)}\left(\lambda\right)=\tfrac{1}{\Gamma\left(r\right)}\left(\tfrac{1-q}{q}\right)^{r}\lambda^{r-1}\exp\left(-\left(\tfrac{1-q}{q}\right)\lambda\right),\:\lambda\in\mathbb{R}_{>0}$
denote the $\textrm{Poisson}\left(\lambda\right)$ PMF and $\textrm{Gamma}\left(r,\tfrac{q}{1-q}\right)$
PDF, respectively. This allows one to show that any Negative Binomial
mixture, $f_{X}\left(x\right)=\int_{0}^{1}f_{X\mid r,q}^{\left(\textrm{NB}\right)}\left(x\right)g_{q\mid r}\left(q\right)dq$,
can be expressed as a unique Poisson mixture, $f_{X}\left(x\right)=$$\int_{0}^{\infty}f_{X\mid\lambda}^{\left(\textrm{P}\right)}\left(x\right)g_{\lambda}\left(\lambda\right)d\lambda$.\medskip{}

\noindent\textbf{Corollary 3.1.2:} For a given frequency, $X\sim f_{X}\left(x\right)$,
if there exists a mixing PDF, $g_{q\mid r}\left(p\right)$, satisfying
$f_{X}\left(x\right)=\int_{0}^{1}f_{X\mid r,q}^{\left(\textrm{NB}\right)}\left(x\right)g_{q\mid r}\left(q\right)dq$
for some $r\in\mathbb{R}_{>0}$, then the function
\begin{equation}
g_{\lambda}\left(\lambda\right)={\displaystyle \int_{0}^{1}}\dfrac{1}{\Gamma\left(r\right)}\left(\dfrac{1-q}{q}\right)^{r}\lambda^{r-1}\exp\left(-\left(\dfrac{1-q}{q}\right)\lambda\right)g_{q\mid r}\left(q\right)dq
\end{equation}
is the unique PDF satisfying $f_{X}\left(x\right)=\int_{0}^{\infty}f_{X\mid\lambda}^{\left(\textrm{P}\right)}\left(x\right)g_{\lambda}\left(\lambda\right)d\lambda$.\medskip{}

\noindent\textbf{Proof:} See Subsection A.12 of Appendix A.\medskip{}

Since $X\mid r,q\sim f_{X\mid r,q}^{\left(\textrm{NB}\right)}\left(x\right)$
converges in distribution to $X\mid\lambda\sim f_{X\mid\lambda}^{\left(\textrm{P}\right)}\left(x\right)$
as $r\rightarrow\infty$ and $rq\rightarrow\lambda$, we can see that
$g_{\lambda}\left(\lambda\right)$ is the limiting distribution of
$\tfrac{rq}{1-q}$ as $r\rightarrow\infty$, where $q\sim g_{q\mid s}\left(q\right)$
lies in the calibrated family $\left[f_{X\mid s,q}^{\left(\textrm{NB}\right)}\left(x\right),g_{q\mid s}\left(q\right)\right]$.

\subsection{Calibrated Severity Mixtures}

\noindent The following analogue of Theorem 3.1 provides the mathematical
formulation of the calibrated severity family, $\left[f_{Y\mid s,\theta}^{\left(\Gamma\right)}\left(y\right),g_{\theta\mid s}\left(\theta\right)\right]$,
for $s\in\left[r,\infty\right)$.\medskip{}

\noindent\textbf{Theorem 3.2:} For a given severity, $Y\sim f_{Y}\left(y\right)$,
if there exists a mixing PDF, $g_{\theta\mid r}\left(\theta\right)$,
satisfying $f_{Y}\left(y\right)=\int_{0}^{\infty}f_{Y\mid r,\theta}^{\left(\Gamma\right)}\left(y\right)g_{\theta\mid r}\left(\theta\right)d\theta$
for some $r\in\mathbb{R}_{>0}$, then:

\noindent (a) $g_{\theta\mid r}\left(\theta\right)$ is unique;

\noindent (b) for all $s\in\left(r,\infty\right)$, the function
\begin{equation}
g_{\theta\mid s}\left(\theta\right)=\dfrac{1}{\mathcal{B}\left(r,s-r\right)}\int_{\theta}^{\infty}\left(\dfrac{\theta}{\omega-\theta}\right)^{r-1}\left(\dfrac{\omega-\theta}{\omega}\right)^{s-2}\left(\dfrac{1}{\omega}\right)g_{\theta\mid r}\left(\omega\right)d\omega
\end{equation}
is the unique PDF satisfying $f_{Y}\left(y\right)=\int_{0}^{\infty}f_{Y\mid s,\theta}^{\left(\Gamma\right)}\left(y\right)g_{\theta\mid s}\left(\theta\right)d\theta$;
and

\noindent (c) for all $s\in\left(0,r\right)$, the function
\[
g_{\theta\mid s}\left(\theta\right)=\dfrac{1}{\left(s-r\right)\mathcal{B}\left(r,s-r\right)}\left\{ \left(s-1\right){\displaystyle \int_{\theta}^{\infty}}\left(\dfrac{\theta}{\omega-\theta}\right)^{r-1}\left(\dfrac{\omega-\theta}{\omega}\right)^{s-1}\left(\dfrac{1}{\omega}\right)g_{\theta\mid r}\left(\omega\right)d\omega\right.
\]
\begin{equation}
\left.-{\displaystyle \int_{\theta}^{\infty}}\left(\dfrac{\theta}{\omega-\theta}\right)^{r-1}\left(\dfrac{\omega-\theta}{\omega}\right)^{s-1}g_{\theta\mid r}^{\prime}\left(\omega\right)d\omega\right\} 
\end{equation}
is either the unique PDF or a quasi-PDF (such that $g_{\theta\mid s}\left(\theta\right)<0$
for some $\theta\in\mathbb{R}_{>0}$) satisfying $f_{Y}\left(y\right)=\int_{0}^{\infty}f_{Y\mid s,\theta}^{\left(\Gamma\right)}\left(y\right)g_{\theta\mid s}\left(\theta\right)d\theta$.\medskip{}

\noindent\textbf{Proof:} See Subsection A.13 of Appendix A.

\medskip{}

As in the frequency case, the determination of whether or not $g_{\theta\mid s}\left(\theta\right)$
is a quasi-PDF (when $s<r$) depends on the behavior of the original
mixing PDF, $g_{q\mid r}\left(q\right)$. In particular, we can see
from the two integrands in (40) that the right-hand side is negative
if and only if
\[
\textrm{E}_{\theta\mid r}\left[\dfrac{\left(s-1\right)}{\theta}-\dfrac{g_{\theta\mid r}^{\prime}\left(\theta\right)}{g_{\theta\mid r}\left(\theta\right)}\right]<0,
\]
a condition more likely to be true if the elasticity function, $e_{\theta\mid r}\left(\theta\right)=\tfrac{\theta g_{\theta\mid r}^{\prime}\left(\theta\right)}{g_{\theta\mid r}\left(\theta\right)}$,
tends to be greater than $s-1$. Consequently, quasi-PDFs occur more
often when $s$ is small and $g_{\theta\mid r}\left(\theta\right)$
tends to be positive sloping. For $r=1$, the following corollary
identifies a sufficient condition entirely analogous to the one stated
in Corollary 3.1.1.\medskip{}

\noindent\textbf{Corollary 3.2: }For a given severity, $Y\sim f_{Y}\left(y\right)$,
and mixing PDF, $g_{\theta\mid r=1}\left(\theta\right)$, satisfying
$f_{Y}\left(y\right)=$\linebreak{}
$\int_{0}^{\infty}f_{Y\mid r=1,\theta}^{\left(\Gamma\right)}\left(y\right)g_{\theta\mid r=1}\left(\theta\right)d\theta$,
let $\underset{\theta\downarrow0}{\lim}\:e_{\theta\mid r=1}\left(\theta\right)>s-1$
for $s\in\left(0,1\right)$. Then the function
\begin{equation}
g_{\theta\mid s}\left(\theta\right)=\left(s-1\right){\displaystyle \int_{\theta}^{\infty}}\left(\dfrac{\omega-\theta}{\omega}\right)^{s-1}\left(\dfrac{1}{\omega}\right)g_{\theta\mid r=1}\left(\omega\right)d\omega-{\displaystyle \int_{\theta}^{\infty}}\left(\dfrac{\omega-\theta}{\omega}\right)^{s-1}g_{\theta\mid r=1}^{\prime}\left(\omega\right)d\omega
\end{equation}
is a quasi-PDF, with $g_{\theta\mid s}\left(q\right)<0$ for all $\theta$
in some neighborhood of 0, satisfying $f_{Y}\left(y\right)=$\linebreak{}
$\int_{0}^{\infty}f_{Y\mid s,\theta}^{\left(\Gamma\right)}\left(y\right)g_{\theta\mid s}\left(\theta\right)d\theta$\medskip{}

\noindent\textbf{Proof:} See Subsection A.14 of Appendix A.\medskip{}

For completeness, we note that there is no analogue of Corollary 3.1.2
in the severity case. This is because $Y\mid r,\theta\sim f_{Y\mid r,\theta}^{\left(\Gamma\right)}\left(y\right)$
converges in distribution to the degenerate random variable $Y\equiv\lambda$
as $r\rightarrow\infty$ and $r\theta\rightarrow\lambda$.

\section{Conclusion}

\noindent In the present work, we presented a new approach to assessing
the heterogeneity of risk factors underlying an empirical frequency
or severity distribution by: (i) fitting canonical mixture models
to observed data; and (ii) estimating the shape of the implied mixing
distribution. This involves paying particular attention to any relatively
large collection of weight at the upper end of a mixing distribution's
sample space, such as those attributable to underwriting and/or risk-classification
errors in the insurance sector.

We began by considering the close mathematical connections between
the Negative Binomial and Gamma distributions, which facilitated the
derivation of certain useful theoretical results. We next constructed
flexible 4-parameter families of mixing distributions for generating
heavy-tailed 4-parameter frequency and severity mixtures from Geometric
and Exponential kernels, respectively. These mixtures then were extended
to arbitrary Negative Binomial and Gamma kernels, yielding 5-parameter
families for detecting and measuring risk heterogeneity. To check
the robustness of such inferences, we demonstrated how a fitted 5-parameter
model may be re-expressed in terms of alternative kernels within an
associated ``calibrated'' family.

Future research should focus on applying the proposed modeling approach
to the analysis of risk heterogeneity within actual insurance frequency
and severity data sets. Such applications would assess the presence
and potential impact of risk heterogeneity through a three-step process:
\begin{enumerate}
\item Modeling the historical data with the relevant heavy-tailed 5-parameter
mixture-distribution families ($\mathcal{F}^{\textrm{HGZY}\left(r\right)}$
and $\mathcal{F}^{\textrm{HGZY}^{\prime}\left(r\right)}$ for frequencies
and $\mathcal{F}^{\textrm{HG}\Sigma\Sigma\left(r\right)}$ for severities),
and identifying the member(s) of those families providing the best
fit.
\item Assessing the member(s) of the mixing-distribution families corresponding
to the best-fitting mixture distributions of Step 1 (i.e., $\mathcal{G}^{\textrm{HG}\Sigma\mathcal{B}}$
and $\mathcal{G}^{\textrm{CHG}\Sigma\mathcal{B}}$ for frequencies
and $\mathcal{G}^{\textrm{IHG}\Sigma\Gamma}$ for severities) for
the presence of relatively large collections of probability mass/density
at the upper end of the sample space.
\item Given evidence of risk heterogeneity in Step 2, using calibrated families
to perform a robustness check to ascertain whether or not the observed
heterogeneity persists for larger values of the $r$ parameter.
\end{enumerate}

\begin{center}
{\Large\textbf{Appendix A}}{\Large\par}
\par\end{center}

\subsection*{{\small A.1 Proof of Lemma 1}}

\noindent{\small Applying the transformations $\theta=\tfrac{q}{1-q}$
and $u=1-e^{-t}$, one can write
\[
\mathcal{L}_{X\mid r,q}\left(t\right)=\mathcal{L}_{Y\mid r,\tfrac{q}{1-q}}\left(1-e^{-t}\right)
\]
\[
=\textrm{E}_{Y\mid r,\tfrac{q}{1-q}}\left[e^{-\left(1-e^{-t}\right)Y}\right]
\]
\[
={\displaystyle \sum_{i=0}^{\infty}}\left(-1\right)^{i}\dfrac{\left(1-e^{-t}\right)^{i}}{i!}\textrm{E}_{Y\mid r,\tfrac{q}{1-q}}\left[Y^{i}\right]
\]
\[
={\displaystyle \sum_{i=0}^{\infty}}\dfrac{\left(e^{-t}-1\right)^{i}}{i!}\dfrac{\Gamma\left(r+i\right)}{\Gamma\left(r\right)}\left(\dfrac{q}{1-q}\right)^{i}
\]
\[
=1+{\displaystyle \sum_{i=1}^{\kappa}}\left[\dfrac{1}{i!}{\displaystyle \sum_{j=0}^{i}}\binom{i}{j}\left(-1\right)^{i-j}e^{-jt}\right]\dfrac{\Gamma\left(r+i\right)}{\Gamma\left(r\right)}\left(\dfrac{q}{1-q}\right)^{i}+{\displaystyle \sum_{i=\kappa+1}^{\infty}}\dfrac{\left(e^{-t}-1\right)^{i}}{i!}\dfrac{\Gamma\left(r+i\right)}{\Gamma\left(r\right)}\left(\dfrac{q}{1-q}\right)^{i}
\]
for any $\kappa\in\mathbb{Z}_{\geq0}.$ It then follows that the $\kappa^{\textrm{th}}$
derivative of $\mathcal{L}_{X\mid r,q}\left(t\right)$ with respect
to $t$ is given by
\[
\mathcal{L}_{X\mid r,q}^{\left(\kappa\right)}\left(t\right)={\displaystyle \sum_{i=1}^{\kappa}}\left[\dfrac{1}{i!}{\displaystyle \sum_{j=0}^{i}}\binom{i}{j}\left(-1\right)^{i-j}\left(-j\right)^{\kappa}e^{-jt}\right]\dfrac{\Gamma\left(r+i\right)}{\Gamma\left(r\right)}\left(\dfrac{q}{1-q}\right)^{i}
\]
\[
+{\displaystyle \sum_{i=\kappa+1}^{\infty}}\dfrac{d^{\kappa}}{dt^{\kappa}}\left[\dfrac{\left(e^{-t}-1\right)^{i}}{i!}\right]\dfrac{\Gamma\left(r+i\right)}{\Gamma\left(r\right)}\left(\dfrac{q}{1-q}\right)^{i},
\]
so that
\[
\textrm{E}_{X\mid r,q}\left[X^{\kappa}\right]=\left(-1\right)^{\kappa}\underset{t\downarrow0}{\lim}\:\mathcal{L}_{X\mid r,q}^{\left(\kappa\right)}\left(t\right)
\]
\[
=\left(-1\right)^{\kappa}\underset{t\downarrow0}{\lim}\:\left\{ {\displaystyle \sum_{i=1}^{\kappa}}\left[\dfrac{1}{i!}{\displaystyle \sum_{j=0}^{i}}\binom{i}{j}\left(-1\right)^{i-j}\left(-j\right)^{\kappa}e^{-jt}\right]\dfrac{\Gamma\left(r+i\right)}{\Gamma\left(r\right)}\left(\dfrac{q}{1-q}\right)^{i}\right.
\]
\[
\left.+{\displaystyle \sum_{i=\kappa+1}^{\infty}}\dfrac{d^{\kappa}}{dt^{\kappa}}\left[\dfrac{\left(e^{-t}-1\right)^{i}}{i!}\right]\dfrac{\Gamma\left(r+i\right)}{\Gamma\left(r\right)}\left(\dfrac{q}{1-q}\right)^{i}\right\} 
\]
\[
={\displaystyle \sum_{i=1}^{\infty}}\left[\dfrac{1}{i!}{\displaystyle \sum_{j=0}^{i}}\binom{i}{j}\left(-1\right)^{i-j}j^{\kappa}\right]\dfrac{\Gamma\left(r+i\right)}{\Gamma\left(r\right)}\left(\dfrac{q}{1-q}\right)^{i},
\]
where
\[
\dfrac{1}{i!}{\displaystyle \sum_{j=0}^{i}}\binom{i}{j}\left(-1\right)^{i-j}j^{\kappa}={\displaystyle S\left(\kappa,i\right)}.\:\blacksquare
\]
}{\small\par}

\subsection*{{\small A.2 Proof of Theorem 2.1}}

\noindent{\small Statement (a) is true if and only if
\[
\textrm{E}_{X\mid r}\left[X^{\kappa}\right]=\infty
\]
\[
\Longleftrightarrow{\displaystyle \int_{0}^{1}\textrm{E}_{X\mid r,q}^{\left(\textrm{NB}\right)}\left[X^{\kappa}\right]g_{q}\left(q\right)dq}=\infty\qquad\qquad\qquad\qquad\textrm{(A1)}
\]
for some $\kappa\in\mathbb{Z}_{\geq1}$, where it is known from Lemma
1 that
\[
\textrm{E}_{X\mid r,q}^{\left(\textrm{NB}\right)}\left[X^{\kappa}\right]={\displaystyle \sum_{i=1}^{\kappa}S\left(\kappa,i\right)\dfrac{\Gamma\left(r+i\right)}{\Gamma\left(r\right)}\left(\dfrac{q}{1-q}\right)^{i}}
\]
\[
={\displaystyle O\left(\left(1-q\right)^{-\kappa}\right)}
\]
as $q\uparrow1$ for all $r\in\mathbb{R}_{>0}$. From statement (b),
we know that, for some $\rho\in\mathbb{R}_{>0}$ and any $L\in\mathbb{R}_{>0}$,
there exists $q_{L}\in\left(0,1\right)$ such that $q\geq q_{L}\Longrightarrow g_{q}\left(q\right)\left(1-q\right)^{1-\rho}\geq L>0$.
Thus,
\[
{\displaystyle \int_{0}^{1}\textrm{E}_{X\mid r,q}^{\left(\textrm{NB}\right)}\left[X^{\kappa}\right]g_{q}\left(q\right)dq}\geq\int_{0}^{q_{L}}\textrm{E}_{X\mid r,q}^{\left(\textrm{NB}\right)}\left[X^{\kappa}\right]g_{q}\left(q\right)dq+\int_{q_{L}}^{1}\textrm{E}_{X\mid r,q}^{\left(\textrm{NB}\right)}\left[X^{\kappa}\right]\left(L\right)O\left(\left(1-q\right)^{\rho-1}\right)dq
\]
\[
=\int_{0}^{q_{L}}\textrm{E}_{X\mid r,q}^{\left(\textrm{NB}\right)}\left[X^{\kappa}\right]g_{q}\left(q\right)dq+L\int_{q_{L}}^{1}O\left(\left(1-q\right)^{-\kappa}\right)O\left(\left(1-q\right)^{\rho-1}\right)dq
\]
\[
=\int_{0}^{q_{L}}\textrm{E}_{X\mid r,q}^{\left(\textrm{NB}\right)}\left[X^{\kappa}\right]g_{q}\left(q\right)dq+L\int_{q_{L}}^{1}O\left(\left(1-q\right)^{\rho-1-\kappa}\right)dq,
\]
where the second term on the right-hand side is infinite for all $\kappa\geq\rho$,
confirming (A1).}{\small\par}

{\small To show that condition (b) is implied by (A1)'s holding true
for some $\kappa\in\mathbb{Z}_{\geq1}$, we first assume the negation
of (b) (i.e., $\underset{q\uparrow1}{\lim}\:g_{q}\left(q\right)\left(1-q\right)^{1-\rho}=L_{\rho}<\infty$
for all $\rho\in\mathbb{R}_{>0}$, since $g_{q}\left(q\right)\left(1-q\right)^{1-\rho}$
cannot diverge by oscillation because $g_{q}\left(q\right)$ does
not oscillate as $q\uparrow1$). This means that, for all $\rho\in\mathbb{R}_{>0}$
and any $\eta\in\mathbb{R}_{>0}$, $\underset{q\uparrow1}{\lim}\:g_{q}\left(q\right)\left(1-q\right)^{\left(1+\eta\right)-\rho}=0$;
and consequently, for any $\varepsilon\in\mathbb{R}_{>0}$, there
exists $q_{\eta,\varepsilon}\in\left(0,1\right)$ such that $q\geq q_{\eta,\varepsilon}\Longrightarrow g_{q}\left(q\right)\left(1-q\right)^{\left(1+\eta\right)-\rho}\leq\varepsilon$.
It then follows that
\[
{\displaystyle \int_{0}^{1}\textrm{E}_{X\mid r,q}^{\left(\textrm{NB}\right)}\left[X^{\kappa}\right]g_{q}\left(q\right)dq}\leq\int_{0}^{q_{\eta,\varepsilon}}\textrm{E}_{X\mid r,q}^{\left(\textrm{NB}\right)}\left[X^{\kappa}\right]g_{q}\left(q\right)dq+{\displaystyle \int_{q_{\eta,\varepsilon}}^{1}\textrm{E}_{X\mid r,q}^{\left(\textrm{NB}\right)}\left[X^{\kappa}\right]\left(\varepsilon\right)O\left(\left(1-q\right)^{\rho-\left(1+\eta\right)}\right)dq}
\]
\[
=\int_{0}^{q_{\eta,\varepsilon}}\textrm{E}_{X\mid r,q}^{\left(\textrm{NB}\right)}\left[X^{\kappa}\right]g_{q}\left(q\right)dq+\varepsilon{\displaystyle \int_{q_{\eta,\varepsilon}}^{1}O\left(\left(1-q\right)^{-\kappa}\right)O\left(\left(1-q\right)^{\rho-\left(1+\eta\right)}\right)dq}
\]
\[
=\int_{0}^{q_{\eta,\varepsilon}}\textrm{E}_{X\mid r,q}^{\left(\textrm{NB}\right)}\left[X^{\kappa}\right]g_{q}\left(q\right)dq+\varepsilon{\displaystyle \int_{q_{\eta,\varepsilon}}^{1}O\left(\left(1-q\right)^{\rho-1-\eta-\kappa}\right)dq},
\]
where both terms on the right-hand side are finite for all $\kappa<\rho-\eta$
for all $\rho\in\mathbb{R}_{>0}$, implying the negation of (A1).}{\small\par}

{\small To demonstrate the equivalence of statements (b) and (c), let
$R\left(q\right)=\tfrac{\ln\left(g_{q}\left(q\right)\right)}{\ln\left(1-q\right)}$
and rewrite (b) as
\[
\underset{q\uparrow1}{\lim}\:\left(1-q\right)^{R\left(q\right)+1-\rho}>0\textrm{ for some }\rho\in\mathbb{R}_{>0}.
\]
Then
\[
\underset{q\uparrow1}{\lim}\:\left(R\left(q\right)+1-\rho\right)\ln\left(1-q\right)>-\infty\textrm{ for some }\rho\in\mathbb{R}_{>0}
\]
\[
\Longleftrightarrow\underset{q\uparrow1}{\lim}\:R\left(q\right)\leq\rho-1\textrm{ for some }\rho\in\mathbb{R}_{>0}
\]
\[
\Longleftrightarrow\underset{q\uparrow1}{\lim}\:\dfrac{\ln\left(g_{q}\left(q\right)\right)}{\ln\left(1-q\right)}<\infty.\:\blacksquare
\]
}{\small\par}

\subsection*{{\small A.3 Proof of Theorem 2.2}}

\noindent{\small Given that
\[
\textrm{E}_{Y\mid r,\theta}^{\left(\Gamma\right)}\left[Y^{\kappa}\right]=\dfrac{\Gamma\left(r+\kappa\right)\theta^{\kappa}}{\Gamma\left(r\right)}
\]
for $r\in\mathbb{R}_{>0}$ and $\kappa\in\mathbb{Z}_{\geq1}$, the
theorem follows from arguments analogous to those in the proof of
Theorem 2.1, with $X\mid r,q$ replaced by $Y\mid r,\theta$, $q\in\left(0,1\right)$
replaced by $\theta\in\mathbb{R}_{>0}$, and limits taken as $q\rightarrow1^{+}$
replaced by limits as $\theta\rightarrow\infty$.$\:\blacksquare$}{\small\par}

\subsection*{{\small A.4 Constant of Integration for $\boldsymbol{g_{q\mid a,b,c,d}^{\left(\textrm{HG}\Sigma\textrm{B}\right)}\left(q\right)}$}}

\noindent{\small To solve for $K$, the constant of integration in
(23), set}{\small\par}

{\small
\[
K{\displaystyle \int}_{0}^{1}\dfrac{q^{ca-1}\left(1-q^{c}\right)^{b}}{\left(1-q^{d}\right)}dq=1.
\]
Substituting $\varrho=q^{c}$ into this integral then yields
\[
K{\displaystyle \int_{0}^{1}}\dfrac{\varrho^{a-1/c}\left(1-\varrho\right)^{b}}{\left(1-\varrho^{d/c}\right)}\left(\dfrac{1}{c}\right)\varrho^{1/c-1}d\varrho=1
\]
\[
\Longleftrightarrow{\displaystyle \int_{0}^{1}}\dfrac{\varrho^{a-1}\left(1-\varrho\right)^{b}}{\left(1-\varrho^{d/c}\right)}d\varrho=\dfrac{c}{K}
\]
\[
\Longleftrightarrow{\displaystyle \int_{0}^{1}}\varrho^{a-1}\left(1-\varrho\right)^{b}\left({\displaystyle \sum_{i=0}^{\infty}\varrho^{id/c}}\right)d\varrho=\dfrac{c}{K}
\]
\[
\Longleftrightarrow\sum_{i=0}^{\infty}\left[{\displaystyle \int_{0}^{1}}\varrho^{a+id/c-1}\left(1-\varrho\right)^{b}d\varrho\right]=\dfrac{c}{K}
\]
\[
\Longleftrightarrow\sum_{i=0}^{\infty}\mathcal{B}\left(a+\dfrac{id}{c},b+1\right)=\dfrac{c}{K}
\]
\[
\Longleftrightarrow K=\dfrac{c}{\Sigma_{\mathcal{B}}\left(\dfrac{d}{c},a,b\right)},
\]
where $\Sigma_{\mathcal{B}}\left(\xi,v,w\right)={\textstyle \sum_{k=0}^{\infty}}\mathcal{B}\left(\xi k+v,w+1\right)$.$\:\blacksquare$}{\small\par}

\subsection*{{\small A.5 Expression for $\boldsymbol{f_{X\mid a,b,c,d}^{\left(\textrm{HGZY}\left(s\right)\right)}\left(x\right)}$}}

\noindent{\small Let
\[
f_{X\mid r,a,b,c,d}^{\left(\textrm{HGZY}\left(r\right)\right)}\left(x\right)={\displaystyle \int_{0}^{1}}f_{X\mid r,q}^{\left(\textrm{NB}\right)}\left(x\right)g_{q\mid a,b,c,d}^{\left(\textrm{HG}\Sigma\mathcal{B}\right)}\left(q\right)dq
\]
\[
={\displaystyle \int_{0}^{1}}\dfrac{\Gamma\left(r+x\right)}{\Gamma\left(r\right)\Gamma\left(x+1\right)}\left(1-q\right)^{r}q^{x}\dfrac{c}{\Sigma_{\mathcal{B}}\left(\dfrac{d}{c},a,b\right)}\dfrac{q^{ca-1}\left(1-q^{c}\right)^{b}}{\left(1-q^{d}\right)}dq
\]
\[
=\dfrac{c}{\Sigma_{\mathcal{B}}\left(\dfrac{d}{c},a,b\right)}\dfrac{\Gamma\left(r+x\right)}{\Gamma\left(r\right)\Gamma\left(x+1\right)}{\displaystyle \int_{0}^{1}}\left[\sum_{\ell=0}^{\infty}\binom{r}{\ell}{\displaystyle \left(-q\right)^{\ell}}\right]\dfrac{q^{ca+x-1}\left(1-q^{c}\right)^{b}}{\left(1-q^{d}\right)}dq.
\]
Substituting $\varrho=q^{c}$ into the integral then yields
\[
f_{X\mid r,a,b,c,d}^{\left(\textrm{HGZY}\left(r\right)\right)}\left(x\right)=\dfrac{c}{\Sigma_{\mathcal{B}}\left(\dfrac{d}{c},a,b\right)}\dfrac{\Gamma\left(r+x\right)}{\Gamma\left(r\right)\Gamma\left(x+1\right)}\sum_{\ell=0}^{\infty}\left[\binom{r}{\ell}\left(-1\right)^{\ell}{\displaystyle \int_{0}^{1}}\dfrac{\varrho^{a+\left(x+\ell-1\right)/c}\left(1-\varrho\right)^{b}}{\left(1-\varrho^{d/c}\right)}\left(\dfrac{1}{c}\right)\varrho^{1/c-1}d\varrho\right]
\]
\[
=\dfrac{1}{\Sigma_{\mathcal{B}}\left(\dfrac{d}{c},a,b\right)}\dfrac{\Gamma\left(r+x\right)}{\Gamma\left(r\right)\Gamma\left(x+1\right)}\sum_{\ell=0}^{\infty}\left[{\displaystyle \binom{r}{\ell}\left(-1\right)^{\ell}}{\displaystyle \int_{0}^{1}}\dfrac{\varrho^{a+\left(x+\ell\right)/c-1}\left(1-\varrho\right)^{b}}{\left(1-\varrho^{d/c}\right)}d\varrho\right]
\]
\[
=\dfrac{1}{\Sigma_{\mathcal{B}}\left(\dfrac{d}{c},a,b\right)}\dfrac{\Gamma\left(r+x\right)}{\Gamma\left(r\right)\Gamma\left(x+1\right)}\sum_{\ell=0}^{\infty}\left[{\displaystyle \binom{r}{\ell}\left(-1\right)^{\ell}}{\displaystyle \int_{0}^{1}}\varrho^{a+\left(x+\ell\right)/c-1}\left(1-\varrho\right)^{b}\left({\displaystyle \sum_{i=0}^{\infty}\varrho^{id/c}}\right)d\varrho\right]
\]
\[
=\dfrac{1}{\Sigma_{\mathcal{B}}\left(\dfrac{d}{c},a,b\right)}\dfrac{\Gamma\left(r+x\right)}{\Gamma\left(r\right)\Gamma\left(x+1\right)}\sum_{\ell=0}^{\infty}\left\{ {\displaystyle \binom{r}{\ell}\left(-1\right)^{\ell}}\sum_{i=0}^{\infty}\left[{\displaystyle \int_{0}^{1}}\varrho^{a+\left(x+\ell+id\right)/c-1}\left(1-\varrho\right)^{b}d\varrho\right]\right\} 
\]
\[
=\dfrac{1}{\Sigma_{\mathcal{B}}\left(\dfrac{d}{c},a,b\right)}\dfrac{\Gamma\left(r+x\right)}{\Gamma\left(r\right)\Gamma\left(x+1\right)}\sum_{\ell=0}^{\infty}\left[{\displaystyle \binom{r}{\ell}\left(-1\right)^{\ell}}\sum_{i=0}^{\infty}\mathcal{B}\left(a+\dfrac{\left(x+\ell+id\right)}{c},b+1\right)\right]
\]
\[
=\dfrac{1}{\Sigma_{\mathcal{B}}\left(\dfrac{d}{c},a,b\right)}\dfrac{\Gamma\left(r+x\right)}{\Gamma\left(r\right)\Gamma\left(x+1\right)}\sum_{\ell=0}^{\infty}{\displaystyle \binom{r}{\ell}\left(-1\right)^{\ell}}\Sigma_{\mathcal{B}}\left(\dfrac{d}{c},a+\dfrac{\left(x+\ell\right)}{c},b\right)
\]
\[
=\dfrac{\Gamma\left(r+x\right)}{\Gamma\left(r\right)\Gamma\left(x+1\right)}{\displaystyle \sum_{\ell=0}^{\infty}\binom{r}{\ell}\left(-1\right)^{\ell}}\dfrac{\Sigma_{\mathcal{B}}\left(\dfrac{d}{c},a+\dfrac{\left(x+\ell\right)}{c},b\right)}{\Sigma_{\mathcal{B}}\left(\dfrac{d}{c},a,b\right)}.\:\blacksquare
\]
}{\small\par}

\subsection*{{\small A.6 Expression for $\boldsymbol{f_{X\mid r,a,b,c,d}^{\left(\textrm{HGZY}^{\prime}\left(r\right)\right)}\left(x\right)}$}}

\noindent{\small Let
\[
f_{X\mid r,a,b,c,d}^{\left(\textrm{HGZY}^{\prime}\left(r\right)\right)}\left(x\right)={\displaystyle \int_{0}^{1}}f_{X\mid r,q}^{\left(\textrm{NB}\right)}\left(x\right)g_{q\mid a,b,c,d}^{\left(\textrm{CHG}\Sigma\mathcal{B}\right)}\left(q\right)dq
\]
\[
={\displaystyle \int_{0}^{1}}\dfrac{\Gamma\left(r+x\right)}{\Gamma\left(r\right)\Gamma\left(x+1\right)}\left(1-q\right)^{r}q^{x}\dfrac{c}{\Sigma_{\mathcal{B}}\left(\dfrac{d}{c},a,b\right)}\dfrac{\left(1-q\right)^{ca-1}\left[1-\left(1-q\right)^{c}\right]^{b}}{\left[1-\left(1-q\right)^{d}\right]}dq
\]
\[
=\dfrac{c}{\Sigma_{\mathcal{B}}\left(\dfrac{d}{c},a,b\right)}\dfrac{\Gamma\left(r+x\right)}{\Gamma\left(r\right)\Gamma\left(x+1\right)}{\displaystyle \int_{0}^{1}}\left(1-q\right)^{r}q^{x}\dfrac{\left(1-q\right)^{ca-1}\left[1-\left(1-q\right)^{c}\right]^{b}}{\left[1-\left(1-q\right)^{d}\right]}dq.
\]
Substituting $\varsigma=1-q$ into the integral then gives
\[
f_{X\mid r,a,b,c,d}^{\left(\textrm{HGZY}^{\prime}\left(r\right)\right)}\left(x\right)=\dfrac{c}{\Sigma_{\mathcal{B}}\left(\dfrac{d}{c},a,b\right)}\dfrac{\Gamma\left(r+x\right)}{\Gamma\left(r\right)\Gamma\left(x+1\right)}{\displaystyle \int_{0}^{1}}\varsigma^{r}\left(1-\varsigma\right)^{x}\dfrac{\varsigma^{ca-1}\left(1-\varsigma^{c}\right)^{b}}{\left(1-\varsigma^{d}\right)}d\varsigma
\]
\[
=\dfrac{c}{\Sigma_{\mathcal{B}}\left(\dfrac{d}{c},a,b\right)}\dfrac{\Gamma\left(r+x\right)}{\Gamma\left(r\right)\Gamma\left(x+1\right)}{\displaystyle \int_{0}^{1}}\left[{\displaystyle \sum_{j=0}^{x}}\binom{x}{j}\left(-\varsigma\right)^{j}\right]\dfrac{\varsigma^{ca+r-1}\left(1-\varsigma^{c}\right)^{b}}{\left(1-\varsigma^{d}\right)}d\varsigma
\]
\[
=\dfrac{c}{\Sigma_{\mathcal{B}}\left(\dfrac{d}{c},a,b\right)}\dfrac{\Gamma\left(r+x\right)}{\Gamma\left(r\right)\Gamma\left(x+1\right)}{\displaystyle \sum_{j=0}^{x}}\left[\binom{x}{j}\left(-1\right)^{j}{\displaystyle \int_{0}^{1}}\dfrac{\varsigma^{ca+r+j-1}\left(1-\varsigma^{c}\right)^{b}}{\left(1-\varsigma^{d}\right)}d\varsigma\right],
\]
and the further substitution of $\varrho=\varsigma^{c}$ into each
integral in the summation yields
\[
f_{X\mid r,a,b,c,d}^{\left(\textrm{HGZY}^{\prime}\left(r\right)\right)}\left(x\right)=\dfrac{c}{\Sigma_{\mathcal{B}}\left(\dfrac{d}{c},a,b\right)}\dfrac{\Gamma\left(r+x\right)}{\Gamma\left(r\right)\Gamma\left(x+1\right)}{\displaystyle \sum_{j=0}^{x}}\left[\binom{x}{j}\left(-1\right)^{j}{\displaystyle \int_{0}^{1}}\dfrac{\varrho^{a+\left(r+j-1\right)/c}\left(1-\varrho\right)^{b}}{\left(1-\varrho^{d/c}\right)}\left(\dfrac{1}{c}\right)\varrho^{1/c-1}d\varrho\right]
\]
\[
=\dfrac{1}{\Sigma_{\mathcal{B}}\left(\dfrac{d}{c},a,b\right)}\dfrac{\Gamma\left(r+x\right)}{\Gamma\left(r\right)\Gamma\left(x+1\right)}{\displaystyle \sum_{j=0}^{x}}\left[\binom{x}{j}\left(-1\right)^{j}{\displaystyle \int_{0}^{1}}\dfrac{\varrho^{a+\left(r+j\right)/c-1}\left(1-\varrho\right)^{b}}{\left(1-\varrho^{d/c}\right)}d\varrho\right].
\]
By rewriting $\left(1-\varrho^{d/c}\right)^{-1}$ as ${\textstyle \sum_{i=0}^{\infty}}\varrho^{id/c}$
as in Subsection A.5, we then obtain
\[
f_{X\mid r,a,b,c,d}^{\left(\textrm{HGZY}^{\prime}\left(r\right)\right)}\left(x\right)=\dfrac{1}{\Sigma_{\mathcal{B}}\left(\dfrac{d}{c},a,b\right)}\dfrac{\Gamma\left(r+x\right)}{\Gamma\left(r\right)\Gamma\left(x+1\right)}{\displaystyle \sum_{j=0}^{x}}\left[\binom{x}{j}\left(-1\right)^{j}\sum_{i=0}^{\infty}\mathcal{B}\left(a+\dfrac{\left(r+j\right)}{c}+\dfrac{id}{c},b+1\right)\right]
\]
\[
=\dfrac{\Gamma\left(r+x\right)}{\Gamma\left(r\right)\Gamma\left(x+1\right)}\sum_{j=0}^{x}\binom{x}{j}\left(-1\right)^{j}\dfrac{\Sigma_{\mathcal{B}}\left(\dfrac{d}{c},a+\dfrac{\left(r+j\right)}{c},b\right)}{\Sigma_{\mathcal{B}}\left(\dfrac{d}{c},a,b\right)}.\:\blacksquare
\]
}{\small\par}

\subsubsection*{{\small A.7 Constant of Integration for $\boldsymbol{g_{\theta\mid\alpha,\beta,\gamma,\delta}^{\left(\textrm{HG}\Sigma\Gamma\right)}}\left(\theta\right)$}}

\noindent{\small To solve for $K$, the constant of integration in
(28), set
\[
K{\displaystyle \int_{0}^{\infty}}\theta^{\gamma\alpha-1}\exp\left(-\beta\theta^{\gamma}\right)e^{\theta/\delta}d\theta=1.
\]
Substituting $\vartheta=\theta^{\gamma}$ into this integral then
yields
\[
{\displaystyle \int_{0}^{\infty}}\vartheta^{\alpha-1/\gamma}e^{-\beta\vartheta}\exp\left(\left(\dfrac{1}{\delta}\right)\vartheta^{1/\gamma}\right)\left(\dfrac{1}{\gamma}\right)\vartheta^{1/\gamma-1}d\vartheta=\dfrac{1}{K}
\]
\[
\Longleftrightarrow{\displaystyle \int_{0}^{\infty}}\vartheta^{\alpha-1}e^{-\beta\vartheta}\exp\left(\left(\dfrac{1}{\delta}\right)\vartheta^{1/\gamma}\right)d\vartheta=\dfrac{\gamma}{K}
\]
\[
\Longleftrightarrow{\displaystyle \int_{0}^{\infty}}\vartheta^{\alpha-1}e^{-\beta\vartheta}\left[{\displaystyle \sum_{i=0}^{\infty}\dfrac{\left(1/\delta\right)^{i}\vartheta^{i/\gamma}}{i!}}\right]d\vartheta=\dfrac{\gamma}{K}
\]
\[
\Longleftrightarrow\sum_{i=0}^{\infty}\left[\dfrac{\left(1/\delta\right)^{i}}{i!}{\displaystyle \int_{0}^{\infty}}\vartheta^{\alpha+i/\gamma-1}e^{-\beta\vartheta}d\vartheta\right]=\dfrac{\gamma}{K}
\]
\[
\Longleftrightarrow\sum_{i=0}^{\infty}\dfrac{\left(1/\delta\right)^{i}}{i!}\dfrac{\Gamma\left(\alpha+\dfrac{i}{\gamma}\right)}{\beta^{\alpha+i/\gamma}}=\dfrac{\gamma}{K}
\]
\[
\Longleftrightarrow K=\dfrac{\gamma\beta^{\alpha}}{\sum_{i=0}^{\infty}\dfrac{\left(\beta^{-1/\gamma}/\delta\right)^{i}}{i!}\Gamma\left(\alpha+\dfrac{i}{\gamma}\right)}
\]
\[
\Longleftrightarrow K=\dfrac{\gamma\beta^{\alpha}}{\Sigma_{\Gamma}\left(\dfrac{1}{\gamma},\alpha,\dfrac{\beta^{-1/\gamma}}{\delta}\right)},
\]
where $\Sigma_{\Gamma}\left(\xi,v,w\right)\equiv{\textstyle \sum_{k=0}^{\infty}}\dfrac{w^{k}}{k!}\Gamma\left(\xi k+v\right).\:\blacksquare$}{\small\par}

\subsubsection*{{\small A.8 Expression for $\boldsymbol{f_{Y\mid r,\alpha,\beta,\gamma,\delta}^{\left(\textrm{HG}\Sigma\Sigma\left(r\right)\right)}\left(y\right)}$}}

\noindent{\small Let
\[
f_{Y\mid r,\alpha,\beta,\gamma,\delta}^{\left(\textrm{HG}\Sigma\Sigma\left(r\right)\right)}\left(y\right)={\displaystyle \int_{0}^{1}}f_{Y\mid r,\theta}^{\left(\Gamma\right)}\left(y\right)g_{\theta\mid\alpha,\beta,\gamma,\delta}^{\left(\textrm{IHG}\Sigma\Gamma\right)}\left(\theta\right)d\theta
\]
\[
={\displaystyle \int_{0}^{\infty}}\dfrac{y^{r-1}e^{-y/\theta}}{\Gamma\left(r\right)\theta^{r}}\dfrac{\gamma\beta^{\alpha}}{\Sigma_{\Gamma}\left(\dfrac{1}{\gamma},\alpha,\dfrac{\beta^{-1/\gamma}}{\delta}\right)}\dfrac{\exp\left(-\beta\theta^{-\gamma}\right)e^{1/\left(\delta\theta\right)}}{\theta^{\gamma\alpha+1}}d\theta.
\]
Substituting $\varphi=\tfrac{1}{\theta}$ into the integral gives
\[
f_{Y\mid r,\alpha,\beta,\gamma,\delta}^{\left(\textrm{HG}\Sigma\Sigma\left(r\right)\right)}\left(y\right)=\dfrac{\gamma\beta^{\alpha}}{\Sigma_{\Gamma}\left(\dfrac{1}{\gamma},\alpha,\dfrac{\beta^{-1/\gamma}}{\delta}\right)}\dfrac{y^{r-1}}{\Gamma\left(r\right)}{\displaystyle \int_{0}^{\infty}}\varphi^{\gamma\alpha+r-1}\exp\left(-\beta\varphi^{\gamma}\right)e^{\left(1/\delta-y\right)\varphi}d\varphi,
\]
and the further substitution $\vartheta=\varphi^{\gamma}$ yields
\[
f_{Y\mid r,\alpha,\beta,\gamma,\delta}^{\left(\textrm{HG}\Sigma\Sigma\left(r\right)\right)}\left(y\right)=\dfrac{\gamma\beta^{\alpha}}{\Sigma_{\Gamma}\left(\dfrac{1}{\gamma},\alpha,\dfrac{\beta^{-1/\gamma}}{\delta}\right)}\dfrac{y^{r-1}}{\Gamma\left(r\right)}{\displaystyle \int_{0}^{\infty}}\vartheta^{\alpha+\left(r-1\right)/\gamma}e^{-\beta\vartheta}\exp\left(\left(\dfrac{1}{\delta}-y\right)\vartheta^{1/\gamma}\right)\left(\dfrac{1}{\gamma}\right)\vartheta^{1/\gamma-1}d\vartheta
\]
\[
=\dfrac{\beta^{\alpha}}{\Sigma_{\Gamma}\left(\dfrac{1}{\gamma},\alpha,\dfrac{\beta^{-1/\gamma}}{\delta}\right)}\dfrac{y^{r-1}}{\Gamma\left(r\right)}{\displaystyle \int_{0}^{\infty}}\vartheta^{\alpha+r/\gamma-1}e^{-\beta\vartheta}\exp\left(\left(\dfrac{1}{\delta}-y\right)\vartheta^{1/\gamma}\right)d\vartheta
\]
\[
=\dfrac{\beta^{\alpha}}{\Sigma_{\Gamma}\left(\dfrac{1}{\gamma},\alpha,\dfrac{\beta^{-1/\gamma}}{\delta}\right)}\dfrac{y^{r-1}}{\Gamma\left(r\right)}{\displaystyle \int_{0}^{\infty}}\vartheta^{\alpha+r/\gamma-1}e^{-\beta\vartheta}\left[{\displaystyle \sum_{i=0}^{\infty}}\dfrac{\left(1/\delta-y\right)^{i}}{i!}\vartheta^{i/\gamma}\right]d\vartheta
\]
\[
=\dfrac{\beta^{\alpha}}{\Sigma_{\Gamma}\left(\dfrac{1}{\gamma},\alpha,\dfrac{\beta^{-1/\gamma}}{\delta}\right)}\dfrac{y^{r-1}}{\Gamma\left(r\right)}{\displaystyle \sum_{i=0}^{\infty}}\left[\dfrac{\left(1/\delta-y\right)^{i}}{i!}{\displaystyle \int_{0}^{\infty}}\vartheta^{\alpha+\left(r+i\right)/\gamma-1}e^{-\beta\vartheta}d\vartheta\right]
\]
\[
=\dfrac{\beta^{\alpha}}{\Sigma_{\Gamma}\left(\dfrac{1}{\gamma},\alpha,\dfrac{\beta^{-1/\gamma}}{\delta}\right)}\dfrac{y^{r-1}}{\Gamma\left(r\right)}{\displaystyle \sum_{i=0}^{\infty}}\dfrac{\left(1/\delta-y\right)^{i}}{i!}\dfrac{\Gamma\left(\alpha+\dfrac{\left(r+i\right)}{\gamma}\right)}{\beta^{\alpha+\left(r+i\right)/\gamma}}
\]
\[
=\dfrac{\beta^{-s/\gamma}}{\Sigma_{\Gamma}\left(\dfrac{1}{\gamma},\alpha,\dfrac{\beta^{-1/\gamma}}{\delta}\right)}\dfrac{y^{r-1}}{\Gamma\left(r\right)}{\displaystyle \sum_{i=0}^{\infty}}\dfrac{\left[\left(1/\delta-y\right)\beta^{-1/\gamma}\right]^{i}}{i!}\Gamma\left(\alpha+\dfrac{\left(r+i\right)}{\gamma}\right)
\]
\[
=\dfrac{\beta^{-s/\gamma}}{\Sigma_{\Gamma}\left(\dfrac{1}{\gamma},\alpha,\dfrac{\beta^{-1/\gamma}}{\delta}\right)}\dfrac{y^{r-1}}{\Gamma\left(r\right)}\Sigma_{\Gamma}\left(\dfrac{1}{\gamma},\alpha+\dfrac{r}{\gamma},\left(\dfrac{1}{\delta}-y\right)\beta^{-1/\gamma}\right).\:\blacksquare
\]
}{\small\par}

\subsubsection*{{\small A.9 Expression for $\boldsymbol{f_{Y\mid r,\alpha,\beta,\gamma,\delta}^{\left(\textrm{HG}\Sigma\Sigma^{\prime}\left(r\right)\right)}\left(y\right)}$}}

\noindent{\small Let
\[
f_{Y\mid r,\alpha,\beta,\gamma,\delta}^{\left(\textrm{HG}\Sigma\Sigma^{\prime}\left(r\right)\right)}\left(y\right)={\displaystyle \int_{0}^{1}}f_{Y\mid r=s,\theta}^{\left(\Gamma\right)}\left(y\right)g_{\theta\mid\alpha,\beta,\gamma,\delta}^{\left(\textrm{HG}\Sigma\Gamma\right)}\left(\theta\right)d\theta
\]
\[
={\displaystyle \int_{0}^{\infty}}\dfrac{y^{r-1}e^{-y/\theta}}{\Gamma\left(r\right)\theta^{r}}\dfrac{\gamma\beta^{\alpha}}{\Sigma_{\Gamma}\left(\dfrac{1}{\gamma},\alpha,\dfrac{\beta^{-1/\gamma}}{\delta}\right)}\theta^{\gamma\alpha-1}\exp\left(-\beta\theta^{\gamma}\right)e^{\theta/\delta}d\theta
\]
\[
=\dfrac{\gamma\beta^{\alpha}}{\Sigma_{\Gamma}\left(\dfrac{1}{\gamma},\alpha,\dfrac{\beta^{-1/\gamma}}{\delta}\right)}\dfrac{y^{r-1}}{\Gamma\left(r\right)}{\displaystyle \int_{0}^{\infty}}e^{-y/\theta}\theta^{\gamma\alpha-r-1}\exp\left(-\beta\theta^{\gamma}\right)e^{\theta/\delta}d\theta
\]
\[
=\dfrac{\gamma\beta^{\alpha}}{\Sigma_{\Gamma}\left(\dfrac{1}{\gamma},\alpha,\dfrac{\beta^{-1/\gamma}}{\delta}\right)}\dfrac{y^{r-1}}{\Gamma\left(r\right)}{\displaystyle \int_{0}^{\infty}}\left[{\displaystyle \sum_{j=0}^{\infty}}\dfrac{\left(-y/\theta\right)^{j}}{j!}\right]\theta^{\gamma\alpha-r-1}\exp\left(-\beta\theta^{\gamma}\right)e^{\theta/\delta}d\theta
\]
\[
=\dfrac{\gamma\beta^{\alpha}}{\Sigma_{\Gamma}\left(\dfrac{1}{\gamma},\alpha,\dfrac{\beta^{-1/\gamma}}{\delta}\right)}\dfrac{y^{r-1}}{\Gamma\left(r\right)}{\displaystyle \sum_{j=0}^{\infty}}\left[\dfrac{\left(-y\right)^{j}}{j!}{\displaystyle \int_{0}^{\infty}}\theta^{\gamma\alpha-r-j-1}\exp\left(-\beta\theta^{\gamma}\right)e^{\theta/\delta}d\theta\right].
\]
Substituting $\vartheta=\theta^{\gamma}$ into each integral in the
infinite series then yields
\[
f_{Y\mid r,\alpha,\beta,\gamma,\delta}^{\left(\textrm{HG}\Sigma\Sigma^{\prime}\left(r\right)\right)}\left(y\right)=\dfrac{\gamma\beta^{\alpha}}{\Sigma_{\Gamma}\left(\dfrac{1}{\gamma},\alpha,\dfrac{\beta^{-1/\gamma}}{\delta}\right)}\dfrac{y^{r-1}}{\Gamma\left(r\right)}{\displaystyle \sum_{j=0}^{\infty}}\left[\dfrac{\left(-y\right)^{j}}{j!}{\displaystyle \int_{0}^{\infty}}\vartheta^{\alpha-\left(r+j+1\right)/\gamma}e^{-\beta\vartheta}\exp\left(\left(\dfrac{1}{\delta}\right)\vartheta^{1/\gamma}\right)\left(\dfrac{1}{\gamma}\right)\vartheta^{1/\gamma-1}d\vartheta\right]
\]
\[
=\dfrac{\beta^{\alpha}}{\Sigma_{\Gamma}\left(\dfrac{1}{\gamma},\alpha,\dfrac{\beta^{-1/\gamma}}{\delta}\right)}\dfrac{y^{r-1}}{\Gamma\left(r\right)}{\displaystyle \sum_{j=0}^{\infty}}\left[\dfrac{\left(-y\right)^{j}}{j!}{\displaystyle \int_{0}^{\infty}}\vartheta^{\alpha-\left(r+j\right)/\gamma-1}e^{-\beta\vartheta}\exp\left(\left(\dfrac{1}{\delta}\right)\vartheta^{1/\gamma}\right)d\vartheta\right]
\]
\[
=\dfrac{\beta^{\alpha}}{\Sigma_{\Gamma}\left(\dfrac{1}{\gamma},\alpha,\dfrac{\beta^{-1/\gamma}}{\delta}\right)}\dfrac{y^{s-1}}{\Gamma\left(s\right)}{\displaystyle \sum_{j=0}^{\infty}}\left\{ \dfrac{\left(-y\right)^{j}}{j!}{\displaystyle \int_{0}^{\infty}}\vartheta^{\alpha-\left(s+j\right)/\gamma-1}e^{-\beta\vartheta}\left[{\displaystyle \sum_{i=0}^{\infty}}\dfrac{\left(1/\delta\right)^{i}}{i!}\vartheta^{i/\gamma}\right]d\vartheta\right\} 
\]
\[
=\dfrac{\beta^{\alpha}}{\Sigma_{\Gamma}\left(\dfrac{1}{\gamma},\alpha,\dfrac{\beta^{-1/\gamma}}{\delta}\right)}\dfrac{y^{r-1}}{\Gamma\left(r\right)}{\displaystyle \sum_{j=0}^{\infty}}\left\{ \dfrac{\left(-y\right)^{j}}{j!}{\displaystyle \sum_{i=0}^{\infty}}\left[\dfrac{\left(1/\delta\right)^{i}}{i!}{\displaystyle \int_{0}^{\infty}}\vartheta^{\alpha-\left(r+j-i\right)/\gamma-1}e^{-\beta\vartheta}d\vartheta\right]\right\} 
\]
\[
=\dfrac{\beta^{\alpha}}{\Sigma_{\Gamma}\left(\dfrac{1}{\gamma},\alpha,\dfrac{\beta^{-1/\gamma}}{\delta}\right)}\dfrac{y^{r-1}}{\Gamma\left(r\right)}{\displaystyle \sum_{j=0}^{\infty}}\left[\dfrac{\left(-y\right)^{j}}{j!}{\displaystyle \sum_{i=0}^{\infty}}\dfrac{\left(1/\delta\right)^{i}}{i!}\dfrac{\Gamma\left(\alpha-\dfrac{\left(r+j-i\right)}{\gamma}\right)}{\beta^{\alpha-\left(r+j-i\right)/\gamma}}\right]
\]
\[
=\dfrac{\beta^{r/\gamma}}{\Sigma_{\Gamma}\left(\dfrac{1}{\gamma},\alpha,\dfrac{\beta^{-1/\gamma}}{\delta}\right)}\dfrac{y^{r-1}}{\Gamma\left(r\right)}{\displaystyle \sum_{j=0}^{\infty}}\left[\dfrac{\left(-y\beta^{1/\gamma}\right)^{j}}{j!}{\displaystyle \sum_{i=0}^{\infty}}\dfrac{\left(\beta^{-1/\gamma}/\delta\right)^{i}}{i!}\Gamma\left(\alpha-\dfrac{\left(r+j-i\right)}{\gamma}\right)\right]
\]
\[
=\dfrac{\beta^{r/\gamma}}{\Sigma_{\Gamma}\left(\dfrac{1}{\gamma},\alpha,\dfrac{\beta^{-1/\gamma}}{\delta}\right)}\dfrac{y^{r-1}}{\Gamma\left(r\right)}{\displaystyle \sum_{j=0}^{\infty}}\dfrac{\left(-y\beta^{1/\gamma}\right)^{j}}{j!}\Sigma_{\Gamma}\left(\dfrac{1}{\gamma},\alpha-\dfrac{\left(r+j\right)}{\gamma},\dfrac{\beta^{-1/\gamma}}{\delta}\right).\:\blacksquare
\]
}{\small\par}

\subsection*{{\small A.10 Proof of Theorem 3.1}}

\noindent{\small For part (a), the uniqueness of $g_{q\mid r}\left(q\right)$
follows immediately from Theorem 1.1.}{\small\par}

{\small For part (b), we first note that $\int_{0}^{1}f_{X\mid s,q}^{\left(\textrm{NB}\right)}\left(x\right)g_{q\mid s}\left(q\right)dq=f_{X}\left(x\right)$
implies
\[
\int_{0}^{1}\dfrac{\Gamma\left(x+s\right)}{\Gamma\left(s\right)\Gamma\left(x+1\right)}\left(1-\omega\right)^{s}\omega^{x}g_{q\mid s}\left(\omega\right)d\omega=\int_{0}^{1}\dfrac{\Gamma\left(x+r\right)}{\Gamma\left(r\right)\Gamma\left(x+1\right)}\left(1-\omega\right)^{r}\omega^{x}g_{q\mid r}\left(\omega\right)d\omega,\qquad\textrm{(A2)}
\]
or equivalently,
\[
\mathcal{B}\left(x,r\right)\int_{0}^{1}\left(1-\omega\right)^{s}\omega^{x}g_{q\mid s}\left(\omega\right)d\omega=\mathcal{B}\left(x,s\right)\int_{0}^{1}\left(1-\omega\right)^{r}\omega^{x}g_{q\mid r}\left(\omega\right)d\omega
\]
\[
\Longleftrightarrow\int_{0}^{1}\eta^{x-1}\left(1-\eta\right)^{r-1}d\eta\int_{0}^{1}\left(1-\omega\right)^{s}\omega^{x}g_{q\mid s}\left(\omega\right)d\omega=\int_{0}^{1}\eta^{x-1}\left(1-\eta\right)^{s-1}d\eta\int_{0}^{1}\left(1-\omega\right)^{r}\omega^{x}g_{q\mid r}\left(\omega\right)d\omega
\]
\[
\Longleftrightarrow\int_{0}^{1}\int_{0}^{1}\dfrac{\left(1-\eta\right)^{r-1}}{\eta}\left(\eta\omega\right)^{x}\left(1-\omega\right)^{s}g_{q\mid s}\left(\omega\right)d\eta d\omega=\int_{0}^{1}\int_{0}^{1}\dfrac{\left(1-\eta\right)^{s-1}}{\eta}\left(\eta\omega\right)^{x}\left(1-\omega\right)^{r}g_{q\mid r}\left(\omega\right)d\eta d\omega.
\]
Substituting $q=\eta\omega$ into both sides of the above equation
yields
\[
\int_{0}^{1}\int_{0}^{\omega}q^{x-1}\left(1-\dfrac{q}{\omega}\right)^{r-1}\left(1-\omega\right)^{s}g_{q\mid s}\left(\omega\right)dqd\omega=\int_{0}^{1}\int_{0}^{\omega}q^{x-1}\left(1-\dfrac{q}{\omega}\right)^{s-1}\left(1-\omega\right)^{r}g_{q\mid r}\left(\omega\right)dqd\omega,
\]
which can be rewritten as
\[
\int_{0}^{1}q^{x-1}\int_{q}^{1}\dfrac{\left(\omega-q\right)^{r-1}\left(1-\omega\right)^{s}}{\omega^{r-1}}g_{q\mid s}\left(\omega\right)d\omega dq=\int_{0}^{1}q^{x-1}\int_{q}^{1}\dfrac{\left(\omega-q\right)^{s-1}\left(1-\omega\right)^{r}}{\omega^{s-1}}g_{q\mid r}\left(\omega\right)d\omega dq
\]
by reversing the order of integration.}{\small\par}

{\small Having isolated all factors not involving $\omega$ outside
the inner integral, we tentatively set
\[
\varphi\left(q\right)=\int_{q}^{1}\dfrac{\left(\omega-q\right)^{r-1}\left(1-\omega\right)^{s}}{\omega^{r-1}}g_{q\mid s}\left(\omega\right)d\omega=\int_{q}^{1}\dfrac{\left(\omega-q\right)^{s-1}\left(1-\omega\right)^{r}}{\omega^{s-1}}g_{q\mid r}\left(\omega\right)d\omega
\]
for all $q\in\left(0,1\right)$, $r\in\mathbb{Z}_{\geq1}$, and $s\in\left(r,\infty\right)$.
Successive differentiation of both integrals with respect to $q$
then gives
\[
\dfrac{d^{r}\varphi\left(q\right)}{dq^{r}}=\left(-1\right)^{r}\left(r-1\right)!\dfrac{\left(1-q\right)^{s}}{q^{r-1}}g_{q\mid s}\left(q\right)=\left(-1\right)^{r}\dfrac{\Gamma\left(s\right)}{\Gamma\left(s-r\right)}\int_{q}^{1}\dfrac{\left(\omega-q\right)^{s-r-1}\left(1-\omega\right)^{r}}{\omega^{s-1}}g_{q\mid r}\left(\omega\right)d\omega,
\]
which can be rearranged into the function given by (34). The validity
of this function then can be confirmed for all $\left(r,s\right)\in\mathbb{R}_{>0}^{2}$
such that $r<s$ by substitution into (A2). Since the right-hand side
of (34) is nonnegative for all $q\in\left(0,1\right)$, one can see
it represents a proper PDF, and thus is unique by Theorem 1.1.}{\small\par}

{\small For part (c), we rewrite (34) by replacing $\left(1-\omega\right)^{r}$
with its binomial expansion and then applying integration by parts
to each term in the resulting summation. This yields (35), whose validity
can be confirmed for all $\left(r,s\right)\in\mathbb{R}_{>0}^{2}$
(including $r>s$) by substitution into (A2). When $r>s$, the right-hand
side of (35) may or may not be nonnegative for $q\in\left(0,1\right)$.
If $g_{q\mid s}\left(q\right)$ is always nonnegative, then it represents
a proper PDF whose uniqueness is ensured by Theorem 1.1. Otherwise,
it constitutes a quasi-PDF that integrates to 1 over $q\in\left(0,1\right)$
but possesses some negative values on the interval. $\blacksquare$}{\small\par}

\subsection*{{\small A.11 Proof of Corollary 3.1.1}}

\noindent{\small First, set $r=1$ in (35) to obtain (36). Then consider
the limit of (36) as $q\rightarrow0^{+}$:
\[
\underset{q\downarrow0}{\lim}\:g_{q\mid s}\left(q\right)=\underset{q\downarrow0}{\lim}\:\dfrac{1}{\left(1-q\right)^{s}}\left[\left(2-s\right){\displaystyle \int_{q}^{1}}\left(\dfrac{\omega-q}{\omega}\right)^{s-1}g_{q\mid r=1}\left(\omega\right)d\omega\right.
\]
\[
\left.+\left(s-1\right){\displaystyle \int_{q}^{1}}\left(\dfrac{\omega-q}{\omega}\right)^{s-1}\left(\dfrac{1}{\omega}\right)g_{q\mid r=1}\left(\omega\right)d\omega-{\displaystyle \int_{q}^{1}}\left(\dfrac{\omega-q}{\omega}\right)^{s-1}\left(1-\omega\right)g_{q\mid r=1}^{\prime}\left(\omega\right)d\omega\right]
\]
\[
=2-s+\underset{q\downarrow0}{\lim}\:\left\{ \left(s-1\right){\displaystyle \int_{q}^{1}}\left(\dfrac{1}{\omega}\right)g_{q\mid r=1}\left(\omega\right)d\omega-\left[\left.\left(1-\omega\right)g_{q\mid r=1}\left(\omega\right)\right|_{q}^{1}+{\displaystyle \int_{q}^{1}}g_{q\mid r=1}\left(\omega\right)d\omega\right]\right\} 
\]
\[
=\underset{q\downarrow0}{\lim}\:\left[\left(s-1\right){\displaystyle \int_{q}^{1}}\left(\dfrac{1}{\omega}\right)g_{q\mid r=1}\left(\omega\right)d\omega-\left(s-1\right)+g_{q\mid r=1}\left(q\right)\right].\qquad\qquad\qquad\qquad\textrm{(A3)}
\]
Clearly, $g_{q\mid s}\left(q\right)$ is a quasi-PDF with negative
values in some neighborhood of 0 if and only if (A3) is negative;
or equivalently,
\[
\underset{q\downarrow0}{\lim}\:{\displaystyle \int_{q}^{1}}\left(\dfrac{1}{\omega}\right)g_{q\mid r=1}\left(\omega\right)d\omega\left[\dfrac{g_{q\mid r=1}\left(q\right)}{{\displaystyle \int_{q}^{1}}\left(\dfrac{1}{\omega}\right)g_{q\mid r=1}\left(\omega\right)d\omega}+s-1\right]<s-1.\qquad\qquad\qquad\textrm{(A4)}
\]
}{\small\par}

{\small If $\underset{q\downarrow0}{\lim}\:g_{q\mid r=1}\left(q\right)=0$
(which implies $\underset{q\downarrow0}{\lim}\:e_{q\mid r=1}\left(q\right)=1$)
then (A7) must hold because}\linebreak{}
{\small$\underset{q\downarrow0}{\lim}\:{\textstyle \int_{q}^{1}}\left(\tfrac{1}{\omega}\right)g_{q\mid r=1}\left(\omega\right)d\omega>1$.
Moreover, if $0<\underset{q\downarrow0}{\lim}\:g_{q\mid r=1}\left(q\right)<\infty$
(which implies $\underset{q\downarrow0}{\lim}\:e_{q\mid r=1}\left(q\right)\geq0$)
then (A4) holds because $\underset{q\downarrow0}{\lim}\:{\textstyle \int_{q}^{1}}\left(\tfrac{1}{\omega}\right)g_{q\mid r=1}\left(\omega\right)d\omega=\infty$.
Finally, if $\underset{q\downarrow0}{\lim}\:g_{q\mid r=1}\left(q\right)=\infty$
(implying $\underset{q\downarrow0}{\lim}\:{\textstyle \int_{q}^{1}}\left(\tfrac{1}{\omega}\right)g_{q\mid r=1}\left(\omega\right)d\omega=\infty$
as well), then (A4) is satisfied if
\[
\underset{q\downarrow0}{\lim}\:\left[\dfrac{g_{q\mid r=1}\left(q\right)}{{\displaystyle \int_{q}^{1}}\left(\dfrac{1}{\omega}\right)g_{q\mid r=1}\left(\omega\right)d\omega}+s-1\right]<0
\]
\[
\Longleftrightarrow\underset{q\downarrow0}{\lim}\:e_{q\mid r=1}\left(q\right)>s-1,
\]
where the last inequality follows from L'Hôpital's rule. $\blacksquare$}{\small\par}

\subsection*{{\small A.12 Proof of Corollary 3.1.2}}

\noindent{\small Substituting the right-hand side of (37) into
\[
f_{X}\left(x\right)=\int_{0}^{1}f_{X\mid r,q}^{\left(\textrm{NB}\right)}\left(x\right)g_{q\mid r}\left(q\right)dq
\]
immediately gives
\[
f_{X}\left(x\right)=\int_{0}^{1}\left[\int_{0}^{\infty}f_{X\mid\lambda}^{\left(\textrm{P}\right)}\left(x\right)g_{\lambda\mid r,\tfrac{q}{1-q}}^{\left(\Gamma\right)}\left(\lambda\right)d\lambda\right]g_{q\mid r}\left(q\right)dq
\]
\[
=\int_{0}^{\infty}f_{X\mid\lambda}^{\left(\textrm{P}\right)}\left(x\right)\left[\int_{0}^{1}g_{\lambda\mid r,\tfrac{q}{1-q}}^{\left(\Gamma\right)}\left(\lambda\right)g_{q\mid r}\left(q\right)dq\right]d\lambda
\]
\[
=\int_{0}^{\infty}f_{X\mid\lambda}^{\left(\textrm{P}\right)}\left(x\right)g_{\lambda}\left(\lambda\right)d\lambda,
\]
which implies (38). The uniqueness of $g_{\lambda}\left(\lambda\right)$
is ensured by the identifiability of Poisson mixtures (see Feller,
1943).$\:\blacksquare$}{\small\par}

\subsection*{{\small A.13 Proof of Theorem 3.2}}

\noindent{\small For part (a), the uniqueness of $g_{\theta\mid r}\left(\theta\right)$
follows immediately from Theorem 1.2.}{\small\par}

{\small For part (b), we first note that $\int_{0}^{\infty}f_{Y\mid s>r,\theta}^{\left(\Gamma\right)}\left(y\right)g_{\theta\mid s>r}\left(\theta\right)d\theta=f_{Y}\left(y\right)$
implies
\[
\int_{0}^{\infty}\dfrac{y^{s-1}e^{-y/\omega}}{\Gamma\left(s\right)\omega^{s}}g_{\theta\mid s>r}\left(\omega\right)d\omega=\int_{0}^{\infty}\dfrac{y^{r-1}e^{-y/\omega}}{\Gamma\left(r\right)\omega^{r}}g_{\theta\mid r}\left(\omega\right)d\omega,\qquad\qquad\qquad\textrm{(A5)}
\]
or equivalently,
\[
\Gamma\left(r\right)\int_{0}^{\infty}\dfrac{y^{s-1}e^{-y/\omega}}{\omega^{s}}g_{\theta\mid s>r}\left(\omega\right)d\omega=\Gamma\left(s\right)\int_{0}^{\infty}\dfrac{y^{r-1}e^{-y/\omega}}{\omega^{r}}g_{\theta\mid r}\left(\omega\right)d\omega
\]
\[
\Longleftrightarrow\int_{0}^{\infty}y^{r}\tau^{r-1}e^{-y\tau}d\tau\int_{0}^{\infty}\dfrac{y^{s-1}e^{-y/\omega}}{\omega^{s}}g_{\theta\mid s>r}\left(\omega\right)d\omega=\int_{0}^{\infty}y^{s}\tau^{s-1}e^{-y\tau}d\tau\int_{0}^{\infty}\dfrac{y^{r-1}e^{-y/\omega}}{\omega^{r}}g_{\theta\mid r}\left(\omega\right)d\omega
\]
\[
\Longleftrightarrow\int_{0}^{\infty}\int_{0}^{\infty}y^{r+s-1}e^{-y\left(\tau+1/\omega\right)}\dfrac{\tau^{r-1}}{\omega^{s}}g_{\theta\mid s>r}\left(\omega\right)d\tau d\omega=\int_{0}^{\infty}\int_{0}^{\infty}y^{r+s-1}e^{-y\left(\tau+1/\omega\right)}\dfrac{\tau^{s-1}}{\omega^{r}}g_{\theta\mid r}\left(\omega\right)d\tau d\omega.
\]
Substituting $\varsigma=\tau+\tfrac{1}{\omega}$ into both sides of
the above equation yields
\[
\int_{0}^{\infty}\int_{1/\omega}^{\infty}y^{r+s-1}e^{-y\varsigma}\dfrac{\left(\varsigma\omega-1\right)^{r-1}}{\omega^{r+s-1}}g_{\theta\mid s>r}\left(\omega\right)d\varsigma d\omega=\int_{0}^{\infty}\int_{1/\omega}^{\infty}y^{r+s-1}e^{-y/\theta}\dfrac{\left(\varsigma\omega-1\right)^{s-1}}{\omega^{r+s-1}}g_{\theta\mid r}\left(\omega\right)d\varsigma d\omega,
\]
which can be rewritten as
\[
\int_{0}^{\infty}y^{r+s-1}e^{-y\varsigma}\int_{1/\varsigma}^{\infty}\dfrac{\left(\varsigma\omega-1\right)^{r-1}}{\omega^{r+s-1}}g_{\theta\mid s>r}\left(\omega\right)d\omega d\varsigma=\int_{0}^{\infty}y^{r+s-1}e^{-y/\theta}\int_{1/\varsigma}^{\infty}\dfrac{\left(\varsigma\omega-1\right)^{s-1}}{\omega^{r+s-1}}g_{\theta\mid r}\left(\omega\right)d\omega d\varsigma
\]
by reversing the order of integration.}{\small\par}

{\small Having isolated all factors not involving $\omega$ outside
the inner integral, we tentatively set
\[
\varphi\left(\varsigma\right)=\int_{1/\varsigma}^{\infty}\dfrac{\left(\varsigma\omega-1\right)^{r-1}}{\omega^{r+s-1}}g_{\theta\mid s>r}\left(\omega\right)d\omega=\int_{1/\varsigma}^{\infty}\dfrac{\left(\varsigma\omega-1\right)^{s-1}}{\omega^{r+s-1}}g_{\theta\mid r}\left(\omega\right)d\omega
\]
for all $\varsigma\in\mathbb{R}_{>0}$, and consider integer values
$r\in\mathbb{Z}_{\geq1}$. Successive differentiation of the above
equation with respect to $\varsigma$ then gives
\[
\dfrac{d^{r}\varphi\left(\varsigma\right)}{d\varsigma^{r}}=\left(-1\right)^{r}\left(r-1\right)!\varsigma^{-r+1}\varsigma^{r+s-1}g_{\theta\mid s>r}\left(\dfrac{1}{\varsigma}\right)=\left(-1\right)^{r}\dfrac{\Gamma\left(s\right)}{\Gamma\left(s-r\right)}\int_{1/\varsigma}^{\infty}\dfrac{\left(\varsigma\omega-1\right)^{s-r-1}\omega^{r}}{\omega^{r+s-1}}g_{\theta\mid r}\left(\omega\right)d\omega,
\]
which implies
\[
\dfrac{\left(r-1\right)!}{\theta^{s-2}}g_{\theta\mid s>r}\left(\theta\right)=\dfrac{\Gamma\left(s\right)}{\Gamma\left(s-r\right)}\int_{\theta}^{\infty}\dfrac{\left(\dfrac{\omega}{\theta}-1\right)^{s-r-1}}{\omega^{s-1}}g_{\theta\mid r}\left(\omega\right)d\omega
\]
for $\theta=\tfrac{1}{\varsigma}$. This equality may be rearranged
to yield (39). The validity of this function then can be confirmed
for all $\left(r,s\right)\in\mathbb{R}_{>0}^{2}$ such that $r<s$
by substitution into (A5). Since the right-hand side of (39) is nonnegative
for all $\theta\in\mathbb{R}_{>0}$, we see it represents a proper
PDF, and so is unique by Theorem 1.2.}{\small\par}

{\small For part (c), we apply integration by parts to (39) to obtain
(40), whose validity can be confirmed for all $\left(r,s\right)\in\mathbb{R}_{>0}^{2}$
(including $r>s$) by substitution into (A5). When $r>s$, the right-hand
side of (40) may or may not be nonnegative for $\theta\in\mathbb{R}_{>0}$.
If $g_{\theta\mid s}\left(\theta\right)$ is always nonnegative, then
it represents a proper PDF whose uniqueness is ensured by Theorem
1.2. Otherwise, it constitutes a quasi-PDF that integrates to 1 over
$\theta\in\mathbb{R}_{>0}$ but possesses some negative values on
the interval. $\blacksquare$}{\small\par}

\subsection*{{\small A.14 Proof of Corollary 3.2}}

\noindent{\small First, set $r=1$ in (40) to obtain (41). Then consider
the limit of (41) as $\theta\rightarrow0^{+}$:
\[
\underset{\theta\downarrow0}{\lim}\:g_{\theta\mid s}\left(\theta\right)=\underset{\theta\downarrow0}{\lim}\:\left[\left(s-1\right){\displaystyle \int_{\theta}^{\infty}}\left(\dfrac{\omega-\theta}{\omega}\right)^{s-1}\left(\dfrac{1}{\omega}\right)g_{\theta\mid r=1}\left(\omega\right)d\omega-{\displaystyle \int_{\theta}^{\infty}}\left(\dfrac{\omega-\theta}{\omega}\right)^{s-1}g_{\theta\mid r=1}^{\prime}\left(\omega\right)d\omega\right]
\]
\[
=\underset{\theta\downarrow0}{\lim}\:\left[\left(s-1\right){\displaystyle \int_{\theta}^{\infty}}\left(\dfrac{1}{\omega}\right)g_{\theta\mid r=1}\left(\omega\right)d\omega-\left.g_{\theta\mid r=1}\left(\omega\right)\right|_{\theta}^{\infty}\right]
\]
\[
=\underset{\theta\downarrow0}{\lim}\:\left[\left(s-1\right){\displaystyle \int_{\theta}^{\infty}}\left(\dfrac{1}{\omega}\right)g_{\theta\mid r=1}\left(\omega\right)d\omega+g_{\theta\mid r=1}\left(\theta\right)\right].\qquad\qquad\qquad\qquad\textrm{(A6)}
\]
Clearly, $g_{\theta\mid s}\left(\theta\right)$ is a quasi-PDF with
negative values in some neighborhood of 0 if and only if (A6) is negative;
or equivalently,
\[
\underset{\theta\downarrow0}{\lim}\:{\displaystyle \int_{\theta}^{\infty}}\left(\dfrac{1}{\omega}\right)g_{\theta\mid r=1}\left(\omega\right)d\omega\left[\dfrac{g_{\theta\mid r=1}\left(\omega\right)}{{\displaystyle \int_{\theta}^{1}}\left(\dfrac{1}{\omega}\right)g_{\theta\mid r=1}\left(\omega\right)d\omega}+s-1\right]<0.
\]
}{\small\par}

{\small The remainder of the proof follows from arguments analogous
to those in the proof of Theorem 3.1.1, with $X\mid r,q$ replaced
by $Y\mid r,\theta$, $q\in\left(0,1\right)$ replaced by $\theta\in\mathbb{R}_{>0}$,
and limits taken as $q\downarrow0$ replaced by limits as $\theta\downarrow0$.}{\small\par}

\begin{singlespace}
\noindent\begin{flushright}
{\small$\blacksquare$}{\small\par}
\par\end{flushright}
\end{singlespace}

\begin{singlespace}
\noindent\pagebreak{}
\end{singlespace}

\begin{center}
{\Large\textbf{Appendix B}}{\Large\par}
\par\end{center}

\begin{singlespace}
\begin{center}
\textbf{Table B1. Frequency Models with Geometric Kernel ($\mathcal{G}^{\textrm{\textrm{HG}\ensuremath{\Sigma}}\mathcal{B}}$
and $\mathcal{F}^{\textrm{HGZY}}$)}
\par\end{center}
\end{singlespace}

\begin{center}
{\scriptsize{}%
\begin{tabular}{|c|c|c|c|}
\hline 
{\scriptsize$\mathcal{G}^{\textrm{\textrm{HG}\ensuremath{\Sigma}}\mathcal{B}}$
Member} & {\scriptsize Mixing PDF} & {\scriptsize$\mathcal{F}^{\textrm{HGZY}}$ Member} & {\scriptsize Mixture PMF}\tabularnewline
\hline 
\hline 
{\scriptsize$\Sigma\mathcal{B}\left(b,c\rightarrow0\right)$} & {\scriptsize$g_{q\mid b,c\rightarrow0}^{\left(\Sigma\mathcal{B}\right)}\left(q\right)=$} & {\scriptsize$\textrm{Zeta}\left(b\right)$} & {\scriptsize$f_{X\mid b}^{\left(\textrm{Z}\right)}\left(x\right)=$}\tabularnewline
{\scriptsize$\left[d=1,\:a=1/c\right]$} & {\scriptsize$\dfrac{\left(-\ln\left(q\right)\right)^{b}}{\zeta\left(b+1\right)\Gamma\left(b+1\right)\left(1-q\right)}$} &  & {\scriptsize$\dfrac{\left(x+1\right)^{-\left(b+1\right)}}{\zeta\left(b+1\right)}$}\tabularnewline
\hline 
{\scriptsize$\textrm{Beta}\left(a=1,b\right)$} & {\scriptsize$g_{q\mid a=1,b}^{\left(\mathcal{B}\right)}\left(q\right)=$} & {\scriptsize$\textrm{Yule}\left(b\right)$} & {\scriptsize$f_{X\mid b}^{\left(\textrm{Y}\right)}\left(x\right)=$}\tabularnewline
{\scriptsize$\left[d=c,\:c=1\right]$} & {\scriptsize$b\left(1-q\right)^{b-1}$} &  & {\scriptsize$b\mathcal{B}\left(x+1,b+1\right)$}\tabularnewline
\hline 
{\scriptsize$\textrm{Kumaraswamy}\left(b=1,c\right)$} & {\scriptsize$g_{q\mid b=1,c}^{\left(\textrm{K}\right)}\left(q\right)=$} & {\scriptsize$\textrm{Quadratic}\left(c\right)$} & {\scriptsize$f_{X\mid c}^{\left(\textrm{Q}\right)}\left(x\right)=$}\tabularnewline
{\scriptsize$\left[d=c,\:a=1\right]$} & {\scriptsize$cq^{c-1}$} &  & {\scriptsize$\dfrac{c}{\left(x+c\right)\left(x+c+1\right)}$}\tabularnewline
\hline 
{\scriptsize$\Sigma\mathcal{B}\left(b,c\right)$} & {\scriptsize$g_{q\mid b,c}^{\left(\Sigma\mathcal{B}\right)}\left(q\right)=$} & {\scriptsize$\textrm{ZY}\left(b,c\right)$} & {\scriptsize$f_{X\mid b,c}^{\left(\textrm{ZY}\right)}\left(x\right)=$}\tabularnewline
{\scriptsize$\left[d=1,\:a=1/c\right]$} & {\scriptsize$\dfrac{c}{\Sigma_{\mathcal{B}}\left(\dfrac{1}{c},\dfrac{1}{c},b\right)}\dfrac{\left(1-q^{c}\right)^{b}}{\left(1-q\right)}$} &  & {\scriptsize$\dfrac{\mathcal{B}\left(\dfrac{\left(x+1\right)}{c},b+1\right)}{\Sigma_{\mathcal{B}}\left(\dfrac{1}{c},\dfrac{1}{c},b\right)}$}\tabularnewline
\hline 
{\scriptsize$\textrm{Beta}\left(a,b\right)$} & {\scriptsize$g_{q\mid a,b}^{\left(\mathcal{B}\right)}\left(q\right)=$} & {\scriptsize$\textrm{Waring}\left(a,b\right)$} & {\scriptsize$f_{X\mid a,b}^{\left(\textrm{W}\right)}\left(x\right)=$}\tabularnewline
{\scriptsize$\left[d=c,\:c=1\right]$} & {\scriptsize$\dfrac{1}{\mathcal{B}\left(a,b\right)}q^{a-1}\left(1-q\right)^{b-1}$} &  & {\scriptsize$\dfrac{\mathcal{B}\left(a+x,b+1\right)}{\mathcal{B}\left(a,b\right)}$}\tabularnewline
\hline 
{\scriptsize$\textrm{Kumaraswamy}\left(b,c\right)$} & {\scriptsize$g_{q\mid b,c}^{\left(\textrm{K}\right)}\left(q\right)=$} & {\scriptsize$\textrm{K-Mix}\left(b,c\right)$} & {\scriptsize$f_{X\mid b,c}^{\left(\textrm{KM}\right)}\left(x\right)=$}\tabularnewline
{\scriptsize$\left[d=c,\:a=1\right]$} & {\scriptsize$bcq^{c-1}\left(1-q^{c}\right)^{b-1}$} &  & {\scriptsize$b\left[\mathcal{B}\left(\dfrac{x}{c}+1,b\right)-\mathcal{B}\left(\dfrac{\left(x+1\right)}{c}+1,b\right)\right]$}\tabularnewline
\hline 
{\scriptsize$\textrm{Generalized}$} & {\scriptsize$g_{q\mid a,b,c}^{\left(\textrm{G}\Sigma\mathcal{B}\right)}\left(q\right)=$} & {\scriptsize$\textrm{Generalized}$} & {\scriptsize$f_{X\mid a,b,c}^{\left(\textrm{GZY}\right)}\left(x\right)=$}\tabularnewline
{\scriptsize$\Sigma\mathcal{B}\left(a,b,c\right)$ $\left[d=1\right]$} & {\scriptsize$\dfrac{c}{\Sigma_{\mathcal{B}}\left(\dfrac{1}{c},a,b\right)}\dfrac{q^{ca-1}\left(1-q^{c}\right)^{b}}{\left(1-q\right)}$} & {\scriptsize$\textrm{ZY}\left(a,b,c\right)$} & {\scriptsize$\dfrac{\mathcal{B}\left(a+\dfrac{x}{c},b+1\right)}{\Sigma_{\mathcal{B}}\left(\dfrac{1}{c},a,b\right)}$}\tabularnewline
\hline 
{\scriptsize$\textrm{Generalized}$} & {\scriptsize$g_{q\mid a,b,c}^{\left(\textrm{G}\mathcal{B}1\right)}\left(q\right)=$} & {\scriptsize$\textrm{Generalized}$} & {\scriptsize$f_{X\mid a,b,c}^{\left(\textrm{GW}2\right)}\left(x\right)=$}\tabularnewline
{\scriptsize$\textrm{Beta 1}\left(a,b,c\right)$ $\left[d=c\right]$} & {\scriptsize$\dfrac{c}{\mathcal{B}\left(a,b\right)}q^{ca-1}\left(1-q^{c}\right)^{b-1}$} & {\scriptsize$\textrm{Waring 2}\left(a,b,c\right)$} & {\scriptsize$\dfrac{\mathcal{B}\left(a+\dfrac{x}{c},b\right)-\mathcal{B}\left(a+\dfrac{\left(x+1\right)}{c},b\right)}{\mathcal{B}\left(a,b\right)}$}\tabularnewline
\hline 
{\scriptsize$\textrm{Hyper-Generalized}$} & {\scriptsize$g_{q\mid a,b,c,d}^{\left(\textrm{HG}\Sigma\mathcal{B}\right)}\left(q\right)=$} & {\scriptsize$\textrm{Hyper-Generalized}$} & {\scriptsize$f_{X\mid a,b,c,d}^{\left(\textrm{HGZY}\right)}\left(x\right)=$}\tabularnewline
{\scriptsize$\Sigma\mathcal{B}\left(a,b,c,d\right)$} & {\scriptsize$\dfrac{c}{\Sigma_{\mathcal{B}}\left(\dfrac{d}{c},a,b\right)}\dfrac{q^{ca-1}\left(1-q^{c}\right)^{b}}{\left(1-q^{d}\right)}$} & {\scriptsize$\textrm{ZY}\left(a,b,c,d\right)$} & {\scriptsize$\dfrac{\Sigma_{\mathcal{B}}\left(\dfrac{d}{c},a+\dfrac{x}{c},b\right)-\Sigma_{\mathcal{B}}\left(\dfrac{d}{c},a+\dfrac{\left(x+1\right)}{c},b\right)}{\Sigma_{\mathcal{B}}\left(\dfrac{d}{c},a,b\right)}$}\tabularnewline
\hline 
\end{tabular}}{\scriptsize\par}
\par\end{center}

\newpage{}
\begin{center}
\textbf{Table B2. Frequency Models with Geometric Kernel ($\mathcal{G}^{\textrm{\textrm{CHG}\ensuremath{\Sigma\textrm{B}}}}$
and $\mathcal{F}^{\textrm{HGZY}^{\prime}}$)}
\par\end{center}

\begin{center}
{\scriptsize{}%
\begin{tabular}{|c|c|c|c|}
\hline 
{\scriptsize$\mathcal{G}^{\textrm{\textrm{CHG}\ensuremath{\Sigma}}\mathcal{B}}$
Member} & {\scriptsize Mixing PDF} & {\scriptsize$\mathcal{F}^{\textrm{HGZY}^{\prime}}$ Member} & {\scriptsize Mixture PMF}\tabularnewline
\hline 
\hline 
{\scriptsize$\textrm{Compl. }\Sigma\mathcal{B}\left(b,c\rightarrow0\right)$} & {\scriptsize$g_{q\mid b,c\rightarrow0}^{\left(\textrm{C}\Sigma\mathcal{B}\right)}\left(1-q\right)=$} & {\scriptsize$\textrm{Zeta Prime}\left(b\right)$} & {\scriptsize$f_{X\mid b}^{\left(\textrm{Z}^{\prime}\right)}\left(x\right)=$}\tabularnewline
{\scriptsize$\left[d=1,\:a=1/c\right]$} & {\scriptsize$\dfrac{\left(-\ln\left(1-q\right)\right)^{b}}{\zeta\left(b+1\right)\Gamma\left(b+1\right)q}$} &  & {\scriptsize$\widetilde{\sum}_{j=0}^{x}\dfrac{\zeta\left(b+1,j+2\right)}{\zeta\left(b+1\right)}\:^{*,\:\dagger}$}\tabularnewline
\hline 
{\scriptsize$\textrm{Compl. Beta}\left(a=1,b\right)$} & {\scriptsize$g_{q\mid a=1,b}^{\left(\textrm{C}\mathcal{B}\right)}\left(q\right)=$} & {\scriptsize$\textrm{Yule Prime}\left(b\right)$} & {\scriptsize$f_{X\mid b}^{\left(\textrm{Y}^{\prime}\right)}\left(x\right)=$}\tabularnewline
{\scriptsize$\left[d=c,\:c=1\right]$} & {\scriptsize$bq^{b-1}$} &  & {\scriptsize$\widetilde{\sum}_{j=0}^{x}b\mathcal{B}\left(j+2,b\right)$}\tabularnewline
\hline 
{\scriptsize$\textrm{Compl. Kumaraswamy}\left(b=1,c\right)$} & {\scriptsize$g_{q\mid b=1,c}^{\left(\textrm{CK}\right)}\left(q\right)=$} & {\scriptsize$\textrm{Quadratic Prime}\left(c\right)$} & {\scriptsize$f_{X\mid c}^{\left(\textrm{Q}^{\prime}\right)}\left(x\right)=$}\tabularnewline
{\scriptsize$\left[d=c,\:a=1\right]$} & {\scriptsize$c\left(1-q\right)^{c-1}$} &  & {\scriptsize$\widetilde{\sum}_{j=0}^{x}\dfrac{c}{\left(j+c+1\right)}$}\tabularnewline
\hline 
{\scriptsize$\textrm{Compl. }\Sigma\mathcal{B}\left(b,c\right)$} & {\scriptsize$g_{q\mid b,c}^{\left(\textrm{C}\Sigma\mathcal{B}\right)}\left(p\right)=$} & {\scriptsize$\textrm{ZY Prime}\left(b,c\right)$} & {\scriptsize$f_{X\mid b,c}^{\left(\textrm{ZY}^{\prime}\right)}\left(x\right)=$}\tabularnewline
{\scriptsize$\left[d=1,\:a=1/c\right]$} & {\scriptsize$\dfrac{c}{\Sigma_{\mathcal{\mathcal{B}}}\left(\dfrac{1}{c},\dfrac{1}{c},b\right)}\dfrac{\left[1-\left(1-q\right)^{c}\right]^{b}}{q}$} &  & {\scriptsize$\widetilde{\sum}_{j=0}^{x}\dfrac{\Sigma_{\mathcal{B}}\left(\dfrac{1}{c},\dfrac{\left(j+2\right)}{c},b\right)}{\Sigma_{\mathcal{B}}\left(\dfrac{1}{c},\dfrac{1}{c},b\right)}$}\tabularnewline
\hline 
{\scriptsize$\textrm{Compl. Beta}\left(a,b\right)$} & {\scriptsize$g_{q\mid a,b}^{\left(\textrm{C}\mathcal{B}\right)}\left(q\right)=$} & {\scriptsize$\textrm{Waring Prime}\left(a,b\right)$} & {\scriptsize$f_{X\mid a,b}^{\left(\textrm{W}^{\prime}\right)}\left(x\right)=$}\tabularnewline
{\scriptsize$\left[d=c,\:c=1\right]$} & {\scriptsize$\dfrac{1}{\mathcal{B}\left(a,b\right)}\left(1-q\right)^{a-1}q^{b-1}$} &  & {\scriptsize$\widetilde{\sum}_{\ell=0}^{x}\dfrac{\mathcal{B}\left(a+\ell+1,b\right)}{\mathcal{B}\left(a,b\right)}$}\tabularnewline
\hline 
{\scriptsize$\textrm{Compl. Kumaraswamy}\left(b,c\right)$} & {\scriptsize$g_{q\mid b,c}^{\left(\textrm{CK}\right)}\left(q\right)=$} & {\scriptsize$\textrm{K-Mix Prime}\left(b,c\right)$} & {\scriptsize$f_{X\mid b,c}^{\left(\textrm{KM}^{\prime}\right)}\left(x\right)=$}\tabularnewline
{\scriptsize$\left[d=c,\:a=1\right]$} & {\scriptsize$bc\left(1-q\right)^{c-1}\left[1-\left(1-q\right)^{c}\right]^{b-1}$} &  & {\scriptsize$\widetilde{\sum}_{j=0}^{x}b\mathcal{B}\left(\dfrac{\left(j+c+1\right)}{c},b\right)$}\tabularnewline
\hline 
{\scriptsize$\textrm{Compl. Generalized}$} & {\scriptsize$g_{q\mid a,b,c}^{\left(\textrm{CG}\Sigma\mathcal{B}\right)}\left(q\right)=$} & {\scriptsize$\textrm{Generalized ZY}$} & {\scriptsize$f_{X\mid a,b,c}^{\left(\textrm{GZY}^{\prime}\right)}\left(x\right)=$}\tabularnewline
{\scriptsize$\Sigma\mathcal{B}\left(a,b,c\right)$ $\left[d=1\right]$} & {\scriptsize$\dfrac{c}{\Sigma_{\mathcal{B}}\left(\dfrac{1}{c},a,b\right)}\dfrac{\left(1-q\right)^{ca-1}\left[1-\left(1-q\right)^{c}\right]^{b}}{q}$} & {\scriptsize$\textrm{Prime}\left(a,b,c\right)$} & {\scriptsize$\widetilde{\sum}_{j=0}^{x}\dfrac{\Sigma_{\mathcal{B}}\left(\dfrac{1}{c},a+\dfrac{\left(j+1\right)}{c},b\right)}{\Sigma_{\mathcal{B}}\left(\dfrac{1}{c},a,b\right)}$}\tabularnewline
\hline 
{\scriptsize$\textrm{Compl. Generalized}$} & {\scriptsize$g_{q\mid a,b,c}^{\left(\textrm{CG}\mathcal{B}1\right)}\left(q\right)=$} & {\scriptsize$\textrm{Generalized Waring 2}$} & {\scriptsize$f_{X\mid a,b,c}^{\left(\textrm{GW}2^{\prime}\right)}\left(x\right)=$}\tabularnewline
{\scriptsize$\textrm{Beta 1}\left(a,b,c\right)$ $\left[d=c\right]$} & {\scriptsize$\dfrac{c}{\mathcal{B}\left(a,b\right)}\left(1-q\right)^{ca-1}\left[1-\left(1-q\right)^{c}\right]^{b-1}$} & {\scriptsize$\textrm{Prime}\left(a,b,c\right)$} & {\scriptsize$\widetilde{\sum}_{j=0}^{x}\dfrac{\mathcal{B}\left(a+\dfrac{\left(j+1\right)}{c},b\right)}{\mathcal{B}\left(a,b\right)}$}\tabularnewline
\hline 
{\scriptsize$\textrm{Compl. Hyper-Generalized}$} & {\scriptsize$g_{q\mid a,b,c,d}^{\left(\textrm{CHG}\Sigma\mathcal{B}\right)}\left(q\right)=$} & {\scriptsize$\textrm{Hyper-Generalized ZY}$} & {\scriptsize$f_{X\mid a,b,c,d}^{\left(\textrm{HGZY}^{\prime}\right)}\left(x\right)=$}\tabularnewline
{\scriptsize$\Sigma\mathcal{B}\left(a,b,c,d\right)$} & {\scriptsize$\dfrac{c}{\Sigma_{\mathcal{B}}\left(\dfrac{d}{c},a,b\right)}\dfrac{\left(1-q\right)^{ca-1}\left[1-\left(1-q\right)^{c}\right]^{b}}{\left[1-\left(1-q\right)^{d}\right]}$} & {\scriptsize$\textrm{Prime}\left(a,b,c,d\right)$} & {\scriptsize$\widetilde{\sum}_{j=0}^{x}\dfrac{\Sigma_{\mathcal{B}}\left(\dfrac{d}{c},a+\dfrac{\left(j+1\right)}{c},b\right)}{\Sigma_{\mathcal{B}}\left(\dfrac{d}{c},a,b\right)}$}\tabularnewline
\hline 
\end{tabular}}{\scriptsize\par}
\par\end{center}

\begin{onehalfspace}
{\scriptsize Notes:}{\scriptsize\par}

{\scriptsize$^{*}\:\zeta\left(\sigma,m\right)\equiv{\textstyle \sum_{k=0}^{\infty}}\left(k+m\right)^{-\sigma}$.}{\scriptsize\par}

{\scriptsize$^{\dagger}\:\widetilde{\sum}_{j=0}^{x}\tau_{j}\equiv{\textstyle \sum_{j=0}^{x}}\tbinom{x}{j}\left(-1\right)^{j}\tau_{j}$.}{\scriptsize\par}
\end{onehalfspace}

\subsection*{{\small\newpage}}
\begin{center}
\textbf{Table B3. Severity Models with Exponential Kernel ($\mathcal{G}^{\textrm{IHG}\Sigma\Gamma}$
and $\mathcal{F}^{\textrm{HG}\Sigma\Sigma}$)}
\par\end{center}

\begin{center}
{\scriptsize{}%
\begin{tabular}{|c|c|c|c|}
\hline 
{\scriptsize$\mathcal{G}^{\textrm{\textrm{IHG}}\Sigma\Gamma}$ Member} & {\scriptsize Mixing PDF} & {\scriptsize$\mathcal{F}^{\textrm{HG}\Sigma\Sigma}$ Member} & {\scriptsize Mixture PDF}\tabularnewline
\hline 
\hline 
{\scriptsize$\textrm{Inverse }\Sigma\Gamma\left(\beta,\gamma\rightarrow0\right)$} & {\scriptsize$\textrm{NOT APPLICABLE}$} & {\scriptsize$\Sigma\Sigma\left(\beta,\gamma\rightarrow0\right)$} & {\scriptsize$\textrm{NOT APPLICABLE}$}\tabularnewline
 & {\scriptsize$\left[\textrm{because }\gamma<1\right]$} &  & {\scriptsize$\left[\textrm{because }\gamma<1\right]$}\tabularnewline
\hline 
{\scriptsize$\textrm{Inverse}$} & {\scriptsize$g_{\theta\mid\alpha=1,\beta}^{\left(\textrm{I}\Gamma\right)}\left(\theta\right)=$} & {\scriptsize$\textrm{Pareto 2}\left(\alpha=1,\beta\right)$} & {\scriptsize$f_{Y\mid\alpha=1,\beta}^{\left(\textrm{P2}\right)}\left(y\right)=$}\tabularnewline
{\scriptsize$\textrm{Gamma}\left(\alpha=1,\dfrac{1}{\beta}\right)$} & {\scriptsize$\beta\dfrac{e^{-\beta/\theta}}{\theta^{2}}$} &  & {\scriptsize$\dfrac{\beta}{\left(\beta+y\right)^{2}}$}\tabularnewline
\hline 
{\scriptsize$\textrm{Inverse}$} & {\scriptsize$g_{\theta\mid\beta=1,\gamma}^{\left(\textrm{IWei}\right)}\left(\theta\right)=$} & {\scriptsize$\textrm{Wei-Mix}\left(\beta=1,\gamma\right)$} & {\scriptsize$f_{Y\mid\beta=1,\gamma}^{\left(\textrm{WeiM}\right)}\left(y\right)=$}\tabularnewline
{\scriptsize$\textrm{Weibull}\left(\beta=1,\gamma\right)$} & {\scriptsize$\gamma\dfrac{e^{-1/\theta^{\gamma}}}{\theta^{\gamma+1}}$} &  & {\scriptsize$\Sigma_{\Gamma}\left(\dfrac{1}{\gamma},1+\dfrac{1}{\gamma},-y\right)$}\tabularnewline
\hline 
{\scriptsize$\textrm{Inverse }\Sigma\Gamma\left(\beta,\gamma\right)$} & {\scriptsize$g_{\theta\mid\beta,\gamma}^{\left(\textrm{I}\Sigma\Gamma\right)}\left(\theta\right)=$} & {\scriptsize$\Sigma\Sigma\left(\beta,\gamma\right)$} & {\scriptsize$f_{Y\mid\beta,\gamma}^{\left(\Sigma\Sigma\right)}\left(y\right)=$}\tabularnewline
{\scriptsize$\left[\delta=1,\:\alpha=1/\gamma\right]$} & {\scriptsize$\dfrac{\gamma\beta^{1/\gamma}}{\Sigma_{\Gamma}\left(\dfrac{1}{\gamma},\dfrac{1}{\gamma},\beta^{-1/\gamma}\right)}\dfrac{e^{-\beta/\theta^{\gamma}}e^{1/\theta}}{\theta^{2}}$} &  & {\scriptsize$\dfrac{\Sigma_{\Gamma}\left(\dfrac{1}{\gamma},\dfrac{2}{\gamma},\left(1-y\right)\beta^{-1/\gamma}\right)}{\beta^{1/\gamma}\Sigma_{\Gamma}\left(\dfrac{1}{\gamma},\dfrac{1}{\gamma},\beta^{-1/\gamma}\right)}$}\tabularnewline
\hline 
{\scriptsize$\textrm{Inverse Gamma}\left(\alpha,\dfrac{1}{\beta}\right)$} & {\scriptsize$g_{\theta\mid\alpha,\beta}^{\left(\textrm{I}\Gamma\right)}\left(\theta\right)=$} & {\scriptsize$\textrm{Pareto 2}\left(\alpha,\beta\right)$} & {\scriptsize$f_{Y\mid\alpha,\beta}^{\left(\textrm{P2}\right)}\left(y\right)=$}\tabularnewline
{\scriptsize$\left[\delta\rightarrow\infty,\:\gamma=1\right]$} & {\scriptsize$\dfrac{\beta^{\alpha}}{\Gamma\left(\alpha\right)}\dfrac{e^{-\beta/\theta}}{\theta^{\alpha+1}}$} &  & {\scriptsize$\dfrac{\alpha}{\beta}\left(\dfrac{\beta}{\beta+y}\right)^{\alpha+1}$}\tabularnewline
\hline 
{\scriptsize$\textrm{Inverse Weibull}\left(\beta,\gamma\right)$} & {\scriptsize$g_{\theta\mid\beta,\gamma}^{\left(\textrm{IWei}\right)}\left(\theta\right)=$} & {\scriptsize$\textrm{Wei-Mix}\left(\beta,\gamma\right)$} & {\scriptsize$f_{Y\mid\beta,\gamma}^{\left(\textrm{WeiM}\right)}\left(y\right)=$}\tabularnewline
{\scriptsize$\left[\delta\rightarrow\infty,\:\alpha=1\right]$} & {\scriptsize$\gamma\beta\dfrac{e^{-\beta/\theta^{\gamma}}}{\theta^{\gamma+1}}$} &  & {\scriptsize$\dfrac{\Sigma_{\Gamma}\left(\dfrac{1}{\gamma},1+\dfrac{1}{\gamma},-y\beta^{-1/\gamma}\right)}{\beta^{1/\gamma}}$}\tabularnewline
\hline 
{\scriptsize$\textrm{Inverse Generalized}$} & {\scriptsize$g_{\theta\mid\alpha,\beta,\gamma}^{\left(\textrm{IG}\Sigma\Gamma\right)}\left(\theta\right)=$} & {\scriptsize$\textrm{Generalized}$} & {\scriptsize$f_{Y\mid\alpha,\beta,\tau}^{\left(\textrm{G}\Sigma\Sigma\right)}\left(y\right)=$}\tabularnewline
{\scriptsize$\Sigma\Gamma\left(\alpha,\beta,\gamma\right)$ $\left[\delta=1\right]$} & {\scriptsize$\dfrac{\gamma\beta^{\alpha}}{\Sigma_{\Gamma}\left(\dfrac{1}{\gamma},\alpha,\beta^{-1/\gamma}\right)}\dfrac{e^{-\beta/\theta^{\gamma}}e^{1/\theta}}{\theta^{\gamma\alpha+1}}$} & {\scriptsize$\Sigma\Sigma\left(\alpha,\beta,\gamma\right)$} & {\scriptsize$\dfrac{\Sigma_{\Gamma}\left(\dfrac{1}{\gamma},\alpha+\dfrac{1}{\gamma},\left(1-y\right)\beta^{-1/\gamma}\right)}{\beta^{1/\gamma}\Sigma_{\Gamma}\left(\dfrac{1}{\gamma},\alpha,\beta^{-1/\gamma}\right)}$}\tabularnewline
\hline 
{\scriptsize$\textrm{Inverse Generalized}$} & {\scriptsize$g_{\theta\mid\alpha,\beta,\gamma}^{\left(\textrm{IG}\Gamma\right)}\left(\theta\right)=$} & {\scriptsize$\textrm{Generalized}$} & {\scriptsize$f_{Y\mid\alpha,\beta,\gamma}^{\left(\textrm{GP2}\right)}\left(y\right)=$}\tabularnewline
{\scriptsize$\textrm{Gamma}\left(\alpha,\beta,\gamma\right)$ $\left[\delta\rightarrow\infty\right]$} & {\scriptsize$\dfrac{\gamma\beta^{\alpha}}{\Gamma\left(\alpha\right)}\dfrac{e^{-\beta/\theta^{\gamma}}}{\theta^{\gamma\alpha+1}}$} & {\scriptsize$\textrm{Pareto 2}\left(\alpha,\beta,\gamma\right)$} & {\scriptsize$\dfrac{\Sigma_{\Gamma}\left(\dfrac{1}{\gamma},\alpha+\dfrac{1}{\gamma},-y\beta^{-1/\gamma}\right)}{\beta^{1/\gamma}\Gamma\left(\alpha\right)}$}\tabularnewline
\hline 
{\scriptsize$\textrm{Inverse Hyper-Generalized}$} & {\scriptsize$g_{\theta\mid\alpha,\beta,\gamma,\delta}^{\left(\textrm{IHG}\Sigma\Gamma\right)}\left(\theta\right)=$} & {\scriptsize$\textrm{Hyper-Generalized}$} & {\scriptsize$f_{Y\mid\alpha,\beta,\gamma,\delta}^{\left(\textrm{HG}\Sigma\Sigma\right)}\left(y\right)=$}\tabularnewline
{\scriptsize$\Sigma\Gamma\left(\alpha,\beta,\gamma,\delta\right)$} & {\scriptsize$\dfrac{\gamma\beta^{\alpha}}{\Sigma_{\Gamma}\left(\dfrac{1}{\gamma},\alpha,\dfrac{\beta^{-1/\gamma}}{\delta}\right)}\dfrac{e^{1/\left(\delta\theta\right)}}{\theta^{\gamma\alpha+1}e^{\beta/\theta^{\gamma}}}$} & {\scriptsize$\Sigma\Sigma\left(\alpha,\beta,\gamma,\delta\right)$} & {\scriptsize$\dfrac{\Sigma_{\Gamma}\left(\dfrac{1}{\gamma},\alpha+\dfrac{1}{\gamma},\left(\dfrac{1}{\delta}-y\right)\beta^{-1/\gamma}\right)}{\beta^{1/\gamma}\Sigma_{\Gamma}\left(\dfrac{1}{\gamma},\alpha,\dfrac{\beta^{-1/\gamma}}{\delta}\right)}$}\tabularnewline
\hline 
\end{tabular}}{\scriptsize\par}
\par\end{center}

\subsection*{{\small\newpage}}
\begin{center}
\textbf{Table B4. Severity Models with Exponential Kernel ($\mathcal{G}^{\textrm{HG}\Sigma\Gamma}$
and $\mathcal{F}^{\textrm{HG}\Sigma\Sigma^{\prime}}$)}
\par\end{center}

\begin{center}
{\scriptsize{}%
\begin{tabular}{|c|c|c|c|}
\hline 
{\scriptsize$\mathcal{G}^{\textrm{\textrm{HG}}\Sigma\Gamma}$ Member} & {\scriptsize Mixing PDF} & {\scriptsize$\mathcal{F}^{\textrm{HG}\Sigma\Sigma^{\prime}}$ Member} & {\scriptsize Mixture PDF}\tabularnewline
\hline 
\hline 
{\scriptsize$\Sigma\Gamma\left(\beta,\gamma\rightarrow0\right)$} & {\scriptsize$\textrm{NOT APPLICABLE}$} & {\scriptsize$\Sigma\Sigma$} & {\scriptsize$\textrm{NOT APPLICABLE}$}\tabularnewline
 & {\scriptsize$\left[\textrm{because }\gamma<1\right]$} & {\scriptsize$\textrm{Prime}\left(\beta,\gamma\rightarrow0\right)$} & {\scriptsize$\left[\textrm{because }\gamma<1\right]$}\tabularnewline
\hline 
{\scriptsize$\textrm{Gamma}\left(\alpha=1,\dfrac{1}{\beta}\right)$} & {\scriptsize$g_{\theta\mid\alpha=1,\beta}^{\left(\Gamma\right)}\left(\theta\right)=$} & {\scriptsize$\textrm{Pareto 2}$} & {\scriptsize$f_{Y\mid\alpha=1,\beta}^{\left(\textrm{P2}^{\prime}\right)}\left(y\right)=$}\tabularnewline
 & {\scriptsize$\beta e^{-\beta\theta}$} & {\scriptsize$\textrm{Prime}\left(\alpha=1,\beta\right)$} & {\scriptsize$2\left(\beta-\dfrac{1}{\delta}\right)K_{0}\left(2\sqrt{\left(\beta-\dfrac{1}{\delta}\right)y}\right)$}\tabularnewline
\hline 
{\scriptsize$\textrm{Weibull}\left(\beta=1,\gamma\right)$} & {\scriptsize$g_{\theta\mid\beta=1,\gamma}^{\left(\textrm{Wei}\right)}\left(\theta\right)=$} & {\scriptsize$\textrm{Wei-Mix}$} & {\scriptsize$f_{Y\mid\beta=1,\gamma}^{\left(\textrm{WeiM}^{\prime}\right)}\left(y\right)=$}\tabularnewline
 & {\scriptsize$\gamma\theta^{\gamma-1}e^{-\theta^{\gamma}}$} & {\scriptsize$\textrm{Prime}\left(\beta=1,\gamma\right)$} & {\scriptsize${\textstyle \widetilde{\sum}_{j=0}^{\infty}}\Gamma\left(1-\dfrac{\left(j+1\right)}{\gamma}\right)$}\tabularnewline
\hline 
{\scriptsize$\Sigma\Gamma\left(\beta,\gamma\right)$} & {\scriptsize$g_{\theta\mid\beta,\gamma}^{\left(\Sigma\Gamma\right)}\left(\theta\right)=$} & {\scriptsize$\Sigma\Sigma$} & {\scriptsize$f_{Y\mid\beta,\gamma}^{\left(\Sigma\Sigma^{\prime}\right)}\left(y\right)=$}\tabularnewline
{\scriptsize$\left[\delta=1,\:\alpha=1/\gamma\right]$} & {\scriptsize$\dfrac{\gamma\beta^{1/\gamma}}{\Sigma_{\Gamma}\left(\dfrac{1}{\gamma},\dfrac{1}{\gamma},\beta^{-1/\gamma}\right)}\dfrac{e^{\theta}}{e^{\beta\theta^{\gamma}}}$} & {\scriptsize$\textrm{Prime}\left(\beta,\gamma\right)$} & {\scriptsize${\textstyle \widetilde{\sum}_{j=0}^{\infty}}\dfrac{\beta^{\left(j+1\right)/\gamma}\Sigma_{\Gamma}\left(\dfrac{1}{\gamma},-\dfrac{j}{\gamma},\beta^{-1/\gamma}\right)}{\Sigma_{\Gamma}\left(\dfrac{1}{\gamma},\dfrac{1}{\gamma},\beta^{-1/\gamma}\right)}$}\tabularnewline
\hline 
{\scriptsize$\textrm{Gamma}\left(\alpha,\dfrac{1}{\beta}\right)$} & {\scriptsize$g_{\theta\mid\alpha,\beta}^{\left(\Gamma\right)}\left(\theta\right)=$} & {\scriptsize$\textrm{Pareto 2}$} & {\scriptsize$f_{Y\mid\alpha,\beta}^{\left(\textrm{P2}^{\prime}\right)}\left(y\right)=$}\tabularnewline
{\scriptsize$\left[\delta\rightarrow\infty,\:\gamma=1\right]$} & {\scriptsize$\dfrac{\beta^{\alpha}}{\Gamma\left(\alpha\right)}\theta^{\alpha-1}e^{-\beta\theta}$} & {\scriptsize$\textrm{Prime}\left(\alpha,\beta\right)$} & {\scriptsize${\textstyle \widetilde{\sum}_{j=0}^{\infty}}\dfrac{\beta^{j+1}\Gamma\left(\alpha-j-1\right)}{\Gamma\left(\alpha\right)}$}\tabularnewline
\hline 
{\scriptsize$\textrm{Weibull}\left(\beta,\gamma\right)$} & {\scriptsize$g_{\theta\mid\beta,\gamma}^{\left(\textrm{Wei}\right)}\left(\theta\right)=$} & {\scriptsize$\textrm{Wei-Mix}$} & {\scriptsize$f_{Y\mid\beta,\gamma}^{\left(\textrm{WeiM}^{\prime}\right)}\left(y\right)=$}\tabularnewline
{\scriptsize$\left[\delta\rightarrow\infty,\:\alpha=1\right]$} & {\scriptsize$\gamma\beta\theta^{\gamma-1}e^{-\beta\theta^{\gamma}}$} & {\scriptsize$\textrm{Prime}\left(\beta,\gamma\right)$} & {\scriptsize${\textstyle \widetilde{\sum}_{j=0}^{\infty}}\beta^{\left(j+1\right)/\gamma}\Gamma\left(1-\dfrac{\left(j+1\right)}{\gamma}\right)$}\tabularnewline
\hline 
{\scriptsize$\textrm{Generalized}$} & {\scriptsize$g_{\theta\mid\alpha,\beta,\gamma}^{\left(\textrm{G}\Sigma\Gamma\right)}\left(\theta\right)=$} & {\scriptsize$\textrm{Generalized }\Sigma\Sigma$} & {\scriptsize$f_{Y\mid\alpha,\beta,\tau}^{\left(\textrm{G}\Sigma\Sigma^{\prime}\right)}\left(y\right)=$}\tabularnewline
{\scriptsize$\Sigma\Gamma\left(\alpha,\beta,\gamma\right)$ $\left[\delta=1\right]$} & {\scriptsize$\dfrac{\gamma\beta^{\alpha}}{\Sigma_{\Gamma}\left(\dfrac{1}{\gamma},\alpha,\beta^{-1/\gamma}\right)}\dfrac{\theta^{\gamma\alpha-1}e^{\theta}}{e^{\beta\theta^{\gamma}}}$} & {\scriptsize$\textrm{Prime}\left(\alpha,\beta,\tau\right)$} & {\scriptsize${\textstyle \widetilde{\sum}_{j=0}^{\infty}}\dfrac{\beta^{\left(j+1\right)/\gamma}\Sigma_{\Gamma}\left(\dfrac{1}{\gamma},\alpha-\dfrac{\left(j+1\right)}{\gamma},\beta^{-1/\gamma}\right)}{\Sigma_{\Gamma}\left(\dfrac{1}{\gamma},\alpha,\beta^{-1/\gamma}\right)}$}\tabularnewline
\hline 
{\scriptsize$\textrm{Generalized}$} & {\scriptsize$g_{\theta\mid\alpha,\beta,\gamma}^{\left(\textrm{G}\Gamma\right)}\left(\theta\right)=$} & {\scriptsize$\textrm{Generalized Pareto 2}$} & {\scriptsize$f_{Y\mid\alpha,\beta,\gamma}^{\left(\textrm{GP2}^{\prime}\right)}\left(y\right)=$}\tabularnewline
{\scriptsize$\textrm{Gamma}\left(\alpha,\beta,\gamma\right)$ $\left[\delta\rightarrow\infty\right]$} & {\scriptsize$\dfrac{\gamma\beta^{\alpha}}{\Gamma\left(\alpha\right)}\theta^{\gamma\alpha-1}e^{-\beta\theta^{\gamma}}$} & {\scriptsize$\textrm{Prime}\left(\alpha,\beta,\gamma\right)$} & {\scriptsize${\textstyle \widetilde{\sum}_{j=0}^{\infty}}\dfrac{\beta^{\left(j+1\right)/\gamma}\Gamma\left(\alpha-\dfrac{\left(j+1\right)}{\gamma}\right)}{\Gamma\left(\alpha\right)}$}\tabularnewline
\hline 
{\scriptsize$\textrm{Hyper-Generalized}$} & {\scriptsize$g_{\theta\mid\alpha,\beta,\gamma,\delta}^{\left(\textrm{HG}\Sigma\Gamma\right)}\left(\theta\right)=$} & {\scriptsize$\textrm{Hyper-Generalized }\Sigma\Sigma$} & {\scriptsize$f_{Y\mid\alpha,\beta,\gamma,\delta}^{\left(\textrm{HG}\Sigma\Sigma^{\prime}\right)}\left(y\right)=$}\tabularnewline
{\scriptsize$\Sigma\Gamma\left(\alpha,\beta,\gamma,\delta\right)$} & {\scriptsize$\dfrac{\gamma\beta^{\alpha}}{\Sigma_{\Gamma}\left(\dfrac{1}{\gamma},\alpha,\dfrac{\beta^{-1/\gamma}}{\delta}\right)}\dfrac{\theta^{\gamma\alpha-1}e^{\theta/\delta}}{e^{\beta\theta^{\gamma}}}$} & {\scriptsize$\textrm{Prime}\left(\alpha,\beta,\gamma,\delta\right)$} & {\scriptsize${\textstyle \widetilde{\sum}_{j=0}^{\infty}}\dfrac{\beta^{\left(j+1\right)/\gamma}\Sigma_{\Gamma}\left(\dfrac{1}{\gamma},\alpha-\dfrac{\left(j+1\right)}{\gamma},\dfrac{\beta^{-1/\gamma}}{\delta}\right)}{\Sigma_{\Gamma}\left(\dfrac{1}{\gamma},\alpha,\dfrac{\beta^{-1/\gamma}}{\delta}\right)}$}\tabularnewline
\hline 
\end{tabular}}{\scriptsize\par}
\par\end{center}

{\scriptsize Note:}{\scriptsize\par}

{\scriptsize$^{*}\:\widetilde{\sum}_{j=0}^{\infty}\tau_{j}\equiv{\textstyle \sum_{j=0}^{\infty}}\dfrac{\left(-y\right)^{j}}{j!}\tau_{j}$.}{\scriptsize\par}
\end{document}